\documentclass[prb,a4paper,twocolumn,floatfix,showpacs,showkeys,amsmath,amssymb,nobibnotes,unsortedaddress,superscriptaddress]{revtex4-2}
\setcitestyle{super}
\usepackage{graphicx}
\usepackage[dvipsnames]{xcolor}
\usepackage[utf8]{inputenc}
\usepackage{amsmath}
\usepackage{upgreek}
\usepackage{subfigure}
\usepackage[percent]{overpic}

\makeatletter
\begin{document}

\title{Charge Photogeneration in Non-Fullerene Organic Solar Cells: Influence of Excess Energy and Electrostatic Interactions}

\author{Maria Saladina}
\affiliation{Institut für Physik, Technische Universität Chemnitz, 09126 Chemnitz, Germany}

\author{Pablo Sim{\'o}n Marqu{\'e}s}
\affiliation{MOLTECH‐Anjou, CNRS UMR 6200, University of Angers, 49045 Angers, France}

\author{Anastasia Markina}
\affiliation{Max Planck Institute for Polymer Research, 55128 Mainz, Germany}

\author{Safakath Karuthedath}
\affiliation{King Abdullah University of Science and Technology (KAUST), KAUST Solar Center (KSC), Physical Sciences and Engineering Division (PSE), Material Science and Engineering Program (MSE), Thuwal 23955-6900, Kingdom of Saudi Arabia}

\author{Christopher W{\"o}pke}
\affiliation{Institut für Physik, Technische Universität Chemnitz, 09126 Chemnitz, Germany}

\author{Clemens G{\"o}hler}
\affiliation{Institut für Physik, Technische Universität Chemnitz, 09126 Chemnitz, Germany}

\author{Yue Chen}
\affiliation{Institut für Physik, Technische Universität Chemnitz, 09126 Chemnitz, Germany}

\author{Magali Allain}
\affiliation{MOLTECH‐Anjou, CNRS UMR 6200, University of Angers, 49045 Angers, France}

\author{Philippe Blanchard}
\affiliation{MOLTECH‐Anjou, CNRS UMR 6200, University of Angers, 49045 Angers, France}

\author{Cl{\'e}ment Cabanetos}
\affiliation{MOLTECH‐Anjou, CNRS UMR 6200, University of Angers, 49045 Angers, France}

\author{Denis Andrienko}
\affiliation{Max Planck Institute for Polymer Research, 55128 Mainz, Germany}

\author{Fr{\'e}d{\'e}ric Laquai}
\affiliation{King Abdullah University of Science and Technology (KAUST), KAUST Solar Center (KSC), Physical Sciences and Engineering Division (PSE), Material Science and Engineering Program (MSE), Thuwal 23955-6900, Kingdom of Saudi Arabia}

\author{Julien Gorenflot}
\affiliation{King Abdullah University of Science and Technology (KAUST), KAUST Solar Center (KSC), Physical Sciences and Engineering Division (PSE), Material Science and Engineering Program (MSE), Thuwal 23955-6900, Kingdom of Saudi Arabia}

\author{Carsten Deibel}
\affiliation{Institut für Physik, Technische Universität Chemnitz, 09126 Chemnitz, Germany}
\email{deibel@physik.tu-chemnitz.de}

\begin{abstract}

In organic solar cells, photogenerated singlet excitons form charge transfer (CT) complexes, which subsequently split into free charge carriers. 
Here, we consider the contributions of excess energy and molecular quadrupole moments to the charge separation process. We investigate charge photogeneration in two separate bulk heterojunction systems consisting of the polymer donor PTB7-Th and two non-fullerene acceptors, ITIC and h-ITIC.
CT state dissociation in these donor--acceptor systems is monitored by charge density decay dynamics obtained from transient absorption experiments. We study the electric field dependence of charge carrier generation at different excitation energies by time delayed collection field (TDCF) and sensitive steady-state photocurrent measurements.
Upon excitation below the optical gap free charge carrier generation becomes less field dependent with increasing photon energy, which challenges the view of charge photogeneration proceeding through energetically lowest CT states. We determine the average distance between electron--hole pairs at the donor--acceptor interface from empirical fits to the TDCF data. The delocalisation of CT states is larger in PTB7-Th:ITIC, the system with larger molecular quadrupole moment, indicating the sizeable effect of the electrostatic potential at the donor--acceptor interface on the dissociation of CT complexes.

\end{abstract}

\keywords{organic solar cells; non-fullerene acceptors; dissociation; excess energy; molecular quadrupole moment}

\maketitle

\section{Introduction}

The field of organic solar cells has shifted its focus to non-fullerene acceptors (NFA) since 2015.\cite{hou2018organic,wadsworth2019critical} Due to the fast development of new molecules and the variety of their chemical structures, understanding of the underlying photophysics in non-fullerene donor--acceptor (D--A) blends is far from complete. Knowledge of the factors that influence photogeneration in these systems is essential to move the field forward.
Charge generation in organic solar cells is a two-step process, owing to their low dielectric constant.\cite{nollau2000dissociation,morteani2004exciton,de2007geminate}
Following the charge transfer from donor to acceptor (or vice versa), a nonrelaxed charge-transfer (CT) state is formed. It undergoes thermalisation in the CT manifold and results in formation of a bound electron--hole pair with initial separation distance $r_0$. The electron--hole pair can subsequently relax to the lowest CT state, and recombine geminately to the ground state or dissociate into free charge carriers.\cite{yokoyama1981mechanism,deibel2010role}

The probability of CT dissociation $\eta_{\text{diss}}$ depends on the energy barrier between the CT and the charge-separated (CS) state.\cite{braun1984electric,rubel2008exact,wojcik2009accuracies} An activation energy for free charge formation is found to be consistently smaller than predicted for a Coulombically bound electron--hole pair,\cite{gerhard2015temperature,kurpiers2018probing,athanasopoulos2019binding,dong2019binding} although the reasons for this discrepancy are still under debate.
This barrier lowering has been explained by energetic disorder effects, interfacial dipoles, entropy etc.\cite{arkhipov2003exciton,gregg2011entropy,baranovskii2012calculating,castet2014charge,tscheuschner2015combined,hood2016entropy}
In fullerene-based organic solar cells, the energetics of CT and CS states are influenced by the quadrupole moment of the donor.\cite{schwarze2019impact} The electrostatic potential at the D--A interface was shown to reduce the dissociation energy required for charge separation.\cite{poelking2015design} The magnitude of the quadrupole moment depends largely on the molecular structure. In efficient non-fullerene organic solar cells, small molecule acceptors with the A--D--A structure usually possess a large quadrupole moment.\cite{markina2020rationale, alamoudi2018impact, perdigon2020barrierless} Surprisingly, its influence on charge separation has received little experimental attention up to now.

\begin{figure*}[t]
    \centering
    \includegraphics[trim = 2cm 19cm 2cm 2cm, clip, width=0.95\textwidth]{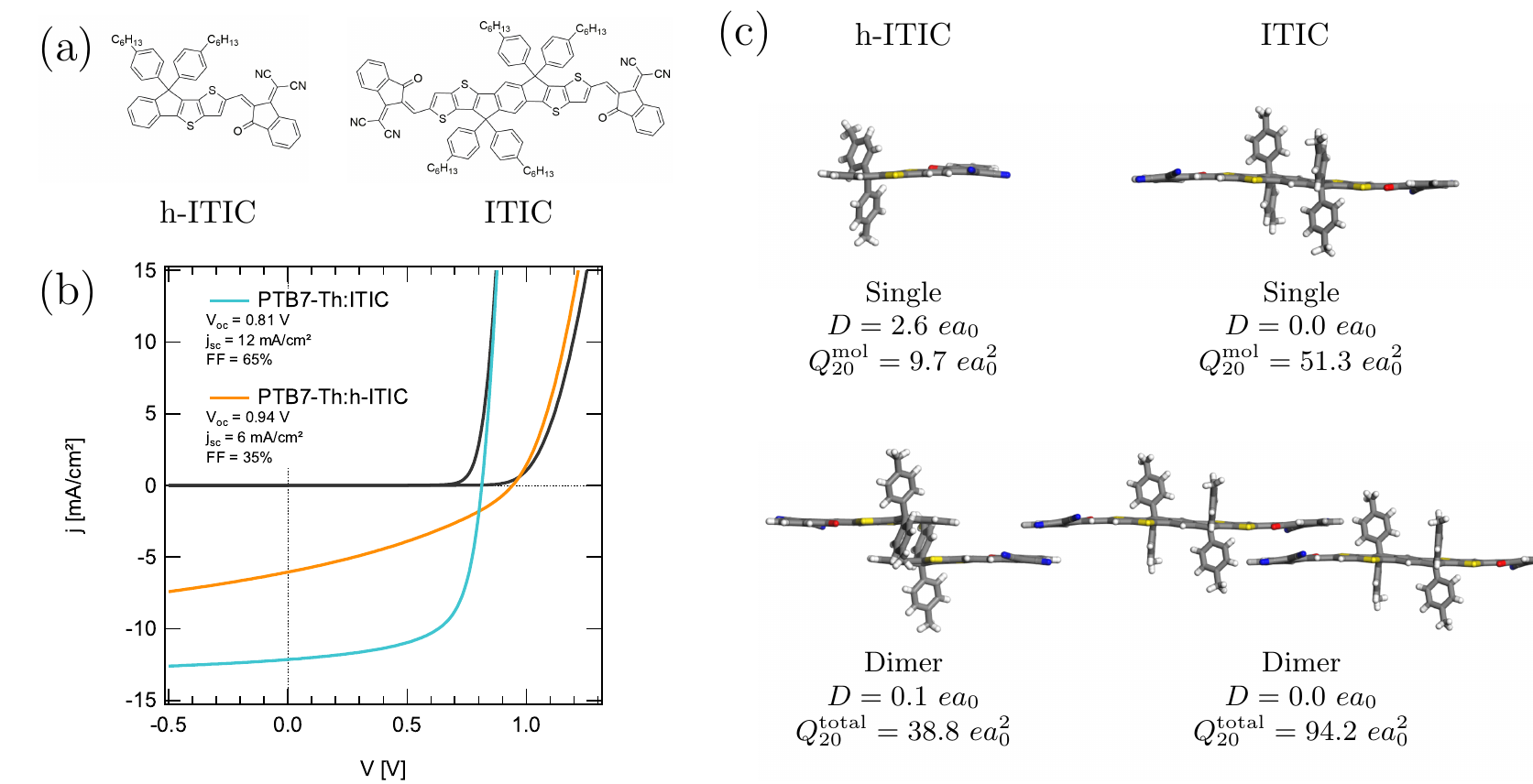}
    \caption{(a) Chemical structures of acceptor molecules. (b) $j(V)$ curves of PTB7-Th:ITIC and PTB7-Th:h-ITIC solar cells in the dark and under one sun illumination intensity. (c) Mutual arrangement of acceptor molecules in a single and dimer configurations and corresponding values of dipole ($D$) and quadrupole moments ($Q_{20}^{total}$--total quadrupole moment of a system, $Q_{20}^{mol}$--quadrupole moment per molecule). Some of the side chains were removed to simplify the representation.}
    \label{Figure1}
\end{figure*}

The binding energy of an electron--hole pair depends on its separation distance.\cite{onsager1934deviations,noolandi1979theory} Greater delocalisation\cite{deibel2009origin} and higher local mobility\cite{veldman2008compositional} of CT states were shown to be important for efficient photogeneration in polymer--fullerene based solar cells. 
It is plausible that during thermalisation in the CT manifold, the electron and hole with excess kinetic energy are driven apart to a larger thermalisation distance $r_0$. Thus excess energy may increase the probability for a CT state to overcome the potential barrier to CS states.\cite{pai1975onsager,yokoyama1981mechanism}

The influence of excess energy on photogeneration yield up to this date was investigated in donor--fullerene systems. Several authors have reported a positive effect of excess photon energy on the yield of free charge formation in these systems. 
Ohkita et al.\ demonstrated that the yield of free charge formation in polythiophene:fullerene blends increased with the free energy difference between the singlet and the CS state.\cite{ohkita2008charge} 
By employing pump--push photocurrent measurements to the polymer BTT-DPP blended with the fullerene acceptor PC\textsubscript{61}BM ([6,6]-phenyl-C\textsubscript{61}-butyric acid methyl ester), Dimitrov et al.\ could observe an increase in the photocurrent yield after applying an infrared push pulse providing bound CT states with excess energy.\cite{dimitrov2012energetic}
Transient absorption spectroscopy (TAS) experiments have shown that the free charge carrier generation yield is dependent on the excitation energy above the optical gap, hence on the excess energy of singlet excitons.\cite{dimitrov2012energetic,grancini2013hot} The results were corroborated by the energy dependent internal quantum efficiency (IQE) above the optical gap. It was proposed that higher energy singlet states transformed into more delocalised hot interfacial charge transfer states that could split more easily.

In contrast, several studies of photogeneration pathways in fullerene systems employing steady-state measurements have reported negligible influence of excess energy of both singlet excitons and charge transfer states on charge separation.\cite{lee2010charge,zusan2014crucial,kurpiers2018probing} 
By using a combination of time delayed collection field (TDCF), electroluminescence, photothermal deflection spectroscopy and sensitive EQE measurements, Vandewal et al.\ found no difference in the IQE or the field dependence of charge generation upon above and below the optical gap excitation in small-molecule and polymer-based organic solar cells.\cite{vandewal2014efficient}

In non-fullerene organic solar cells the influence of excess energy on the yield of photogeneration has been investigated in only one material system. Recently, Perdigón-Toro et al.\ studied a highly efficient PM6:Y6 non-fullerene system, in which free charge formation was probed by TDCF and found to proceed through relaxed CT states virtually independent of excitation energy.\cite{perdigon2020barrierless} However, at this point little is known about the generality of this result. 

In this work, we aim to determine the influence of photogeneration pathways and electrostatic potential at the D--A interface on the yield of free charge carrier formation in non-fullerene organic solar cells.
For this purpose, we employ TDCF measurements in two NFA systems at excitation energies above and below the optical gap, allowing us to assess the efficiency of electron--hole pair separation at the interface between donor and acceptor.
To support our analysis, we examine photogeneration and recombination in these systems with sensitive photocurrent and TAS measurements. 

Through careful assessment of the CT dissociation efficiency at various excitation energies, we demonstrate that non-fullerene organic solar cells benefit from photogeneration via nonrelaxed CT states exhibiting higher dissociation yield. 
Using the quantum chemical calculations and the kinetic model of photogeneration we relate the quadrupole moments of acceptor molecules to the CT binding energy.
We find that the solar cells comprising of the small molecule acceptor with a larger molecular quadrupole moment show higher CT separation yield due to the reduced activation energy for free charge carrier formation.

\newpage
\section{Results}

We focus on two non-fullerene acceptor systems, PTB7-Th:ITIC and PTB7-Th:h-ITIC. h-ITIC was designed as the dipolar analogue of the reference acceptor ITIC (Figure~\ref{Figure1}(a)). To that end, the new acceptor presents a push--pull $\pi$-conjugated structure, with an indenothienothiophene as electron-donating group connected to only one electron-withdrawing moiety, namely dicyanovynilindanone (DCI). In contrast, ITIC exhibits an indenodithienothiophene central core, which is functionalised with two DCI groups, showing lateral symmetry.\cite{lin2015electron} Therefore, both compounds open an excellent scenario for comparing two analogues, one with a quadrupole moment (ITIC) and its counterpart with a dipole moment (h-ITIC).
The polymer donor PTB7-Th was chosen to complement the absorption spectrum of h-ITIC molecule (Figure~S10), while PTB7-Th and ITIC absorb in the same region.

We perform gas-phase quantum chemical calculations to obtain the dipole and quadrupole moments of the ITIC and h-ITIC small molecule acceptors.
The mutual arrangement of molecules, dipole ($D$) and quadrupole moments ($Q_{20}^\text{total}$, $Q_{20}^\text{mol}$) for single and dimer configurations are shown in Figure~\ref{Figure1}(c).
ITIC molecules have zero charge and dipole moment, while their quadrupole moments are non-zero due to the A--D--A architecture. Both single and dimer states yield approximately the same quadrupole moment per molecule ($Q_{20}^\text{mol}\approx49.2\,ea_0^2$). 
h-ITIC acceptor is an asymmetric molecule with an A--D architecture. Electro-negative and electro-positive areas are distributed in such a way that they lead to the formation of a non-zero dipole moment. 
However, this dipole moment is cancelled out at a certain molecular orientation in a dimer, leading to the rise in a quadrupole moment value. 

Indeed, in Figure~\ref{Figure1}(c) we see that in the dimer configuration, the dipole moments are significantly reduced, while the quadrupole moment is non-zero ($Q_{20}^\text{total}\approx38.8\,ea_0^2$). Overall dipole moment of the unit cell is nearly zero $D = 0.04\,ea_0$ (cf.\ Figure~S12). However, the quadrupole moment of h-ITIC dimer is lower than the quadrupole moment of a single ITIC molecule while the effective van der Waals volume is higher. Using a lattice model approximation of point quadrupoles placed on a grid one can estimate electrostatic contribution in both cases of h-ITIC and ITIC. Lower quadrupole moments and higher grid space lead to the lower electrostatic potential in the case of h-ITIC (cf.\ Figure~S13). When blended with a donor polymer, this difference in electrostatic potential of the acceptors influences the device performance through charge separation at the D--A interface and charge transport.\cite{d2016electrostatic}

\begin{figure}[t]
    \centering
    \begin{overpic}[width=0.5\textwidth]{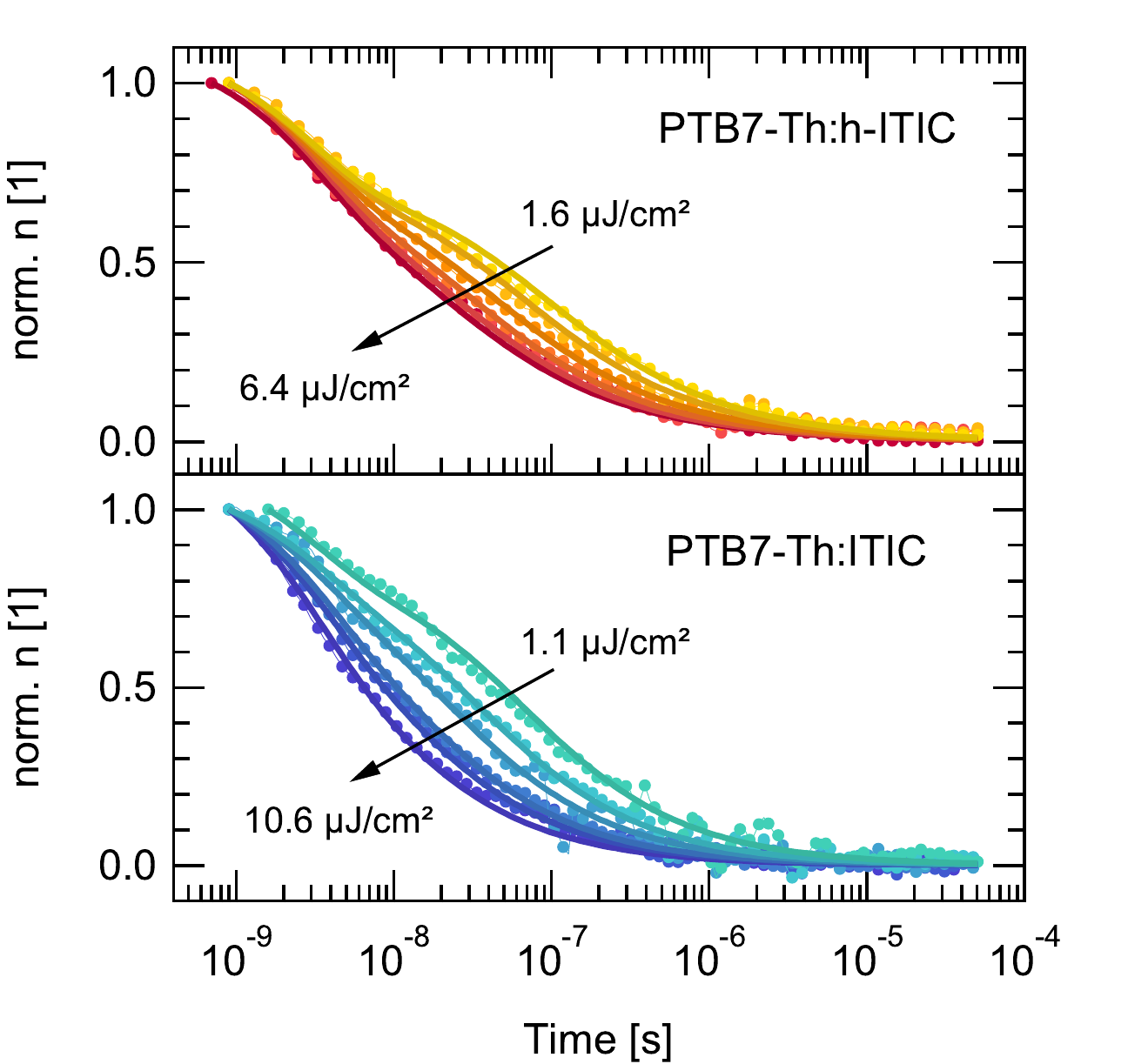}\put(79,63){\Large(a)}\put(79,25){\Large(b)}\end{overpic}
    \caption{Kinetics of charge carriers in (a) PTB7-Th:h-ITIC and (b) PTB7-Th:ITIC (symbols) at different fluences and global fits (solid lines) according to Equation~\eqref{eq_TA}.}
    \label{Figure2}
\end{figure}

Current--voltage characteristics of the devices in Figure~\ref{Figure1}(b) reveal that h-ITIC blend devices suffer from lower short-circuit current ($j_\text{sc}$) and fill factor ($FF$), but have higher open circuit voltage ($V_\text{oc}$) than ITIC devices. 
The extremely low fill factor in PTB7-Th:h-ITIC solar cells points to substantial geminate recombination and collection efficiency losses in this material system.

\begin{table}[b] \normalsize \begin{center}
\caption{Parameters extracted from the fits with Equation~\eqref{eq_TA} to the charge carrier density kinetics from TAS.} \label{table_TA}
\begin{tabular}{ |c|c|c| } \hline
Parameter & PTB7-Th:ITIC & PTB7-Th:h-ITIC\\\hline
$f$ & 0.82 & 0.71 \\
$\tau$ [ns] & 2.80 & 2.83 \\
$\lambda + 1$ & 2.40 & 2.64 \\
$\gamma$ [(cm$^3)^\lambda$s$^{-1}$] & $6.6\cdot10^{-17}$ & $4.2\cdot10^{-21}$ \\
$\gamma_\text{eff}$ [cm$^{3}$s$^{-1}$] & $1.7\cdot10^{-10}$ & $6.3\cdot10^{-11}$ \\\hline
\end{tabular}\end{center}\end{table}

We employ TAS to determine the relative importance of geminate and nongeminate recombination in ITIC and h-ITIC blend films in the absence of an external electric field. In Figure~\ref{Figure2} we plot the normalised decay dynamics of the charge carrier density $n$ at the excitation energy of 2.33\,eV for different pump fluences with pump delays from 1\,ns to 70\,$\upmu$s.\ The data is fitted globally assuming that the charge carrier population consists of bound electron--hole pairs and free charge carriers.\cite{howard2010effect}
\begin{equation} \label{eq_TA}
n(t) = \left(1-f\right) n_0 \exp{\left(- \frac{t}{\tau}\right)} + \left[\lambda \gamma t + \left(fn_0\right)^{-\lambda}\right]^{-1/\lambda},
\end{equation}
where $n_0$ is the initial charge carrier density, $\tau$ is the CT lifetime, $\gamma$ is the nongeminate recombination rate constant and $\lambda +1$ the recombination order.\cite{gohler2018nongeminate} The fractions of CS and CT states are denoted $f$ and $1-f$, respectively. The effective Langevin recombination rate constant was calculated according to $\gamma_\text{eff} = \gamma n^{\lambda -1}$ with $n = 1\cdot10^{16}$\,cm$^{-3}$. $n_0$ was estimated from TDCF measurements at similar conditions (cf.\ Figure~S15).

From the fits we obtain recombination parameters for the PTB7-Th:ITIC and PTB7-Th:h-ITIC blend films, listed in Table~\ref{table_TA}. The fraction of geminate recombination is 18\% and 29\%, respectively, with a typical CT lifetime of 2.8\,ns in both cases, thus the dissociation of electron--hole pairs at the D--A interface is less efficient in PTB7-Th:h-ITIC. The faster nongeminate recombination in PTB7-Th:ITIC blend is overcompensated by better collection efficiency due to higher and more balanced charge carrier mobilities,\cite{mackel2018determination,heiber2015encounter} as shown in Figures~S16 and S17. In PTB7-Th:h-ITIC the collection efficiency limits the fill factor, in addition to the primary limitation, the field dependent dissociation of CT states.

\textbf{Field and excess energy dependence of CT state dissociation efficiency.} 
In the following, we examine the influence of an external electric field on the CT state dissociation efficiency of our NFA systems. In addition, we aim to determine whether charge separation depends on the excitation energy of the CT state. 
If photogeneration proceeds through low energy relaxed CT states, we expect similar field dependence of CT dissociation yield for all the photon energies, as the hot CT states will thermally relax to the lowest lying CT state before separation. If on the other hand, dissociation of CT states is assisted by excess kinetic energy, then an applied external field would have a weaker effect on the hot than on the relaxed CT states.

We use the TDCF technique to investigate the field dependence of charge carrier photogeneration in PTB7-Th:ITIC and PTB7-Th:h-ITIC bulk heterojunction solar cells upon photoexcitation above and below the optical gap, down to the relaxed CT states. 
TDCF is a transient method with an optical pump and an electrical probe which is sensitive at pump fluences as low as several nJ\,cm$^{-2}$. This allows us to work with low photogenerated charge carrier densities during the experiment to minimise the influence of nongeminate recombination. The photogenerated charge carriers are collected at high reverse bias that further decreases the chance of their recombination. Consequently, when the system is kept at a certain prebias voltage during photogeneration, the field dependence of TDCF signal is governed by the changes in CT dissociation.

\begin{figure}[t]
    \centering
    \begin{overpic}[width=0.5\textwidth]{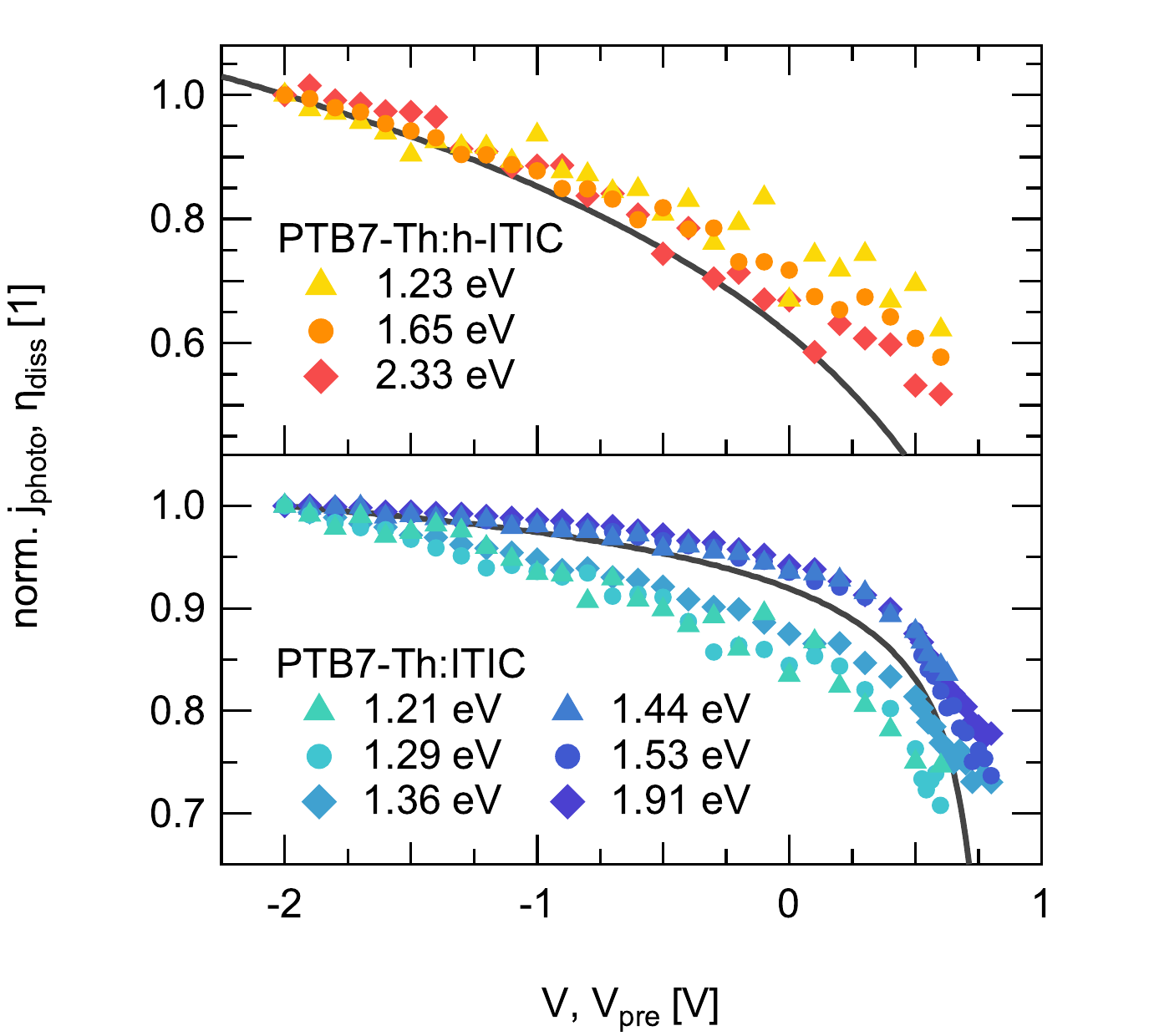}\put(79,77){\Large(a)}\put(79,42){\Large(b)}\end{overpic}
    \caption{Dissociation yield $\eta_{\text{diss}}$ from TDCF (symbols) as a function of prebias voltage $V_{\text{pre}}$ for different excitation photon energies for (a) PTB7-Th:h-ITIC and (b) PTB7-Th:ITIC, compared to $j_\text{photo}$ at one sun illumination intensity, normalised to the value at -2.0\,V (black solid lines).}
    \label{Figure3}
\end{figure}

TDCF measurements on PTB7-Th:ITIC were performed at six excitation photon energies $E_\text{ph}$ ranging from 1.21\,eV to 1.91\,eV. The lower limit of $E_\text{CT}$ in this NFA system is 1.19\,eV, as determined from the temperature dependent $V_\text{oc}$ measurements in Figure~S18. PTB7-Th:h-ITIC devices were excited with $E_\text{ph}$ of 1.23\,eV (lowest energy with detectable TDCF signal), 1.65\,eV and 2.33\,eV, last two corresponding to donor and acceptor absorption. The optimum experimental conditions for charge photogeneration studies were determined by applying different laser pulse fluences and delay times during TDCF measurements (see Figures~S19 and S20). The excitation density was adjusted to remain in a charge carrier density range of $1.5\cdot10^{14} - 3.5\cdot10^{15}$\,cm$^{-3}$ for all excitations (cf.\ Figure~S21). 

We measured the total extracted charge as a function of applied prebias $V_\text{pre}$ during photogeneration in PTB7-Th:ITIC and PTB7-Th:h-ITIC. 
Using the evaluation method employed by Zusan et al., \cite{zusan2014crucial} we quantify the efficiency of CT state dissociation in Figure~\ref{Figure3} as $\eta_{\text{diss}} = \text{EGE}(V_\text{pre})/\text{EGE}(-2.0\,\text{V})$, where EGE denotes the external generation efficiency. The dissociation is assumed to be equal to unity for PTB7-Th:ITIC at $V_\text{pre} = -2.0$\,V. 
In the PTB7-Th:h-ITIC blend the photocurrent does not saturate at -2.0\,V and the normalised EGE gives a higher limit of $\eta_{\text{diss}}$ in this system. EGE is calculated as the charge carrier density normalised to the incident photon density, which is estimated from the fluence at the sample at given wavelength.

The dissociation efficiency $\eta_{\text{diss}}$ in PTB7-Th:ITIC in Figure~\ref{Figure3}(b) follows the photocurrent at higher photon energies (1.44\,eV, 1.53\,eV and 1.91\,eV).
At low photon energies (1.21\,eV, 1.29\,eV and 1.36\,eV) the charge generation yield is lower and has a stronger field dependence, indicating that relaxed CT states in this system dissociate less efficiently. 
There is no gradual change from one regime to the other, the data rather forms two clusters. These results suggest that there is a limit to photogeneration in PTB7-Th:ITIC below the absorption edge, but this limit is at higher energy than the relaxed CT energy determined by linear extrapolation of $V_\text{oc}(T)$ to 0\,K.
In contrast, the electric field dependence of $\eta_{\text{diss}}$ in PTB7-Th:h-ITIC in Figure~\ref{Figure3}(a) does not change with excitation energy, indicating that charge separation in this system does not benefit from the excess photon energy.

Stronger bias dependence of the photocurrent in PTB7-Th:h-ITIC as compared to PTB7-Th:ITIC is a sign of a stronger field dependence of CT dissociation at the D--A interface. In the presence of an electric field, geminate recombination is suppressed and the number of free charge carriers in this system increases.
Comparing the values at a forward bias of 0.5\,V to the ones at a reverse bias of -2.0\,V, around 20\% of charge carriers are lost to geminate recombination in PTB7-Th:ITIC, while in PTB7-Th:h-ITIC geminate losses are much larger---more than 40\%, a sign of less efficient splitting of electron--hole pairs at the D--A interface. For both systems the lower photocurrent yield around the maximum power point is most likely originating from the transport resistance caused by imbalanced electron and hole mobilities (see Figures~S16 and S17), which is more pronounced in the PTB7-Th:h-ITIC device.
All in all, the results from TDCF confirm the findings from TAS and indicate that the difference in molecular structure of ITIC and h-ITIC has a decisive influence on the process of charge separation.

\textbf{Internal quantum efficiency.} 
The differences in the charge separation yield should be reflected in the IQE of the devices. 
Recently, Felekidis et al.\ have used EQE measured under short circuit and reverse bias to approximate IQE spectra of polymer--fullerene solar cells.\cite{felekidis2020role} We apply the same method and obtain excitation energy dependent IQE of PTB7-Th:ITIC and PTB7-Th:h-ITIC by normalising the photocurrent yield to the value at -2\,V and -4\,V, respectively. Due to low illumination intensities, the influence of nongeminate recombination on the change in photocurrent is minimised and the wavelength dependence of IQE at short circuit is predominantly governed by CT dissociation, as corroborated by the comparison to TDCF at delay times of only 5\,ns in Figure~S22.

\begin{figure}[t]
    \centering
    \includegraphics[width=0.5\textwidth]{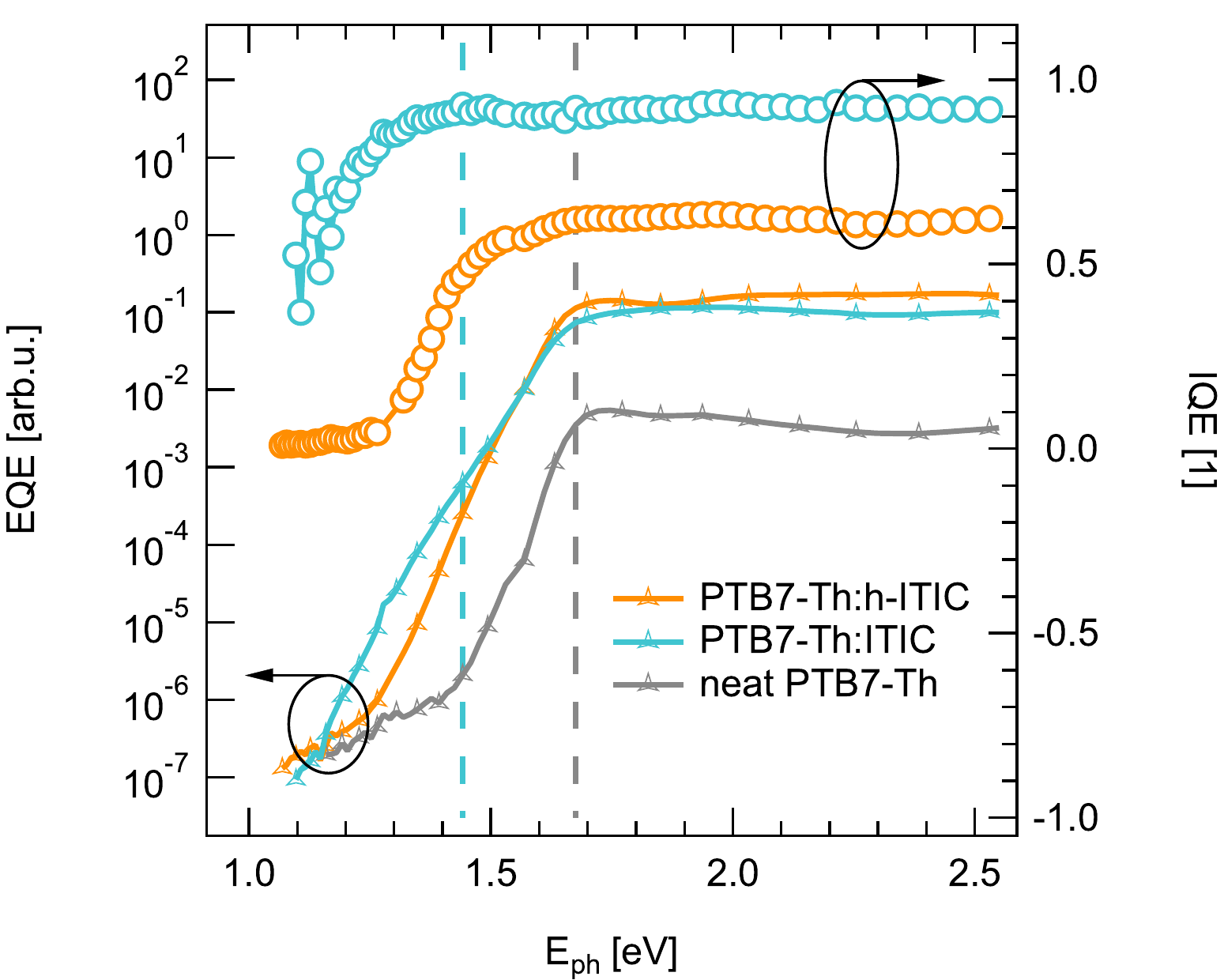}
    \caption{Sensitive EQE and IQE of PTB7-Th:ITIC and PTB7-Th:h-ITIC. EQE of neat PTB7-Th is plotted for comparison. IQE is calculated from the photocurrent normalised to the value at high reverse bias as in ref.\ \onlinecite{felekidis2020role}. IQE of h-ITIC blend is photon energy dependent below the optical gap of the donor 1.68\,eV (grey dashed line), IQE of ITIC blend is energy dependent below 1.44\,eV (cyan dashed line).}
    \label{Figure4}
\end{figure}

The IQE along with the sensitive EQE data is presented in Figure~\ref{Figure4}. 
Both PTB7-Th:ITIC and PTB7-Th:h-ITIC blends share the same optical gap, as shown in Figure~S10. 
For photon energies above this gap, the IQE remains constant for both blends with the mean value of 92\% for ITIC and 62\% for h-ITIC. 
In the subgap region, the field dependence of the IQE becomes stronger as the dissociation efficiency drops, verifying the findings from TDCF measurements. 
In the h-ITIC blend, the IQE becomes photon energy dependent roughly below the optical gap (grey dashed line), while in PTB7-Th:ITIC the IQE decline occurs below 1.44\,eV (cyan dashed line) as determined from the normalised IQE spectra at different bias voltages. This transition occurs 0.25\,eV above the relaxed CT energy of 1.19\,eV. 

The EQE of the h-ITIC blend in the subgap region resembles that of neat PTB7-Th, the former being slightly redshifted, probably due to larger energetic disorder in the blend.
In the EQE spectra of the ITIC blend, there is a broadening in the subgap region due to additional states, whether from the acceptor or the charge transfer states. The difference in wavelength dependence of the IQE of both systems below the optical gap suggests that a contribution from these states is the reason for better charge photogeneration in ITIC compared to h-ITIC. At the same time, these lower energy states could be the cause of lower $V_\text{oc}$ in PTB7-Th:ITIC devices. 

\section{Discussion}

We now consider the charge separation properties of the ITIC and h-ITIC small molecule acceptors by examining their molecular structure. 
In the absence of a dipole moment, quadrupole moments are the main contribution to the electrostatics of ITIC and h-ITIC molecules. 
The electrostatic potential induces a shift of energy levels of the highest occupied (HOMO) and the lowest unoccupied molecular orbitals (LUMO) of both acceptors. At the D--A interface the density of acceptor molecules is smaller than in the bulk, which gives rise to the bending of the electrostatic potential. Therefore, the shift of the HOMO and LUMO energy levels is smaller at the interface with a donor, and the energy gap between the CT and the CS states is reduced.\cite{poelking2015design} The A--D--A architecture of the ITIC molecules provide higher quadrupole moment than h-ITIC molecules. As was shown before for small molecule:C$_{60}$ bulk heterojunction solar cells,\cite{schwarze2019impact} a larger quadrupole moment leads to a lower energy barrier for free charge formation.  
The lower electric field dependence of $\eta_\text{diss}$ in PTB7-Th:ITIC blend is likely the consequence of such a quadrupole-induced barrier lowering, leading to higher $FF$ and $j_\text{sc}$ of the solar cell devices. Additionally, charge splitting is more favourable in the A--D--A architecture, due to charge delocalisation along the backbone reducing the Coulomb interaction. Therefore, ITIC molecules with their higher quadrupole moment per unit volume in comparison to h-ITIC are better candidates for charge photogeneration, as verified by our experimental results. 

We now discuss the influence of singlet exciton energy in excess of the optical gap on the photogeneration yield. Above the optical gap, the IQE and the field dependence of $\eta_{\text{diss}}$ are independent of the photon energy in both PTB7-Th:ITIC and PTB7-Th:h-ITIC. 
These findings indicate that photogeneration is not dependent on the singlet exciton excess energy, suggesting that free charge carriers are rather originating from a common intermediate precursor. Consistent with our reasoning, previous investigations of photogeneration pathways in fullerene\cite{lee2010charge,vandewal2014efficient} and non-fullerene\cite{perdigon2020barrierless} systems inferred negligible effect of excitation energy above the optical gap on photogeneration. The absence of $\eta_{\text{diss}}$ dependence on the singlet exciton excess energy supports the notion of two-step photogeneration and makes the concept of direct photogeneration from singlet excitons to free charge carriers rather improbable.

Next, we turn to the question whether charge photogeneration proceeds through the energetically lowest CT states or not. We argue that the distinct field dependence of $\eta_{\text{diss}}$ in PTB7-Th:ITIC upon excitation above and below 1.44\,eV in Figure~\ref{Figure3}(b) corresponds to two pools of relatively hot and relatively relaxed CT states. We expect the existence of a threshold energy level (in this case 1.44\,eV), above which dissociation of CT states into free charges is more efficient. The electric field dependence of $\eta_{\text{diss}}$ in PTB7-Th:ITIC is comparable to $j_\text{photo}$ only when exciting the system with photon energies above the threshold. This implies that the main contribution to photogeneration in PTB7-Th:ITIC is from the pool of CT states with energy in excess of the relaxed CT. At excitation below the threshold of 1.44\,eV, the pool of relaxed CT states dissociates into free charges less efficiently. Interestingly, the IQE becomes energy dependent also below 1.44\,eV. We note that the energy dependence of IQE is only obvious when plotted on the linear scale.

\begin{figure}[!ht] \centering
    \subfigure{\begin{overpic}[trim = 0cm 0cm 0cm 0cm, clip, width=0.5\textwidth]{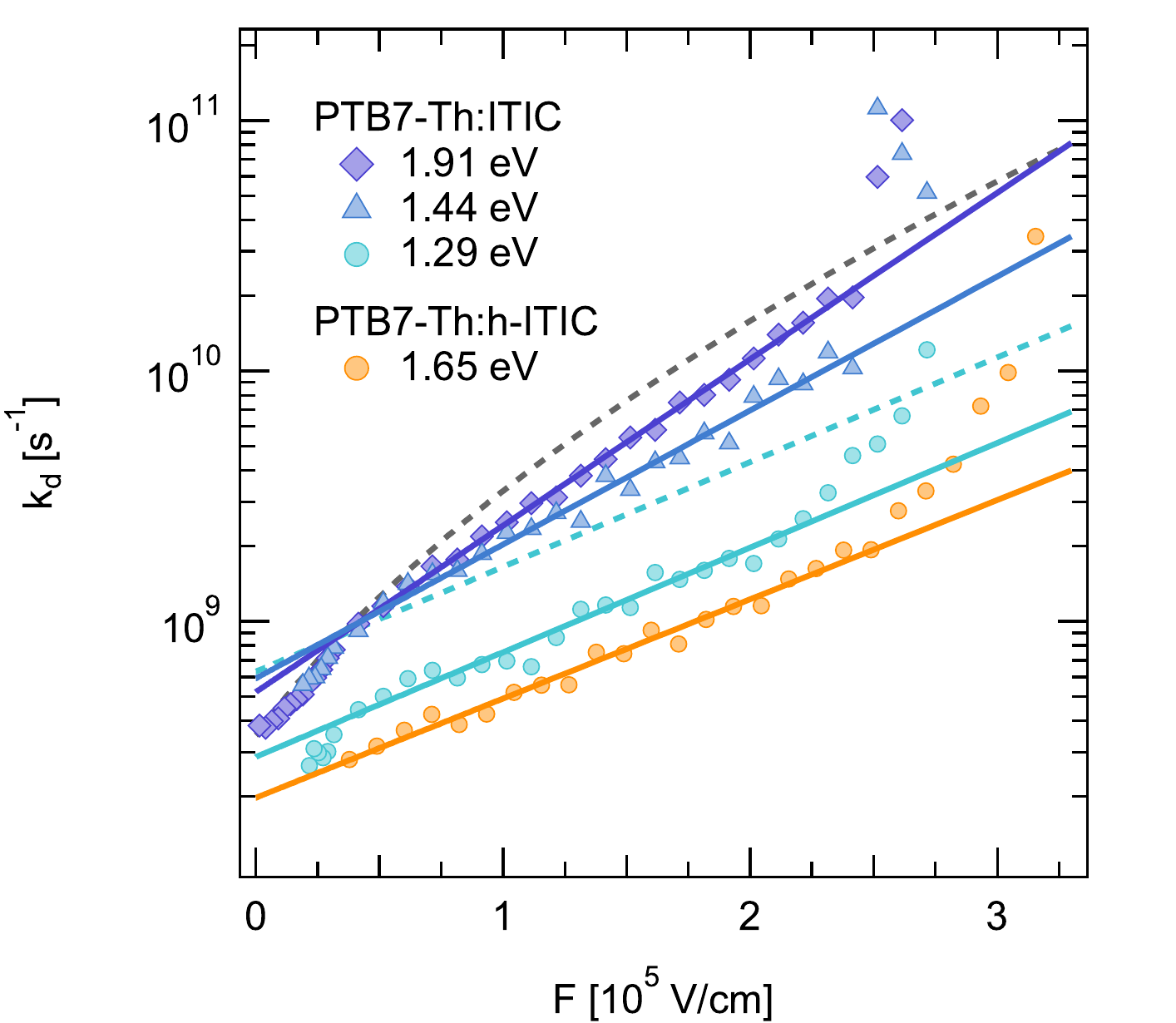}\put(0,80){\Large(a)}\end{overpic}}
    \subfigure{\begin{overpic}[trim = 0cm 0cm 0cm 0cm, clip, width=0.5\textwidth]{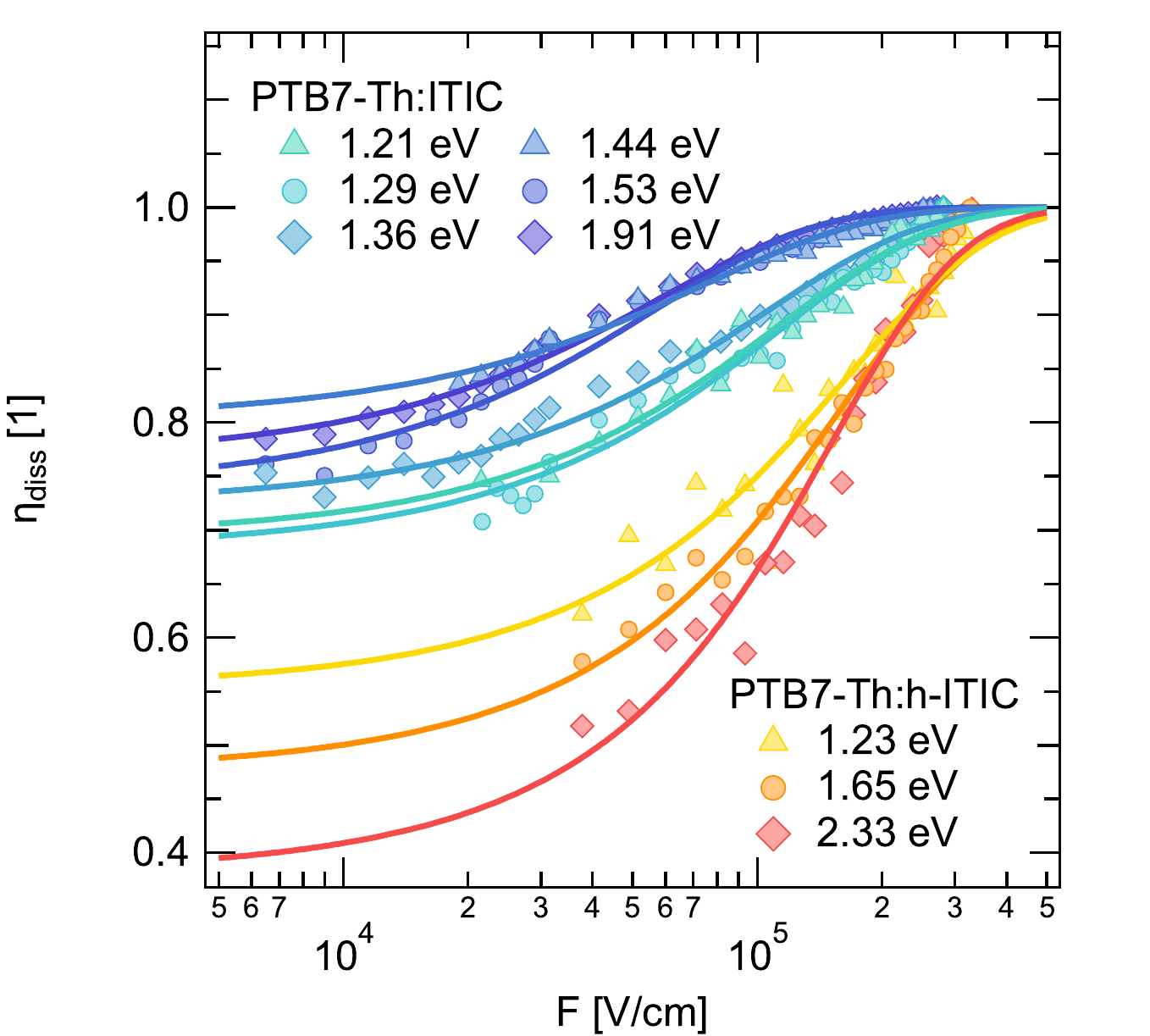}\put(0,83){\Large(b)}\end{overpic}}
    \caption{(a) $k_{d}(F)$ of PTB7-Th:ITIC at $E_\text{ph}$ of 1.29\,eV, 1.44\,eV and 1.91\,eV, and PTB7-Th:h-ITIC at $E_\text{ph}$ of 1.65\,eV (the rest omitted for clarity). The fits with Equation~\eqref{eq_model2} (solid lines) reveal an exponential dependence of $k_{d}(F)$. The fit to $k_{d}(F)$ of ITIC at 1.29\,eV is scaled to illustrate the difference in the slopes (cyan dashed line). The Onsager--Braun model has fixed slope with $r_\text{c} = \text{const}$ (grey dashed line). (b) $\eta_\text{diss}(F)$ of PTB7-Th:ITIC and PTB7-Th:h-ITIC, and fits to the data with Equations~\eqref{eq_model1} and \eqref{eq_model2}.} \label{Figure5}
\end{figure}

In contrast to PTB7-Th:ITIC, the electric field dependence of $\eta_{\text{diss}}$ in PTB7-Th:h-ITIC 
is independent of the excitation energy. 
In view of the low IQE and strongly field dependent charge separation, the photogeneration in the h-ITIC blend is likely originating from the pool of CT states below the threshold. Taken together, these findings indicate that the barrier for free charge formation is too high in this system, 
even for the states with energy in excess of the relaxed CT. Thus, we propose that the primary limitation of CT separation in NFA organic solar cells stems from the energy barrier between the CT and CS states, which is influenced by the electrostatic potential. Only when this barrier is low enough, as in the case of PTB7-Th:ITIC, separation of CT states can benefit from the excitation energy in excess of the threshold.

To further develop our understanding of the difference in separation of relaxed and hot CT states, we evaluate the parameters influencing $\eta_{\text{diss}}$ directly from TDCF data. Considering a kinetic competition between charge separation and recombination, the yield of CT dissociation is given by
\begin{equation} \label{eq_model1}
\eta_{\text{diss}}(F) = \frac{k_{d}(F)}{k_{f} + k_{d}(F)}
\end{equation}
where the dissociation rate constant $k_{d}$ depends strongly on the electric field $F$, while the rate constant of recombination to the ground state $k_{f}$ is assumed to be field independent. 
Due to the distribution of separation distances between electrons and holes at the interface, $\eta_{\text{diss}}$ represents the effective dissociation probability.\cite{kirchartz2009mobility}

We find that the field dependence of
\begin{equation}
k_{d}(F)=\frac{\eta_{\text{diss}}(F)}{1-\eta_{\text{diss}}(F)}\cdot k_{f}
\end{equation} 
can be described by an exponential function in the electric field range relevant to working conditions of organic solar cells, as indicated by a linear slope in a semi-log plot of $k_{d}$ vs $F$ in Figure~\ref{Figure5}(a). At higher fields, above $\approx2.5\cdot10^5$\,V\,cm$^{-1}$, $k_{d}(F)$ exhibits a second exponential behaviour as $\eta_{\text{diss}}(F)$ approaches unity.

Interestingly, the slope of $k_{d}(F)$ changes with excitation energy $E_\text{ph}$. 
The commonly used Onsager function for the field dependent increase of $k_{d}$ depends on the Coulomb radius $r_\text{c}$,\cite{onsager1934deviations,braun1984electric} i.e.\ the distance at which electron--hole binding energy is equal to the thermal energy. Thus, as the slope of $k_{d}$ changes, this function does not fit the data at all the photon energies with a fixed value of $r_\text{c}$.

To describe the electric field dependent increase of the CT dissociation yield we use a model that was previously implemented by Popovic et al.\ for fitting the field-induced fluorescence quenching in organic semiconductors.\cite{popovic1979electric} 
According to the model, the probability of dissociation is increased due to an applied electric field $F$ as
\begin{equation} \label{eq_model2}
k_{d}(F) = k_{d}(0)\exp{\left(\frac{e F r_0}{kT}\right)}.
\end{equation}
where $k_{d}(0)$ denotes the dissociation rate constant at zero field ($F=0$), $e$ the elementary charge, $k$ the Boltzmann constant and $T$ the temperature. 
The parameter $r_0$ represents a mean initial separation distance between an electron and a hole after thermalisation in the CT manifold and before dissociation into free charges. 

We plot $\eta_{\text{diss}}$ against field $F$ in Figure~\ref{Figure5}(b) and fit the data with Equations~\eqref{eq_model1} and \eqref{eq_model2}. From the fit we obtain the zero field dissociation rate constant $k_{d}(0)$ and thermalisation length $r_0$. The field is approximated as $F\approx(V-V_\text{oc})/d$ with $d$ being the active layer thickness. 
For $k_{f}$ we use the relation $k_{f} = \tau^{-1} - k_{d}(0)$, with $\tau$ determined from TAS (see Table~\ref{table_TA}) and $k_{d}(0)$ from fitting $\eta_{\text{diss}}(F)$ at lowest excitation pulse energies.
The geminate recombination rate constant $k_{f}$ obtained this way is $1.1\cdot10^8$\,s$^{-1}$ in the ITIC blend and $2.0\cdot10^8$\,s$^{-1}$ in the h-ITIC blend, respectively. 

\begin{figure}[t]
    \centering
    \begin{overpic}[width=0.5\textwidth]{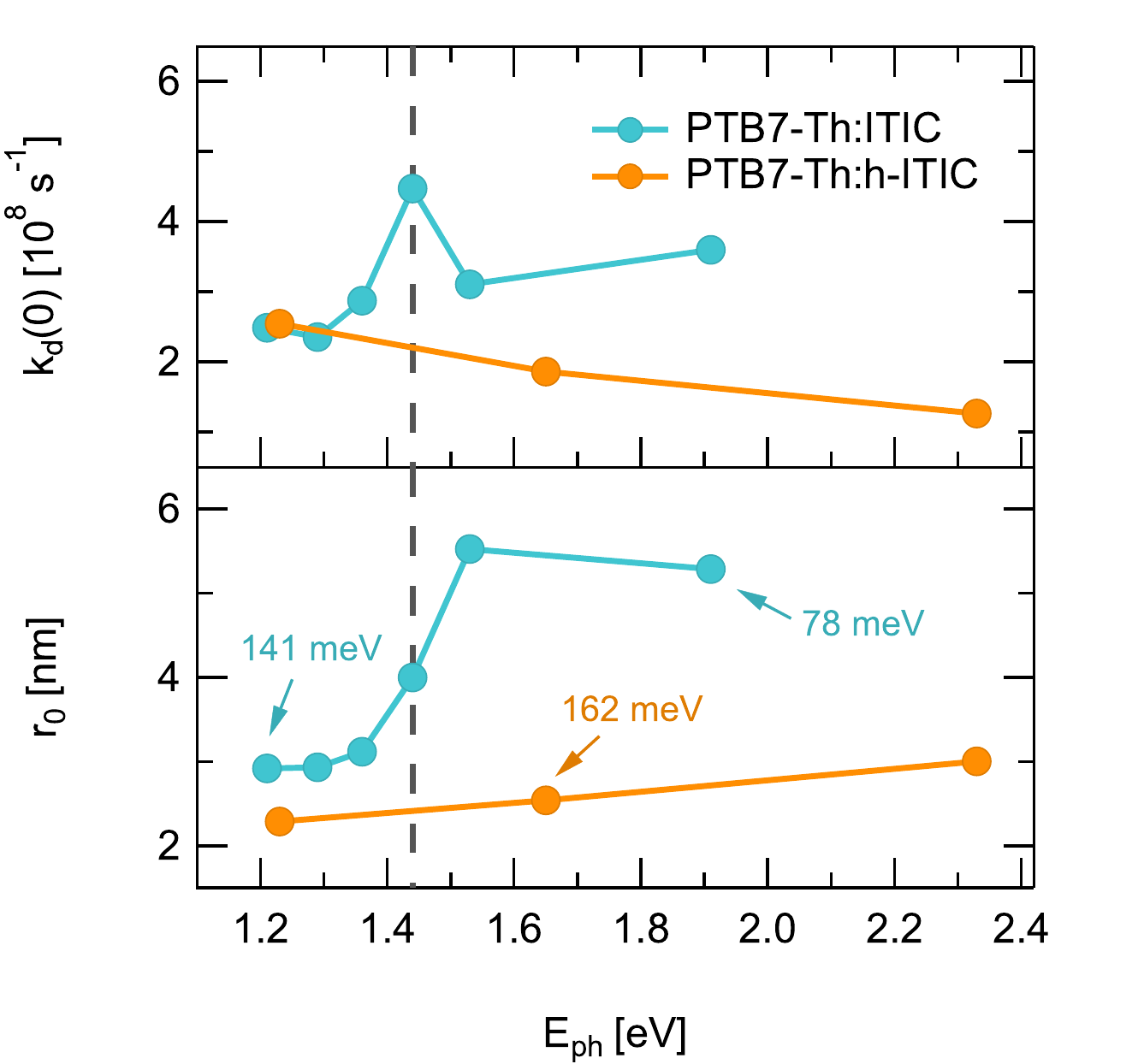}\put(22,79){\Large(a)}\put(22,44){\Large(b)}\end{overpic}
    \caption{(a) $k_{d}(0)$ and (b) $r_0$ of CT states in PTB7-Th:h-ITIC and PTB7-Th:ITIC obtained from the fits to $\eta_{\text{diss}}(F)$ with Equations~\eqref{eq_model1} and \eqref{eq_model2}, plotted against $E_\text{ph}$. Binding energies for several $r_0$ are calculated according to $E_0 = q^2/4\pi\epsilon\epsilon_0 r_0$ with $\epsilon = 3.5$, where $\epsilon \epsilon_0$ is the permittivity of the organic semiconductor material. The vertical dashed line indicates transition between sub- and supergap characteristic field dependence in PTB7-Th:ITIC.}
    \label{Figure6}
\end{figure}

Given that the effective CT lifetimes $\tau$ were similar for both systems, dissociation is expected to be slower in h-ITIC.
Indeed, the zero-field dissociation rate constant $k_{d}(0)$ in PTB7-Th:h-ITIC is on average lower than in PTB7-Th:ITIC (Figure~\ref{Figure6}(a)) and comparable to $k_{f}$. 
As seen from the slope in Figure~\ref{Figure5}(a), the electric field dependence of $k_{d}$ in h-ITIC and low energy subgap ITIC is similar, which implies comparable thermalisation length $r_0$.
At the lowest excitation energies (1.23\,eV and 1.21\,eV, respectively) dissociation in both donor--acceptor systems is equal as $k_{d}(0) = 2.5\cdot10^8$\,s$^{-1}$. 
The evidently much stronger electric field dependence of $\eta_\text{diss}$ in h-ITIC at this photon energy in Figure~\ref{Figure5}(b) is thus primarily a consequence of faster geminate recombination. 
Consistent with the kinetic Monte Carlo simulations performed by Felekidis et al.,\cite{felekidis2020role} the simultaneous blueshift and downshift of the IQE spectrum in h-ITIC relative to ITIC in Figure~\ref{Figure4} can also be explained by the higher $k_{f}$.

In the PTB7-Th:ITIC solar cell kinetic energy in excess of the relaxed CT states provides a stronger delocalisation of electron--hole pairs at the D--A interface.
As shown in Figure~\ref{Figure6}(b), the thermalisation distance $r_0$ between electron--hole pairs in this system, extracted from the fits to $\eta_{\text{diss}}(F)$, increases from 2.9\,nm at 1.21\,eV excitation energy to 5.5\,nm at excitation of 1.53\,eV. 
The latter agrees well with a recent report of electron--hole separation distance of 5.1\,nm in another NFA blend PM6:Y6.\cite{zhang2020delocalization} 
Due to the larger separation of electron--hole pairs in PTB7-Th:ITIC at higher photon energies, their Coulomb binding energy decreases from 141\,meV to 75\,meV. Thus, the activation energy for free charge carrier formation is reduced for the CT states with excess kinetic energy. 
This finding confirms previous reports of CT excess energy assisting charge separation in organic solar cells.\cite{ohkita2008charge,dimitrov2012energetic}

The drastic rise of $k_{d}(0)$ in PTB7-Th:ITIC at 1.44\,eV excitation, where the thermalisation length $r_0$ is $4.0$\,nm, might signify the transition between relaxed and hot CT dissociation in the PTB7-Th:ITIC system, supporting our previous argument about the existence of a threshold energy level. Excess energy is driving an electron--hole pair apart during its thermalisation in the CT manifold. 
At the threshold excitation energy, corresponding to a certain separation distance after the thermalisation process, the CT exciton binding is small enough to generate free charges barrierless.

\section{Conclusion}
In summary, we have investigated the influence of photon energy and molecular quadrupole moments on the yield of CT dissociation for two non-fullerene acceptor systems, PTB7-Th:ITIC and PTB7-Th:h-ITIC. Quantum chemical calculations reveal that in the more stable dimer configuration the molecular quadrupole moment of the h-ITIC acceptor is $38.8\,ea_0^2$, lower than the molecular quadrupole moment of a single ITIC molecule ($51.3~ea_0^2$). In combination with the larger effective van der Waals volume of the h-ITIC dimer, this lower molecular quadrupole moment leads to a weaker electrostatic potential at the donor--acceptor interface. As a consequence, the observed yield of free charge carrier generation in the PTB7-Th:h-ITIC blend is inferior to its counterpart with a large molecular quadrupole moment PTB7-Th:ITIC, as consistently shown by TAS, TDCF and sensitive EQE measurements. In PTB7-Th:ITIC the electric field dependence of the CT dissociation yield decreases with increasing photon energy when exciting the system below the optical gap, while in PTB7-Th:h-ITIC the field dependence of charge separation efficiency is independent of excitation energy. From empirical fits to the TDCF data the separation length of electron--hole pairs in PTB7-Th:h-ITIC is under 3\,nm, roughly independent of photon energy. In PTB7-Th:ITIC, the CT separation increases from 2.9\,nm to 5.5\,nm when exciting the system with photon energies of 1.21\,eV and 1.53\,eV, respectively. Our results establish that the free charge carrier generation in PTB7-Th:ITIC proceeds through nonrelaxed more delocalised CT states with reduced Coulomb binding of 75\,meV. Whereas in PTB7-Th:h-ITIC, due to its lower electrostatic potential and more localised CT states, the barrier for free charge formation is ca.\ 160\,meV even for the CT states with excess energy. 

\section{Experimental Section}

\textbf{Materials.} Poly[4,8-bis(5-(2-ethylhexyl)thiophen-2-yl)benzo[1,2-$b$;4,5-$b^{\prime}$]dithiophene-2,6-diyl-alt-(4-(2-ethyl\\hexyl)-3-fluorothieno[3,4-$b$]thiophene-)-2-carboxylate-2-\\6-diyl)] (PTB7-Th) and 3,9-bis(2-methylene-(3-(1,1-dicyanomethylene)-indanone))-5,5,11,11-tetrakis(4-hexyl\\phenyl)-dithieno[2,3-$d$:2$^{\prime}$,3$^{\prime}$-$d^{\prime}$]-s-indaceno[1,2-$b$:5,6-$b^{\prime}$]di\\thiophene (ITIC) were purchased from 1-Material. 
(Z)-2-(2-((9,9-bis(4-hexylphenyl)-9$H$-indeno[1,2-$b$]thieno[2,3-\\$d$]thiophen-2-yl)methylene)-3-oxo-2,3-dihydro-1$H$-inden-1-ylidene)malononitrile (h-ITIC) was synthetised as described in the Supporting Information. 
All reagents and chemicals from commercial sources were used without further purification unless specified. Solvents were dried and purified using standard techniques. Flash chromatography was performed with analytical-grade solvents using Sigma-Aldrich silica gel (technical grade, pore size 60\,Å, 230-400 mesh particle size). Flexible plates Alugram Xtra SIL G UV254 from Macherey-Nagel were used for TLC. Compounds were detected by UV irradiation (Bioblock Scientific). NMR spectra were recorded with a Bruker Avance III 300 ($^{1}$H, 300 MHz and $^{13}$C, 75MHz) or a Bruker Avance DRX500 ($^{1}$H, 500 MHz; $^{13}$C, 125 MHz). Chemical shifts are given in ppm relative to TMS and coupling constants $J$ in Hz. IR spectra were recorded on a Bruker spectrometer Vertex 70 and UV-Vis spectra with a Perkin Elmer 950 spectrometer. High-resolution mass spectrometry (HRMS) was performed with a JEOL JMS-700 B/E. Single-crystals of h-ITIC were obtained from solvent diffusion of hexanes in a solution of the named compound in chloroform. X-ray diffraction data were collected and is gathered in the Supporting Information.

\textbf{Device fabrication and characterisation.} 
The active layer blend solutions were prepared with 20\,mg\,ml$^{-1}$ concentration of PTB7-Th:ITIC (1:1.2) and PTB7-Th:h-ITIC (1:2.4) in chlorobenzene. The solutions were stirred overnight at 50\,$^{\circ}$C in the glovebox. Bathocuproine (Ossila) solution (0.5\,mg\,ml$^{-1}$) in methanol was left under stirring at 50\,$^{\circ}$C in the glovebox overnight, then filtered with 0.2\,$\upmu$m PTFE membrane filter and used directly afterwards.
Solar cells were fabricated in regular architecture on glass substrates coated with pre-patterned indium tin oxide (ITO). The substrates were cleaned in ultrasonic bath with detergent, acetone, isopropanol and deionised water. The substrates were then dried with nitrogen and exposed to low-pressure oxygen plasma for 5\,min. After that, a 35\,nm layer of poly(3,4-ethylenedioxythiophene) polystyrene sulfonate (PEDOT:PSS, Clevios AI~4083) was spin-coated on ITO and annealed at 140\,$^{\circ}$C for 10\,min.
The substrates were then transferred to a nitrogen-filled glovebox, where they were heated again for 10\,min. The blend solutions were subsequently spin-coated to yield an active layer thickness of ca.\ 100\,nm, followed by the filtered bathocuproine solution giving a layer thickness of 8\,nm. The devices were completed by 100\,nm of thermally evaporated Ag with a base pressure below 10$^{-6}$\,mbar through shadow masks with active areas of 4.0\,mm$^2$ for steady-state measurements and 0.5\,mm$^2$ for TDCF. Current--voltage characteristics were measured using a Keithley~236 source measure unit. White light LED intensity was matched to the short circuit current calculated from the EQE spectra. The EQE measurements were performed using Bentham~TM300 monochromator and a Si reference photodiode. The output current was measured using Stanford Research SR830~DSP lock-in amplifier.

\textbf{Gas-phase quantum chemical calculations.}
DFT calculations were done using B3LYP functional and 6-311g(d,p) basis set as implemented in the GAUSSIAN package. Singlet and dimer states were extracted from the available crystal packing of ITIC and h-ITIC unit cells (see ref.\ \onlinecite{han2017terminal} and Figure~S9). Obtained molecule configurations were used to calculate dipole and quadrupole moments. See Supporting Information for more details.

\textbf{Transient absorption spectroscopy.} 
TAS was carried out using a previously described custom pump--probe setup.\cite{karuthedath2019impact} The output of a titanium:sapphire amplifier (Coherent LEGEND DUO, 4.5\,mJ, 3\,kHz, 100\,fs) reduced to 2\,mJ/pulse by splitting, pumped an optical parametric amplifiers (OPA) (Light Conversion TOPAS Prime) to generate 1300\,nm pulses. This beam was in turn used to seed a calcium fluoride (CaF$_2$) crystal mounted on a continuously moving stage, thereby generating a white-light super continuum from 350 to 1100\,nm. The excitation light (pump pulse) was provided by an actively Q-switched Nd:YVO$_4$ laser (InnoLas picolo AOT) frequency-doubled to provide ca.\ 800\,ps wide pulses at 532\,nm. The pump laser was triggered by an electronic delay generator (Stanford Research Systems DG535) itself triggered by the transistor–transistor logic (TTL) sync from the Legend DUO, allowing control of the delay from pump and probe with a jitter of roughly 100\,ps. Pulse length and repetition rate allowed for the delay between slightly less than 1\,ps to 300\,$\upmu$s. The sample was kept under a dynamic vacuum of $<10^{-5}$\,mbar in a cryostat (Optistat CFV, OXFORD Instruments). The transmitted fraction of the white light was guided to a custom-made prism spectrograph (Entwicklungsbüro Stresing) where it was dispersed by a prism onto a 512 pixel NMOS linear image sensor (Hamamatsu S8380-512DA). The probe pulse repetition rate was 3\,kHz, while the excitation pulses were directly generated at 1.5\,kHz frequency, while the detector array was read out at 3\,kHz. Adjacent diode readings corresponding to the transmission of the sample after excitation and in the absence of an excitation pulse were used to calculate $\Delta T/T$. Measurements were averaged over several thousand shots to obtain a good signal-to-noise ratio. The delay at which pump and probe arrive simultaneously on the sample (i.e.\ zero time) was determined from the point of maximum positive slope of the TA signal rise, which is expected to correspond to the maximum of the pump pulse.

\textbf{Time delayed collection field.} 
For TDCF, the sample was excited monochromatically using a Light Conversion PHAROS femtosecond laser (400\,$\upmu$J, 5\,kHz, 290\,fs, 50/50 beam splitter) together with an Light Conversion ORPHEUS optical parametric amplifier enabling an output beam of variable wavelength 315\,nm to 2600\,nm. The repetition rate of a laser was reduced to 5\,kHz by a pulse picker. To ensure monochromatic irradiation, longpass filters were used to block the residual radiation emitted by the laser. The fluence arriving at the sample was set by a combination of two Thorlabs FW102C/FW112C filter wheels. Fluence and laser stability were monitored with a Newport 818-BB-40 photodetector. During the experiment, the sample was mounted in a self-built sample holder. Prebias and collection bias voltages were produced by a Keysight 81160A pattern generator, its output voltage was applied to the sample through a unity gain amplifier. The current from the solar cell was measured as a voltage drop over a 10\,$\Upomega$ resistor connected in series to the device. This voltage drop was amplified using a differential amplifier and subsequently measured with a GaGe CS121G2 digitizer. During photogeneration experiment the time delay of electrical probe was 5\,ns. Collection bias was at least -4.0\,V for efficient extraction of all the remaining mobile charge carriers. Total extracted charge for all the excitation wavelengths was of the same order of magnitude (Figure~S21). Fluences were varying from 3\,nJ\,cm$^{-2}$ to 13\,$\upmu$J\,cm$^{-2}$, being higher when exciting sub-gap states.

\textbf{Sensitive EQE.} 
Sensitive EQE was measured with a home-built monochromatic excitation setup. A MSH-300D double monochromator (LOT Quantum Design) with an 100\,W quartz-tungsten halogen lamp was used as monochromatic light source, with additional optical bandpass filters to reduce stray light from higher harmonic wavelengths. The photocurrent from the solar cells was measured with a Zurich HF2LI lock-in-amplifier in combination with a variable Zurich HF2TA trans-impedance amplifier. External bias to the solar cell was provided by an output of the latter, where applicable. A small fraction of the monochromatic light was recorded using a Hamamatsu K1718-B two-color photodiode involving a Silicon and an Indium-Gallium-Arsenide detector, and measured with a SR-DSP830 lock-in amplifier and the second input of the HF2LI, respectively. During the measurements, the solar cells  were held in a Helium atmosphere in a custom built closed-cycle liquid helium cryostat (Cryostat). Sensitive EQE measurements were performed at bias voltages ranging from -3.0\,V to +0.5\,V for PTB7-Th:ITIC and from -4.0\,V to +0.5\,V for PTB7-Th:h-ITIC. Photocurrent of PTB7-Th:ITIC saturated at -2.0\,V. In the case of PTB7-Th:h-ITIC there was no saturation, and EQE at -4.0\,V gives a higher limit of the calculated IQE.

\begin{acknowledgements}

The work of M.S., P.S.M.\ and C.D.\ has received funding from the European Union's Horizon 2020 research and innovation programme under the Marie Skłodowska-Curie Grant Agreement No.\ 722651 (SEPOMO). The work of C.W.\ and C.G.\ was funded by the DFG (project DE 830/19-1). The work of S.K., F.L.\ and J.G.\ was supported by funding from King Abdullah University of Science and Technology (KAUST), S.K.\ specifically acknowledges the Office of Sponsored Research (OSR) under Award No: OSR-2018-CARF/CCF-3079 and Award No. OSR-CRG2018-3746. D.A.\ acknowledges KAUST for funding his sabbatical stay. A.M.\ acknowledges the financial support received from Marie Curie Individual Fellowship (SMOLAC). The authors thank the MATRIX SFR of the University of Angers. 

\end{acknowledgements}

\appendix

\section{Corresponding Author}
Email: deibel@physik.tu-chemnitz.de

\section{Notes}
The authors declare no competing financial interest.

\bibliographystyle{apsrev4-2}
\bibliography{mainlit}

\end{document}


\title{Supporting Information\\ Charge Photogeneration in Non-Fullerene Organic Solar Cells: Influence of Excess Energy and Electrostatic Interactions}
\author{Maria Saladina}
\affiliation{Institut für Physik, Technische Universität Chemnitz, 09126 Chemnitz, Germany}
\author{Pablo Sim{\'o}n Marqu{\'e}s}
\affiliation{MOLTECH‐Anjou, CNRS UMR 6200, University of Angers, 49045 Angers, France}
\author{Anastasia Markina}
\affiliation{Max Planck Institute for Polymer Research, 55128 Mainz, Germany}
\author{Safakath Karuthedath}
\affiliation{King Abdullah University of Science and Technology (KAUST), KAUST Solar Center (KSC), Physical Sciences and Engineering Division (PSE), Material Science and Engineering Program (MSE), Thuwal 23955-6900, Kingdom of Saudi Arabia}
\author{Christopher W{\"o}pke}
\affiliation{Institut für Physik, Technische Universität Chemnitz, 09126 Chemnitz, Germany}
\author{Clemens G{\"o}hler}
\affiliation{Institut für Physik, Technische Universität Chemnitz, 09126 Chemnitz, Germany}
\author{Yue Chen}
\affiliation{Institut für Physik, Technische Universität Chemnitz, 09126 Chemnitz, Germany}
\author{Magali Allain}
\affiliation{MOLTECH‐Anjou, CNRS UMR 6200, University of Angers, 49045 Angers, France}
\author{Philippe Blanchard}
\affiliation{MOLTECH‐Anjou, CNRS UMR 6200, University of Angers, 49045 Angers, France}
\author{Cl{\'e}ment Cabanetos}
\affiliation{MOLTECH‐Anjou, CNRS UMR 6200, University of Angers, 49045 Angers, France}
\author{Denis Andrienko}
\affiliation{Max Planck Institute for Polymer Research, 55128 Mainz, Germany}
\author{Fr{\'e}d{\'e}ric Laquai}
\affiliation{King Abdullah University of Science and Technology (KAUST), KAUST Solar Center (KSC), Physical Sciences and Engineering Division (PSE), Material Science and Engineering Program (MSE), Thuwal 23955-6900, Kingdom of Saudi Arabia}
\author{Julien Gorenflot}
\affiliation{King Abdullah University of Science and Technology (KAUST), KAUST Solar Center (KSC), Physical Sciences and Engineering Division (PSE), Material Science and Engineering Program (MSE), Thuwal 23955-6900, Kingdom of Saudi Arabia}
\author{Carsten Deibel}
\affiliation{Institut für Physik, Technische Universität Chemnitz, 09126 Chemnitz, Germany}
\email{deibel@physik.tu-chemnitz.de}

\maketitle

\newpage
\section{Material synthesis}

\begin{figure}[h]
    \centering
    \includegraphics[trim = 0.5cm 3cm 0.5cm 2cm, clip, width=0.95\textwidth]{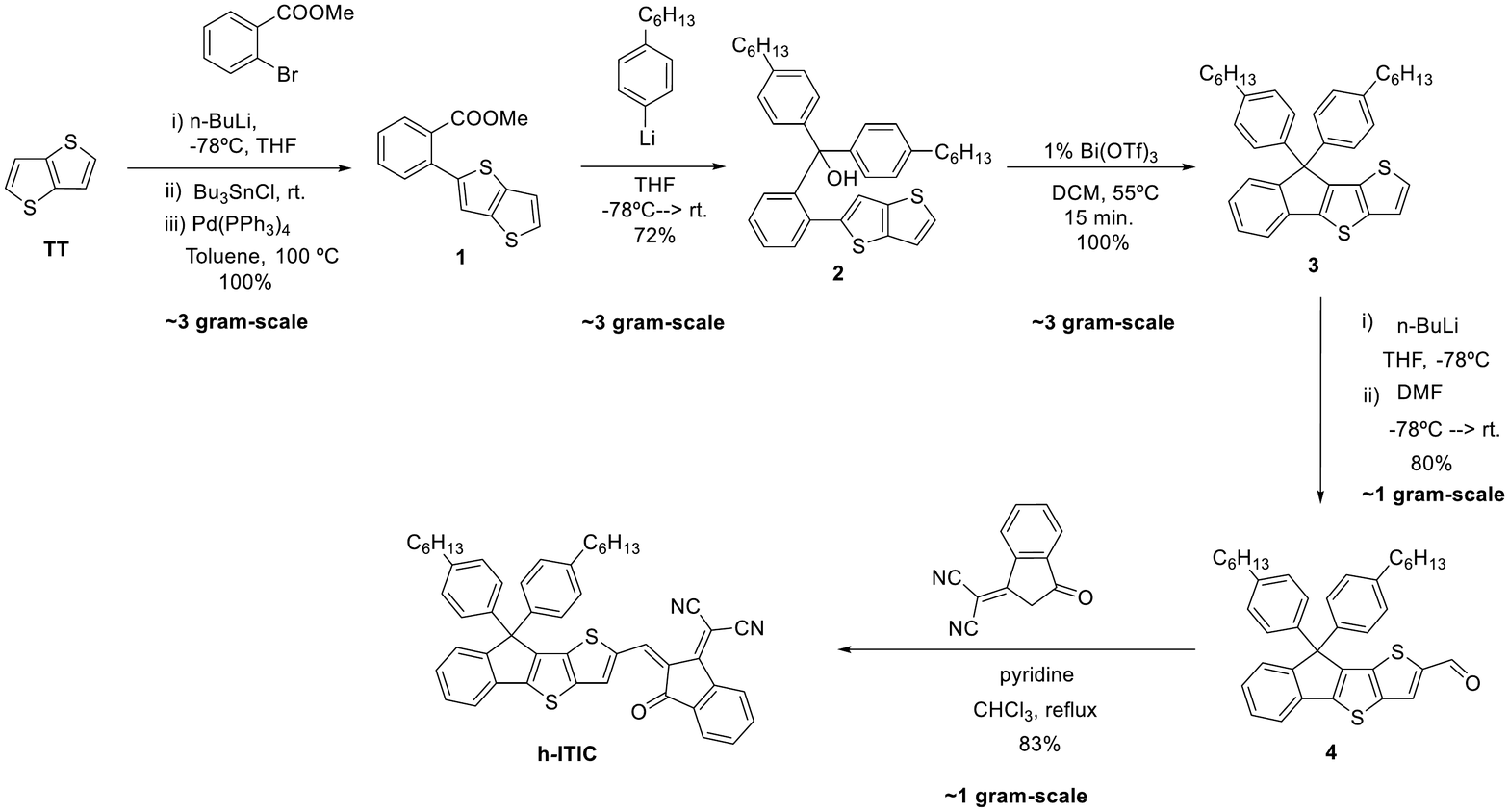}
    \caption{Synthetic route for the preparation of h-ITIC.}
    \label{SI_FigureS01}
\end{figure}

\noindent The synthetic route for h-ITIC is depicted in Figure~\ref{SI_FigureS01}. The pathway started with an efficient Stille coupling between methyl 2-bromobenzoate and thienothiophene (TT), formerly stannylated in one-pot. Then compound \textbf{1} was treated with (4-hexylphenyl)lithium to furnish the benzyl alcohol \textbf{2}. To avoid the classic hard conditions for cyclization, which consists in reflux of acetic and sulfuric acid, tertiary-alcohol Lewis-acid activation was studied to furnish the indenothienothiophene core in mild conditions. Consequently, \textbf{3} was obtained in quantitative yield after catalytic addition of \ch{Bi(OTf)3} and 15\,min of reaction. Finally, formylation and Knoevenagel condensation provided h-ITIC with an overall yield of 46\% in 5 steps.

\noindent \textbf{Methyl 2-(thieno[3,2-\textit{b}]thiophen-2-yl)benzoate (1)}: anhydrous toluene (20\,mL) was added to a mixture of methyl-2-bromobenzoate (2.2\,g, 10.03\,mmol) and tributyl(thieno[3,2-\textit{b}]thiophen-2-yl)stannane (6.6\,g, 15.35 mmol). \ch{Pd(PPh3)4} (236\,mg, 0.20\,mmol) was further added before refluxing the reaction mixture for 16\,h. The latter was then cooled to room temperature and the solvent removed under vacuum. Purification of the crude was performed by column chromatography on silica gel (eluent: petroleum ether/dichloromethane, 3:2) affording 2.8\,g of yellow oil (quantitative). \textbf{$^1$H-NMR} (300\,MHz, \ch{CDCl3}): $\updelta$ 7.75 (dt, $J = 7.6$, 0.8\,Hz, 1H), $7.55 - 7.48$ (m, 2H), 7.43 (ddd, $J = 7.6$, 5.8, 3.1\,Hz, 1H), 7.37 (t, $J = 5.2$\,Hz, 1H), 7.22 (d, $J = 0.7$\,Hz, 1H), 3.75 (s, 3H). \textbf{$^{13}$C-NMR} (75\,MHz, \ch{CDCl3}): $\updelta$ 169.1, 144.0, 139.7, 139.4, 134.4, 131.8, 131.4, 131.2, 129.7, 128.2, 126.9, 119.6, 118.6, 52.5. \textbf{MS} (EI+) m/z: 274.0 [M+]. \textbf{HRMS} (EI+): calculated for \ch{C14H10O2S2} 274.0117, found 274.0121.
\vspace{0.3cm}

\noindent \textbf{Bis(4-hexylphenyl)(2-(thieno[3,2-\textit{b}]thiophen-2-yl)phenyl)methanol (2)}: to a solution of 1-bromo-4-hexyl-
benzene (4.3\,g, 17.64\,mmol) in distilled THF (10\,mL) was added dropwise $n$-BuLi (2.5\,M in hexane, 7.06\,mL, 17.64\,mmol) at -78\,$^\circ$C. After 1\,h, a diluted solution of \textbf{1} (2.2\,g, 8.02\,mmol) in dry THF (10\,mL) was slowly added at -78\,$^\circ$C. The mixture was then stirred for 16\,h at room temperature before being poured in water (75\,mL). The aqueous layer was extracted with EtOAc (100\,mL $\times2$), the organic phase was dry over \ch{MgSO4} and the solvent removed under reduced pressure. The reaction mixture was purified by column chromatography (eluent: petroleum ether/dichloromethane 7:3), achieving 3.25\,g of a colourless solid that was used without further purification (72\%). \textbf{$^1$H-NMR} (300\,MHz, \ch{CDCl3}): $\updelta$ 7.39 – 7.19 (m, 4H), 7.16 (dd, $J = 5.3$, 0.7\,Hz, 1H), 7.10 (s, 8H), 6.84 (dd, $J = 7.9$, 1.5\,Hz, 1H), 6.22 (s, 1H), 3.44 (d, $J = 0.7$\,Hz, 1H), 2.69 – 2.56 (m, 4H), 1.70 – 1.58 (m, 4H), 1.39 – 1.28 (m, $J = 5.3$, 4.4\,Hz, 12H), 0.91 – 0.86 (m, 6H).

\clearpage
\noindent \textbf{9,9-bis(4-hexylphenyl)-9\textit{H}-indeno[1,2-\textit{b}]thieno[2,3-\textit{d}]thiophene (3)}: \textbf{2} (3.25\,g, 5.73\,mmol) was solved in dichloromethane (18\,mL) under inert atmosphere and degassed for 30\,min by Ar bubbling. Then, \ch{Bi(OTf)3} (38\,mg, 0.06\,mmol) was added and the solution was stirred at 55\,$^\circ$C, following the reaction by $^1$H-NMR. After 15\,min, the solvent was removed in vacuum and the crude purified by column chromatography (eluent: petroleum ether/dichloromethane 9:1), giving 3.15\,g of colourless oil (100\%). \textbf{$^1$H-NMR} (300\,MHz, \ch{CDCl3}): $\updelta$ 7.45 (t, $J = 7.5$\,Hz, 2H), 7.30 (m, 3H), 7.19 (dt, $J = 6.6$\,Hz, 1.0\,Hz, 1H), 7.13 (d, $J = 8.2$\,Hz, 4H), 7.04 (d, $J = 8.2$\,Hz, 4H), 2.58 – 2.49 (m, 4H), 1.60 – 1.51 (m, 4H), 1.34 – 1.24 (m, 12H), 0.88 - 0.84 (m, 6H). \textbf{$^{13}$C-NMR} (75\,MHz, \ch{CDCl3}): $\updelta$ 153.4, 146.2, 143.1, 142.1, 141.8, 140.5, 138.1, 133.9, 128.5, 128.0, 127.7, 126.7, 126.1, 126.0, 120.5, 119.3, 63.3, 35.7, 31.8, 31.4, 29.3, 22.7, 14.2. \textbf{MS} (MALDI$-$dctb$+$) m/z: 548.0 [M$+$]. \textbf{HRMS} (EI$+$): calculated for C$_{37}$H$_{40}$S$_2$ 548.2566, found 548.2565.
\vspace{0.3cm}

\noindent \textbf{9,9-bis(4-hexylphenyl)-9\textit{H}-indeno[1,2-\textit{b}]thieno[2,3-\textit{d}]thiophene-2-carbaldehyde (4)}: \textbf{3} (600\,mg, 1.09\,mmol) was dissolved in distilled THF (10\,mL) under argon, then $n$-BuLi (2.5\,M in hexane, 0.52\,mL, 1.31\,mmol) was added dropwise at -78\,$^\circ$C and the reaction was stirred for 1\,h 30\,min, giving a dark blue solution. \textit{N,N}-dimethylformamide (0.17\,mL, 2.19\,mmol) was then added before warming the reaction mixture to room temperature. The overnight--stirred yellow solution was quenched with water and extracted with dichloromethane, the organic phase dried over \ch{Mg2SO4} and the solvent evaporated under vacuum. The resulting crude was purified by column chromatography (eluent: petroleum ether/dichloromethane 1:1), affording 500\,mg of the yellow aldehyde (80\%). \textbf{$^1$H-NMR} (300\,MHz, \ch{CDCl3}): $\updelta$ 9.89 (s, 1H), 7.94 (s, 1H), 7.54 (dd, $J = 7.5$, 1.3\,Hz, 1H), 7.49 (dd, $J = 7.5$, 1.3\,Hz, 1H), 7.36 (td, $J = 7.5$, 1.3\,Hz, 1H), 7.29 (td, $J = 7.5$, 1.3\,Hz, 1H), 7.10 (d, $J = 8.5$\,Hz, 4H), 7.05 (d, $J = 8.5$\,Hz, 4H), 2.57 – 2.49 (m, 4H), 1.60 – 1.51 (m, 4H), 1.33 – 1.23 (m, 12H), 0.89 – 0.84 (m, 6H). \textbf{$^{13}$C-NMR} (75\,MHz, \ch{CDCl3}): $\updelta$ 183.1, 154.3, 149.9, 146.1, 144.3, 142.2, 141.7, 140.5, 139.6, 137.0, 130.1, 128.7, 128.0, 127.9, 127.6, 126.4, 120.4, 63.4, 35.7, 31.8, 31.4, 29.2, 22.7, 14.2. \textbf{IR} (neat): $\nu = 3039-3018$\,cm$^{-1}$ (\ch{C_{sp2}-H}), $2951-2850$\,cm$^{-1}$ (\ch{C_{sp3}-H}), 1658\,cm$^{-1}$ (\ch{C=O}), 1493 (\ch{C_{sp2}=C_{sp2}}). \textbf{MS} (EI$+$) m/z: 576.3 [M$+$]. \textbf{HRMS} (EI$+$): calculated for \ch{C38H40OS2} 576.2515, found 576.2516.
\vspace{0.3cm}

\noindent \textbf{(Z)-2-(2-((9,9-bis(4-hexylphenyl)-9\textit{H}-indeno[1,2-\textit{b}]thieno[2,3-\textit{d}]thiophen-2-yl)methylene)-3-oxo-2,3-di-
hydro-1\textit{H}-inden-1-ylidene)malononitrile (h-ITIC)}: to a suspension of aldehyde \textbf{4} (450\,mg, 0.78\,mmol) and 2-(3-oxo-2,3-dihydro-1\textit{H}-inden-1-ylidene)malononitrile (454\,mg, 2.34\,mmol) in chloroform, 0.1\,mL of pyridine were added before refluxing the reaction overnight. Thereafter, the solvent was evaporated in vacuum and the crude purified by column chromatography (eluent: dichloromethane). The resulting solid was triturated in petroleum ether, filtrated, washed with petroleum ether and distilled pentane. Eventually, 486\,mg of the final compound were achieved as a green solid (83\%). \textbf{$^1$H-NMR} (300\,MHz, \ch{CDCl3}): $\updelta$ 8.76 (s, 1H), 8.53 (dd, $J = 6.7$, 1.8\,Hz, 1H), 7.91 (dd, $J = 6.5$, 2.3\,Hz, 1H), 7.76 (s, 1H), 7.74 – 7.65 (m, 2H), 7.57 – 7.48 (m, 2H), 7.43 – 7.31 (m, 2H), 7.18 (d, $J = 8.4$\,Hz, 4H), 7.12 (d, $J = 8.4$\,Hz, 4H), 2.55 (t, $J = 7.7$\,Hz, 4H), 1.60-1.54 (m, 4H), 1.32-1.24 (m, 12H), 0.90 – 0.76 (m, 6H). \textbf{$^{13}$C-NMR} (75\,MHz, \ch{CDCl3}): 188.2, 160.2, 155.0, 153.7, 147.6, 146.6, 143.3, 142.4, 140.0, 139.5, 139.2, 138.7, 137.5, 136.9, 135.2, 134.6, 128.9, 128.5, 128.2, 128.0, 126.6, 125.3, 123.8, 122.3, 120.9, 114.8, 114.6. \textbf{IR} (neat): $\nu = 3067$\,cm$^{-1}$ (\ch{C_{sp2}-H}, Ar), $2958-2855$\,cm$^{-1}$ (\ch{C_{sp3}-H}), 2220\,cm$^{-1}$ (\ch{C+N}), 1701\,cm$^{-1}$ (\ch{C=O}), $1546-1391$\,cm$^{-1}$ (\ch{C=C}, Ar). \textbf{UV-Vis} (\ch{CH2Cl2}): $\lambda_{\text{max}}(\varepsilon)$ = 568\,nm (52000\,L$\cdot$mol$^{-1}\cdot$cm$^{-1}$), 305\,nm (20000\,L$\cdot$mol$^{-1}\cdot$cm$^{-1}$). \textbf{MS} (MALDI $-$dit$+$) m/z: 752.3 [M$+$]. \textbf{HRMS} (EI$+$): calculated for \ch{C50H44N2OS2} 752.2890, found 752.2891.

\newpage
\section{Material characterisation}
\subsection{$^1$H-NMR spectra}

\begin{figure}[h]
    \centering
    \subfigure{\begin{overpic}%
    [trim = 0cm 0cm 0cm 0cm, clip, width=0.45\textwidth]%
    {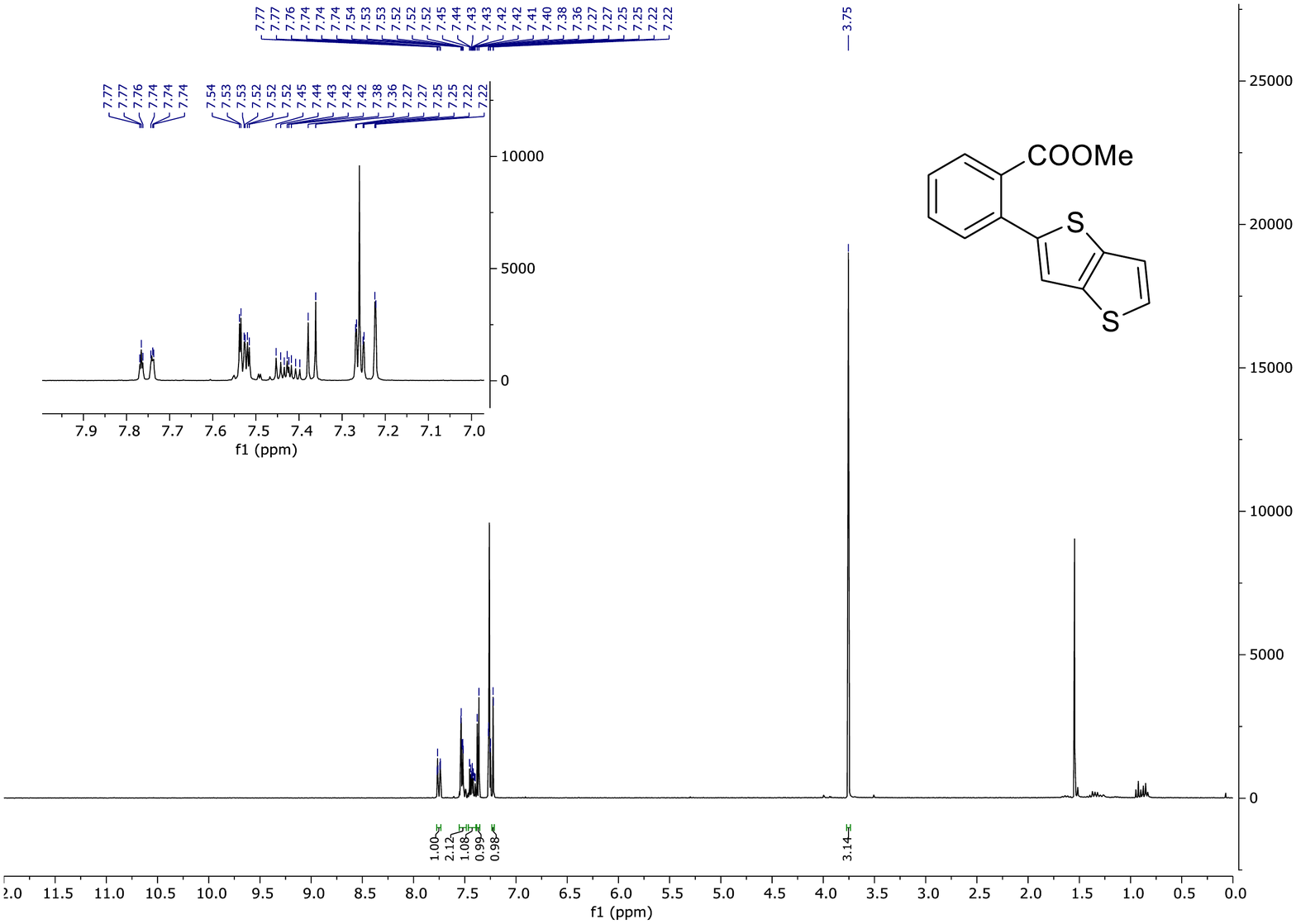}%
    \put(3,17){\Large(a)}%
    \put(112,17){\Large(b)}%
    \end{overpic}}
    \qquad
    \subfigure{\begin{overpic}%
    [trim = 0cm 0cm 0cm 0cm, clip, width=0.45\textwidth]%
    {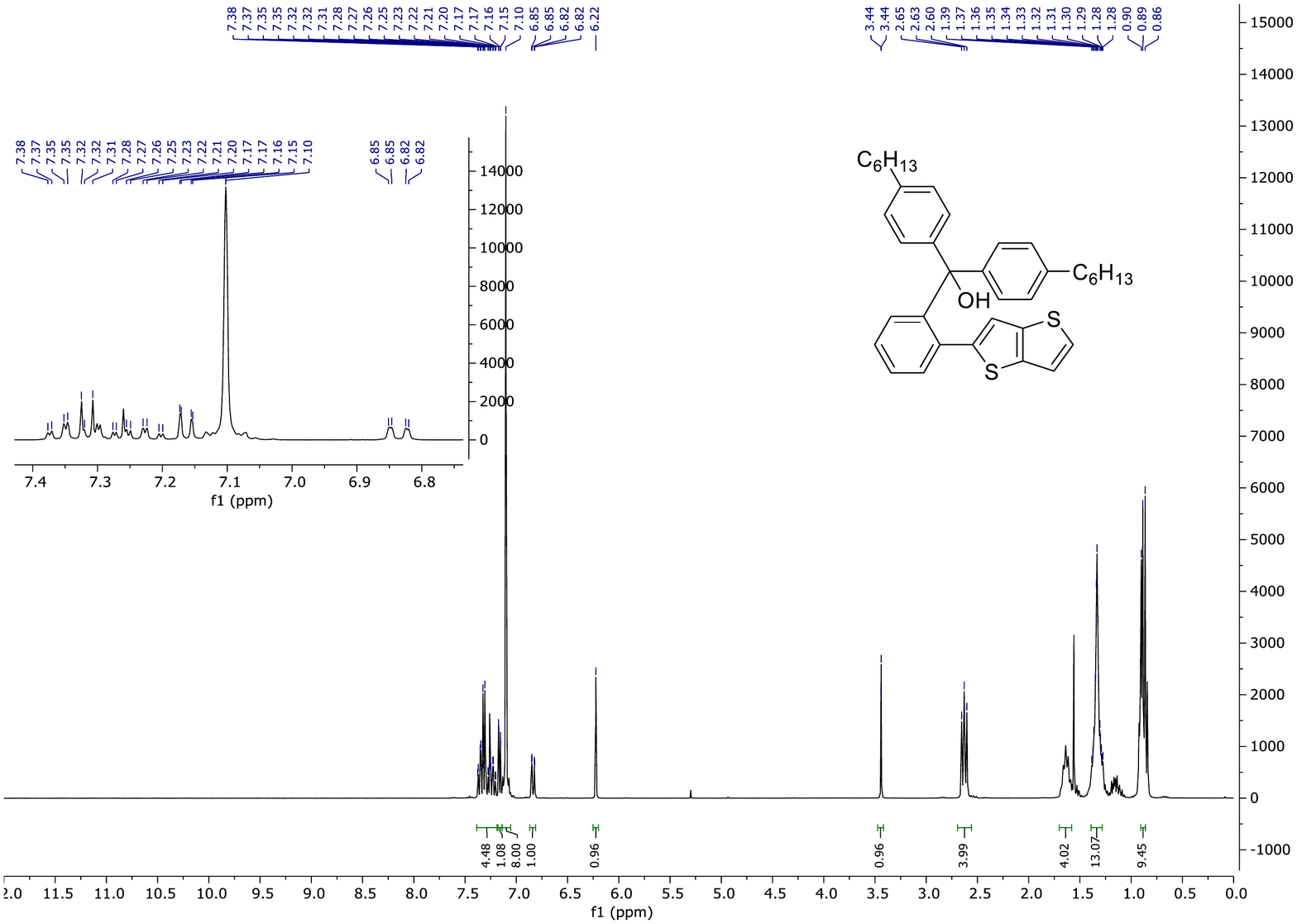}%
    \end{overpic}}
    \par
    \centering
    \subfigure{\begin{overpic}%
    [trim = 0cm 0cm 0cm 0cm, clip, width=0.45\textwidth]%
    {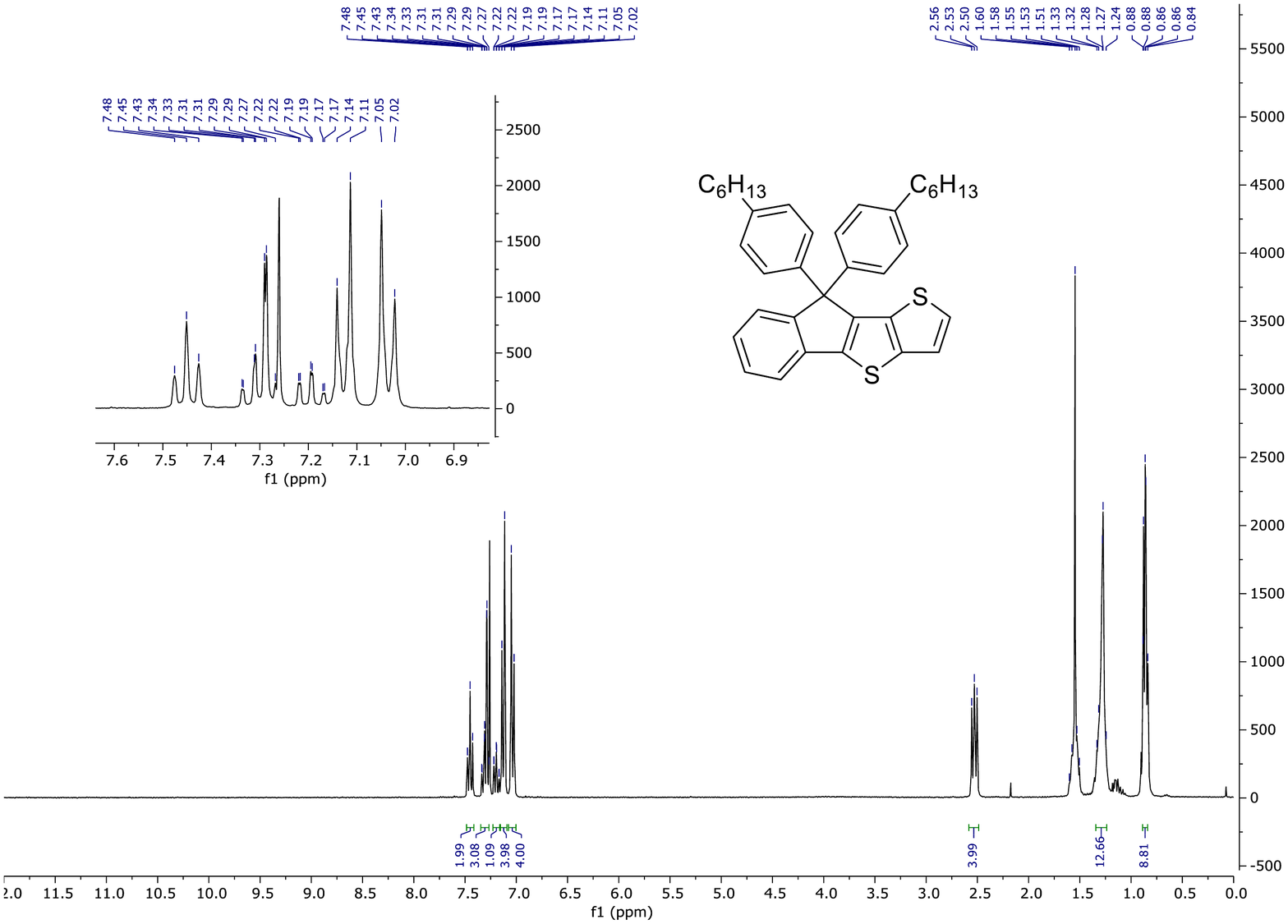}%
    \put(3,17){\Large(c)}%
    \end{overpic}}
    \par
    \centering
    \subfigure{\begin{overpic}%
    [trim = 0cm 0cm 0cm 0cm, clip, width=0.45\textwidth]%
    {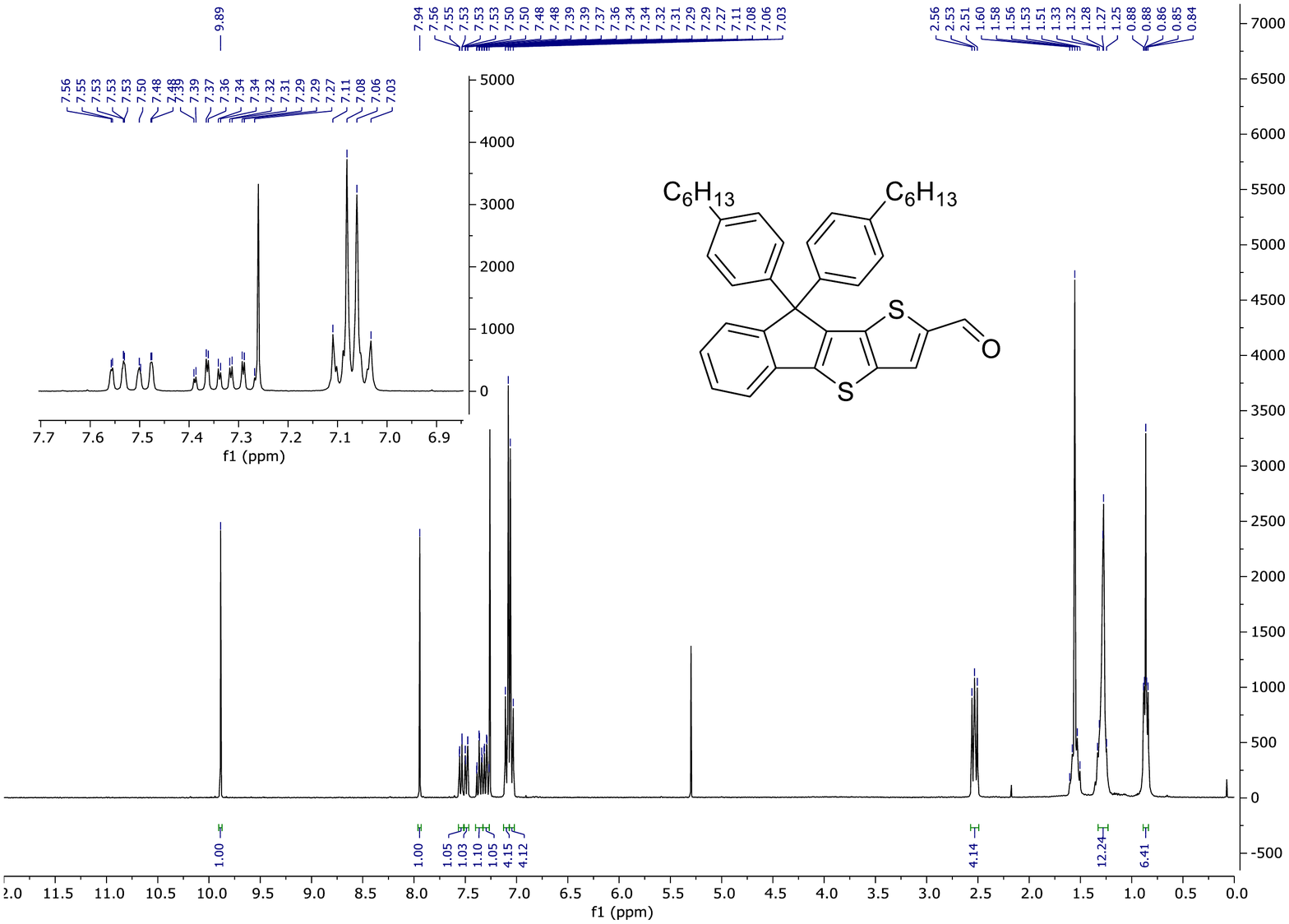}%
    \put(3,17){\Large(d)}%
    \put(112,17){\Large(e)}%
    \end{overpic}}
    \qquad
    \subfigure{\begin{overpic}%
    [ trim = 0cm 0cm 0cm 0cm, clip, width=0.45\textwidth]%
    {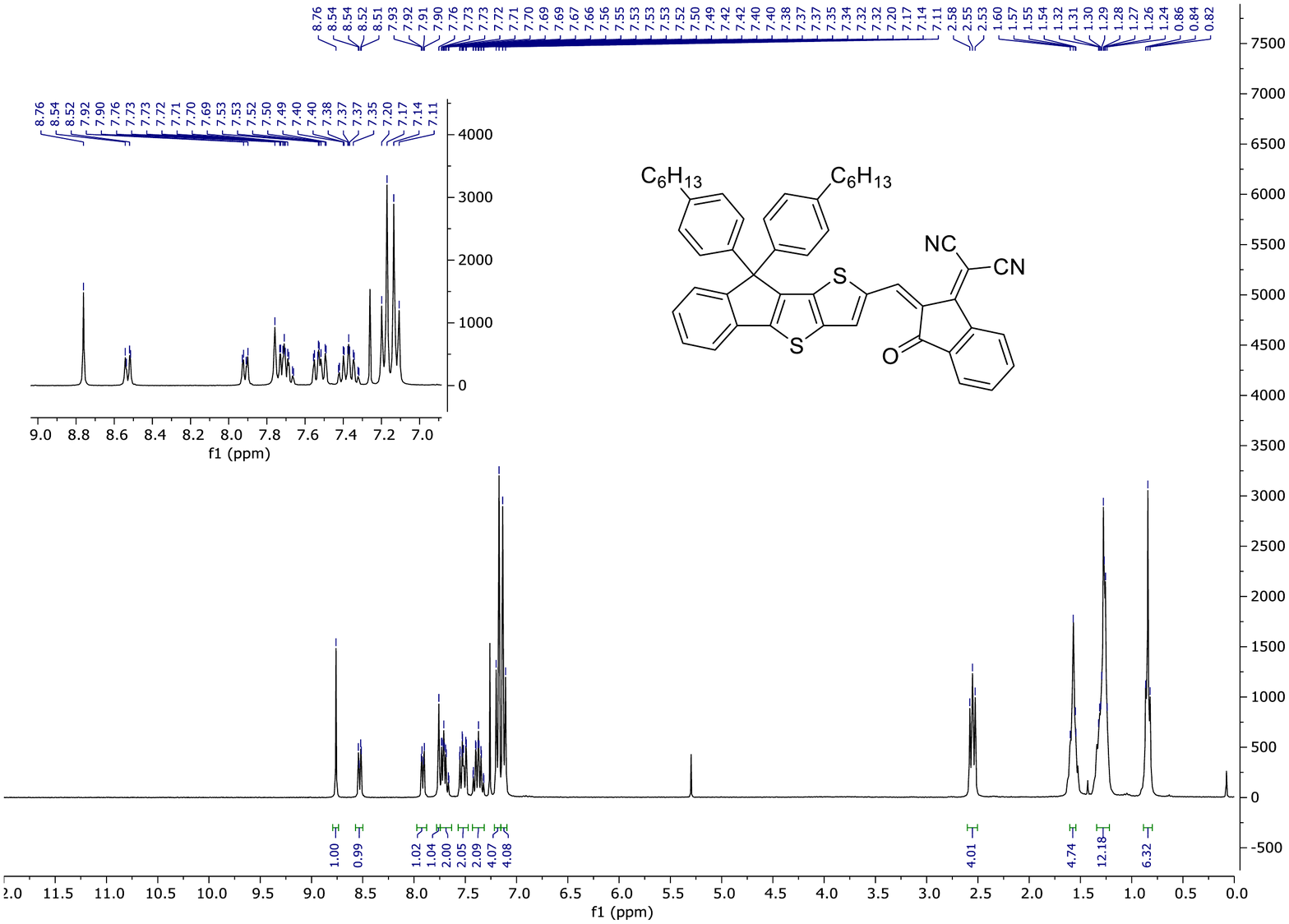}%
    \end{overpic}}
    \caption{$^1$H-NMR (\ch{CDCl3}, 300\,MHz) spectra of (a) \textbf{1}, (b) \textbf{2}, (c) \textbf{3} (d) \textbf{4} and (e) \textbf{h-ITIC}.}
    \label{SI_FigureS02}
\end{figure}

\clearpage
\subsection{$^{13}$C-NMR spectra}

\begin{figure}[h]
    \centering
    \subfigure{\begin{overpic}%
    [trim = 0cm 0cm 0cm 0cm, clip, width=0.45\textwidth]%
    {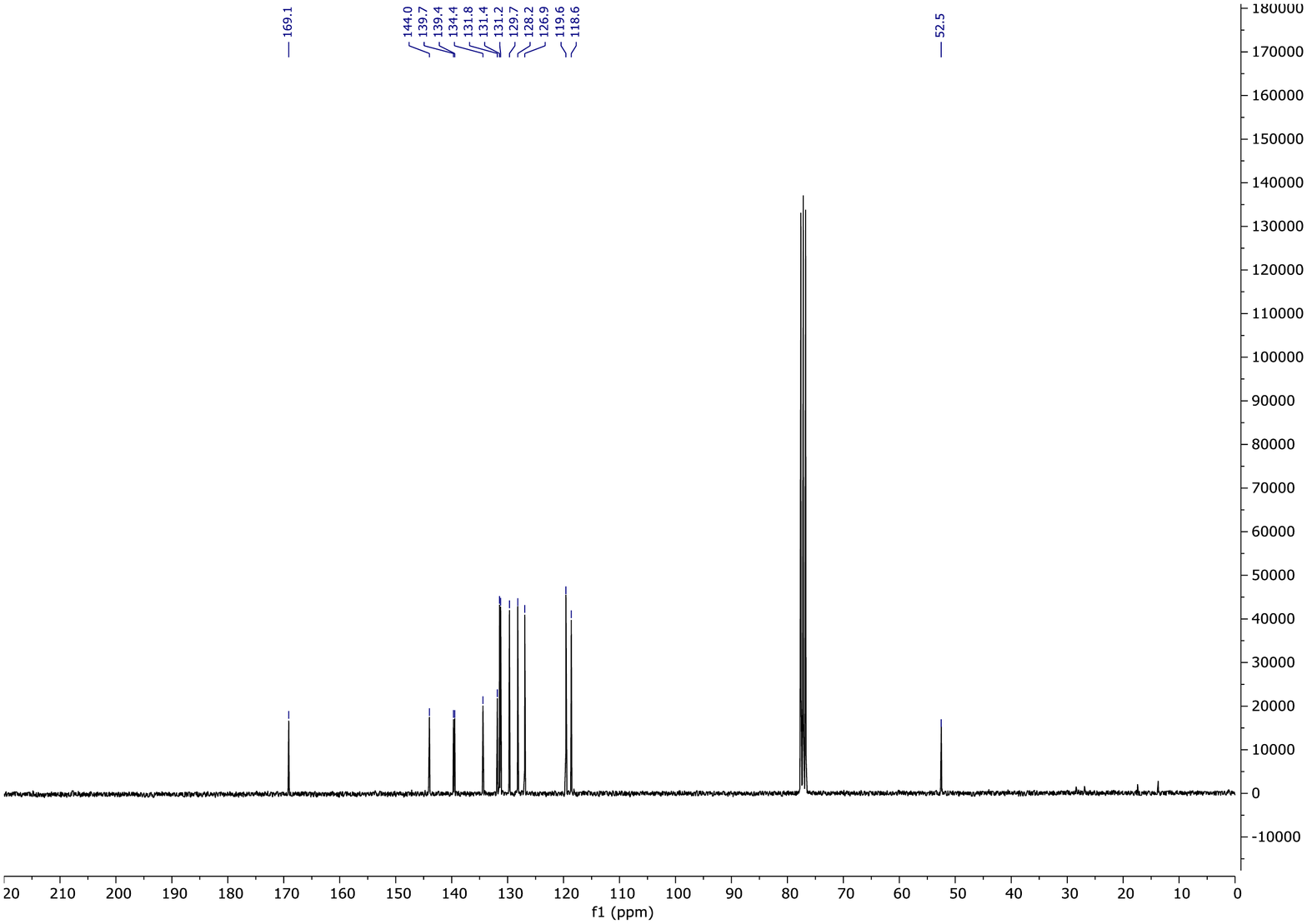}%
    \put(3,60){\Large(a)}%
    \put(112,60){\Large(b)}%
    \end{overpic}}
    \qquad
    \subfigure{\begin{overpic}%
    [trim = 0cm 0cm 0cm 0cm, clip, width=0.45\textwidth]%
    {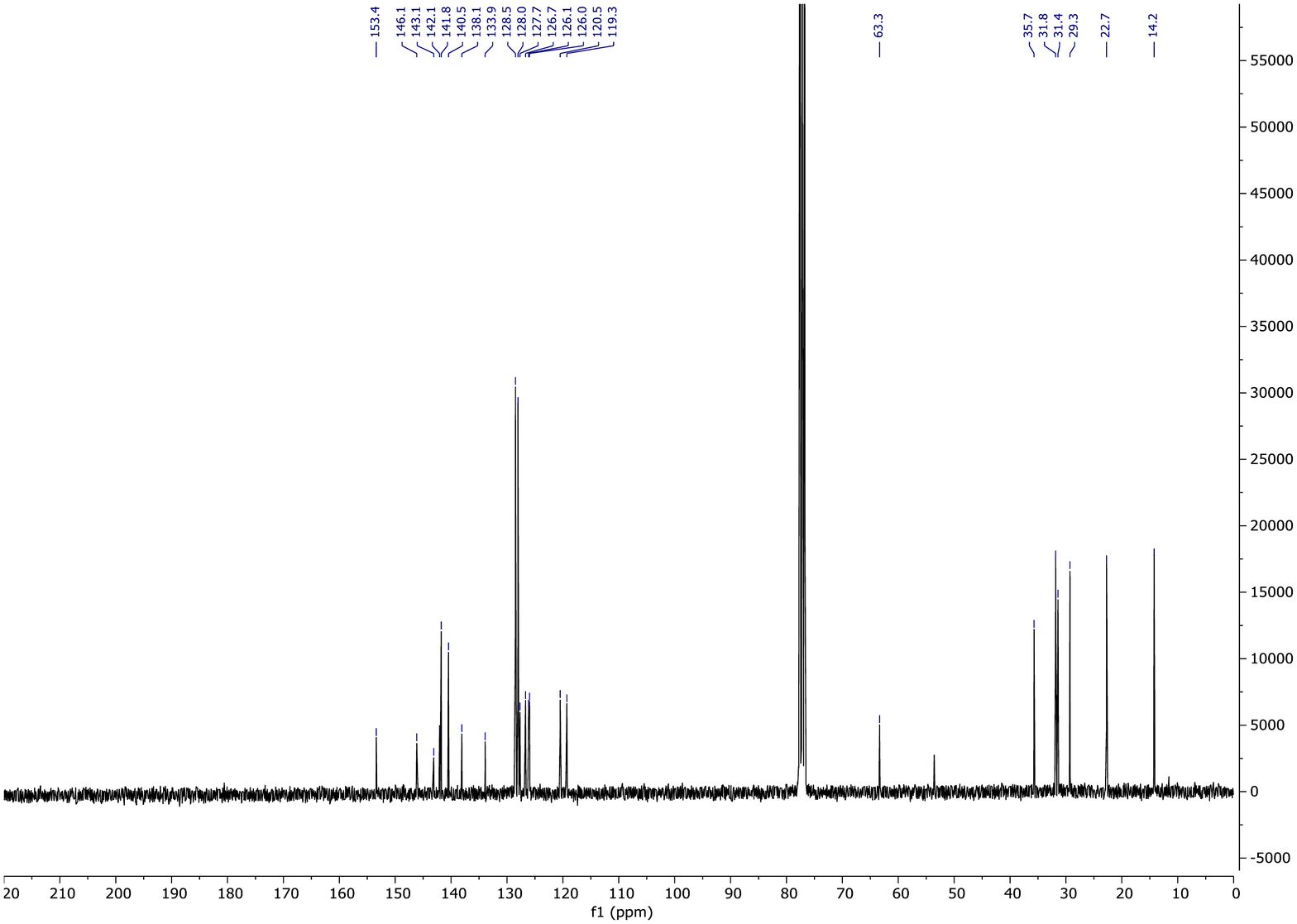}%
    \end{overpic}}
    \par
    \centering
    \subfigure{\begin{overpic}%
    [trim = 0cm 0cm 0cm 0cm, clip, width=0.45\textwidth]%
    {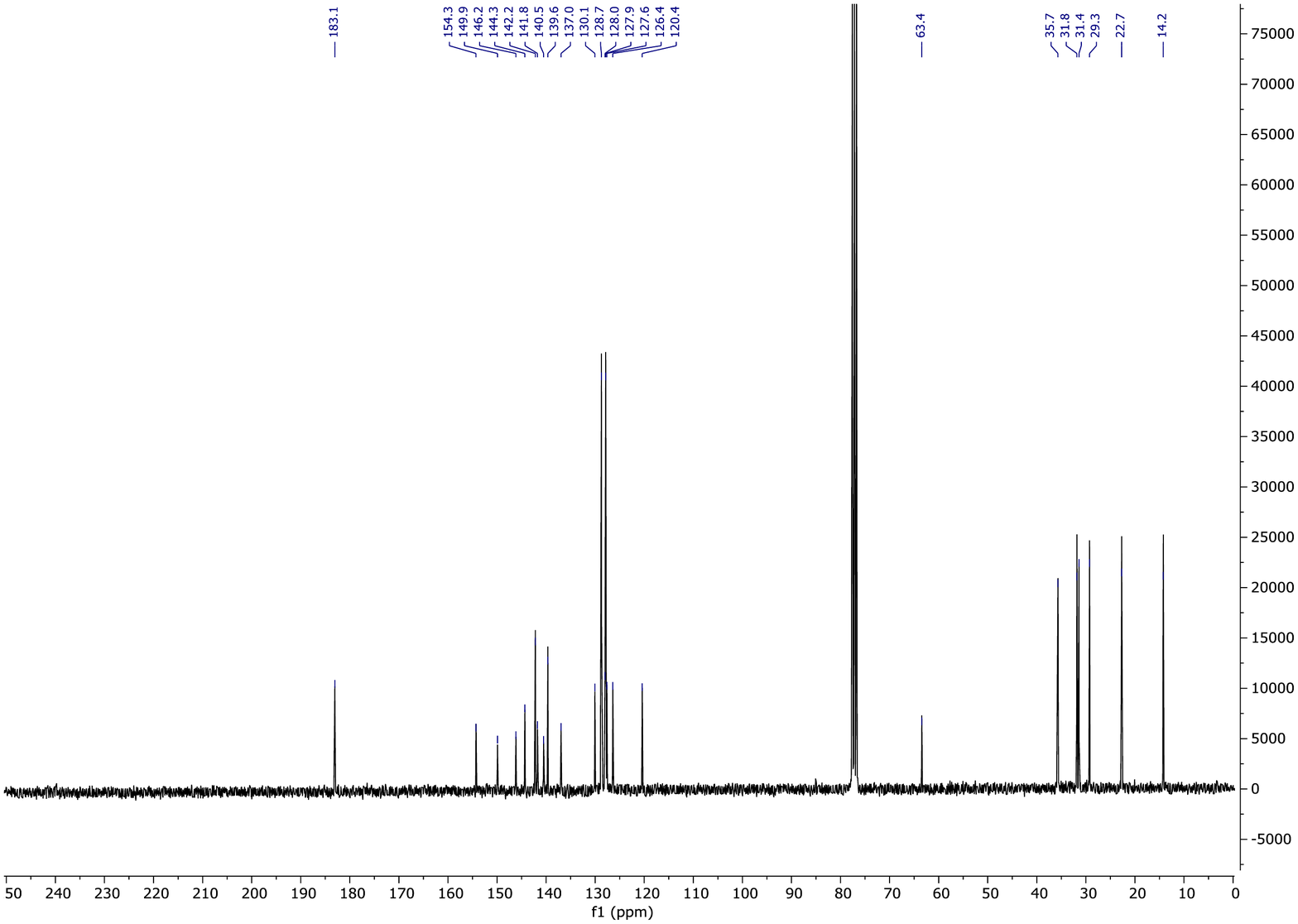}%
    \put(3,60){\Large(c)}%
    \put(112,60){\Large(d)}%
    \end{overpic}}
    \qquad
    \subfigure{\begin{overpic}%
    [trim = 0cm 0cm 0cm 0cm, clip, width=0.45\textwidth]%
    {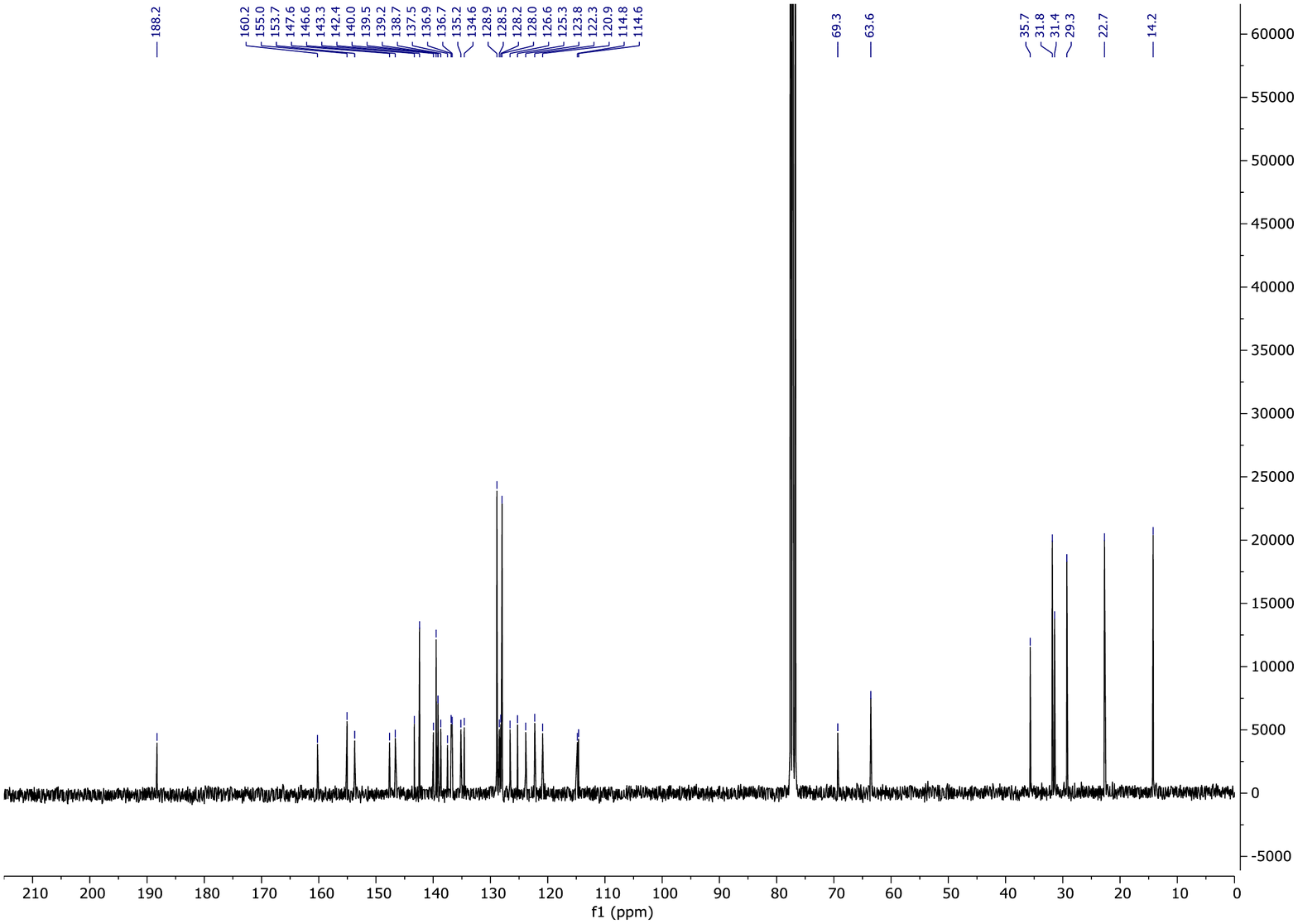}%
    \end{overpic}}
    \caption{$^{13}$C-NMR (\ch{CDCl3}, 300\,MHz) spectra of (a) \textbf{1}, (b) \textbf{3}, (c) \textbf{4} and (d) \textbf{h-ITIC}.}
    \label{SI_FigureS03}
\end{figure}

\newpage
\subsection{MS and HRMS spectra}

\begin{figure}[h]
    \centering
    \subfigure{\begin{overpic}%
    [trim = 0cm 2cm 0cm 4.3cm, clip, width=0.7\textwidth]%
    {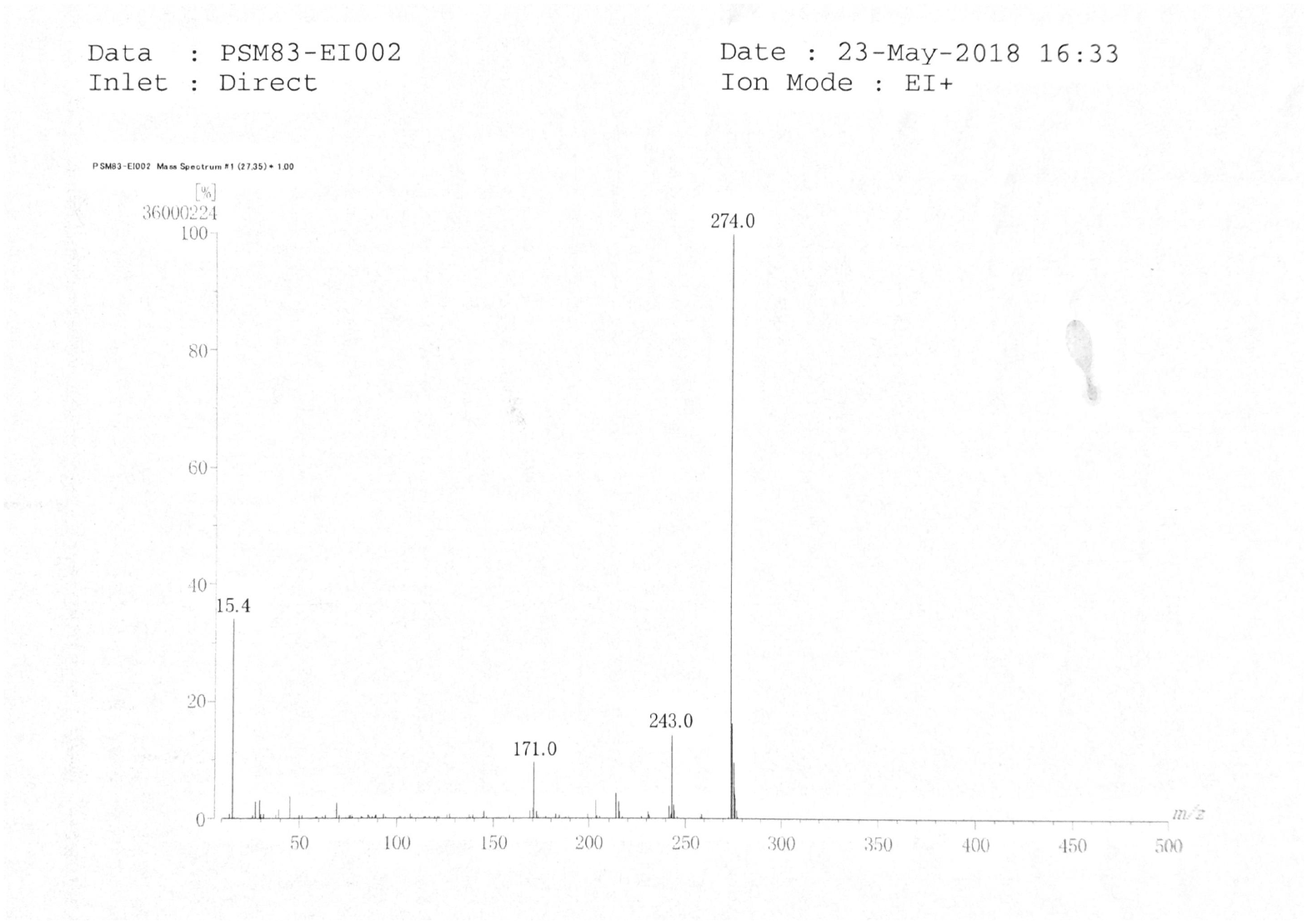}%
    \end{overpic}}
    \par\vfill
    \subfigure{\begin{overpic}%
    [trim = 1.1cm 1cm 0cm 2cm, clip, width=0.7\textwidth]%
    {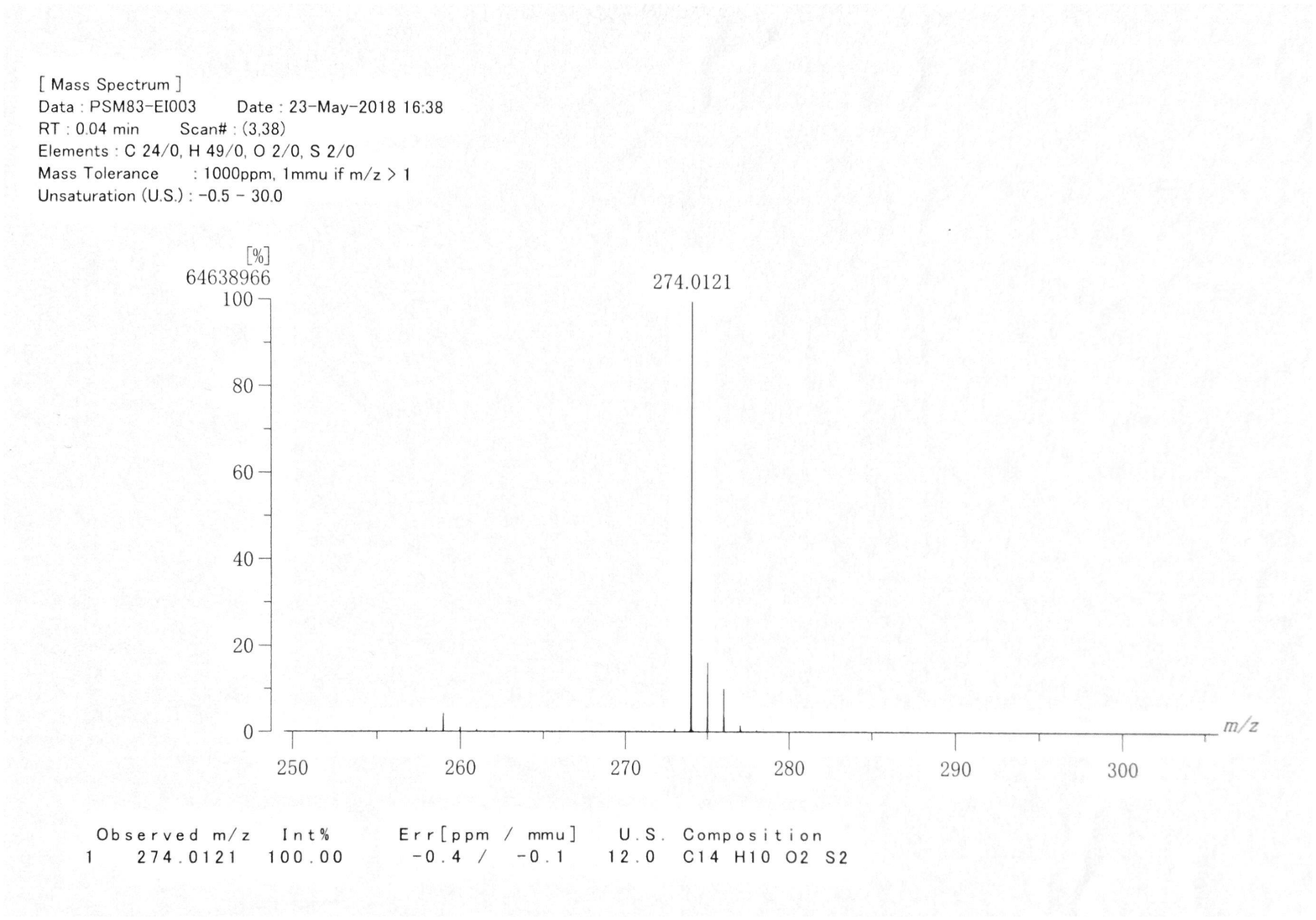}%
    \put(80,108){\Large(a)}%
    \put(80,45){\Large(b)}%
    \end{overpic}}
    \caption{(a) MS (EI$+$) and (b) HRMS (EI$+$) of \textbf{1}.}
    \label{SI_FigureS04}
\end{figure}

\begin{figure}[h]
    \centering
    \subfigure{\begin{overpic}%
    [trim = 0cm 5.5cm 0cm 0cm, clip, width=0.7\textwidth]%
    {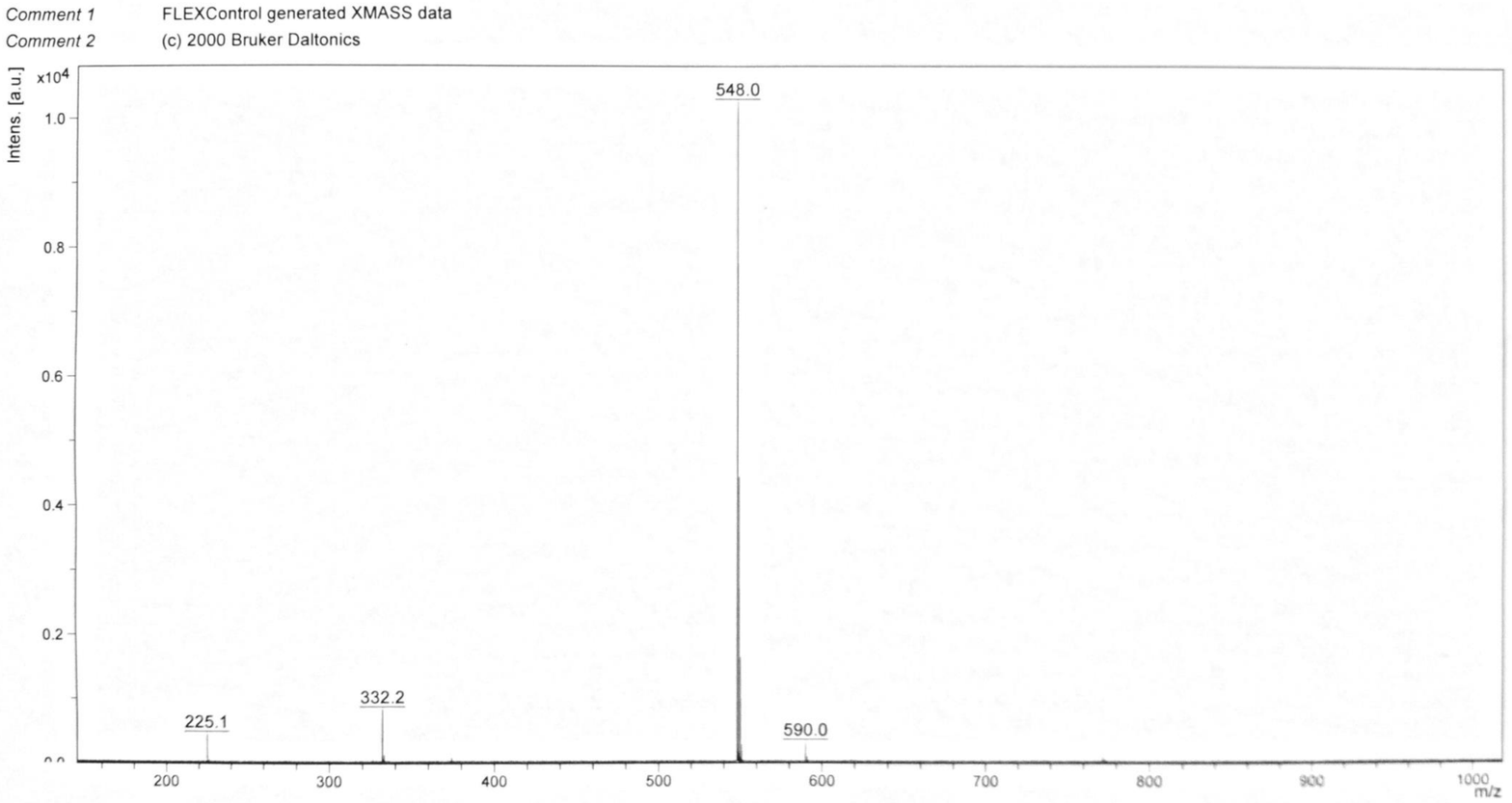}%
    \end{overpic}}
    \par\vfill
    \subfigure{\begin{overpic}%
    [trim = 1cm 2cm 0cm 1.5cm, clip, width=0.7\textwidth]%
    {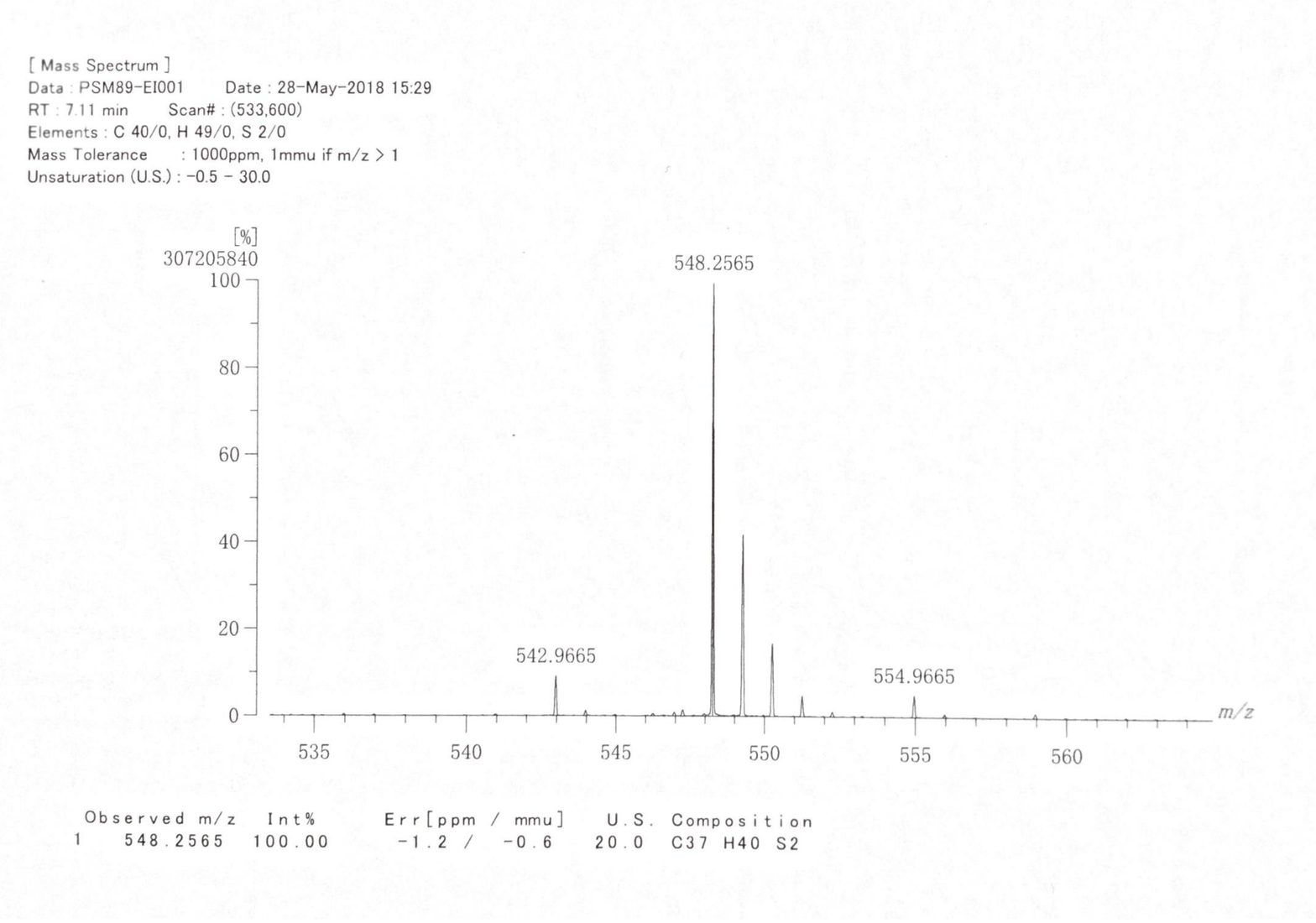}%
    \put(80,102){\Large(a)}%
    \put(80,42){\Large(b)}%
    \end{overpic}}
    \caption{(a) MALDI-TOF MS (dctb$+$) and (b) HRMS (EI$+$) of \textbf{3}.}
    \label{SI_FigureS05}
\end{figure}

\begin{figure}[h]
    \centering
    \subfigure{\begin{overpic}%
    [trim = 0cm 1cm 0cm 4.8cm, clip, width=0.7\textwidth]%
    {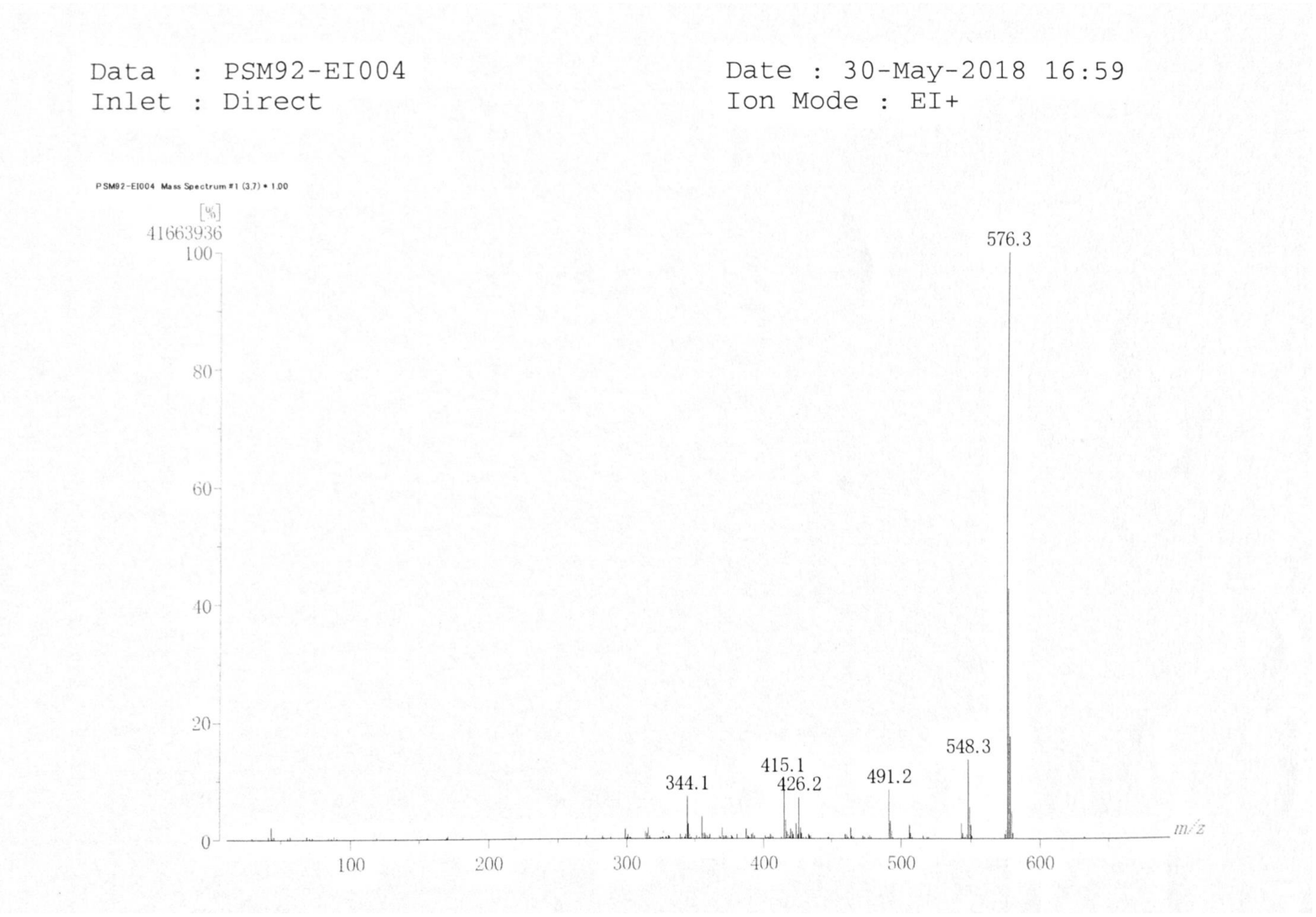}%
    \end{overpic}}
    \par\vfill
    \subfigure{\begin{overpic}%
    [trim = 1.5cm 2cm 0cm 1.5cm, clip, width=0.7\textwidth]%
    {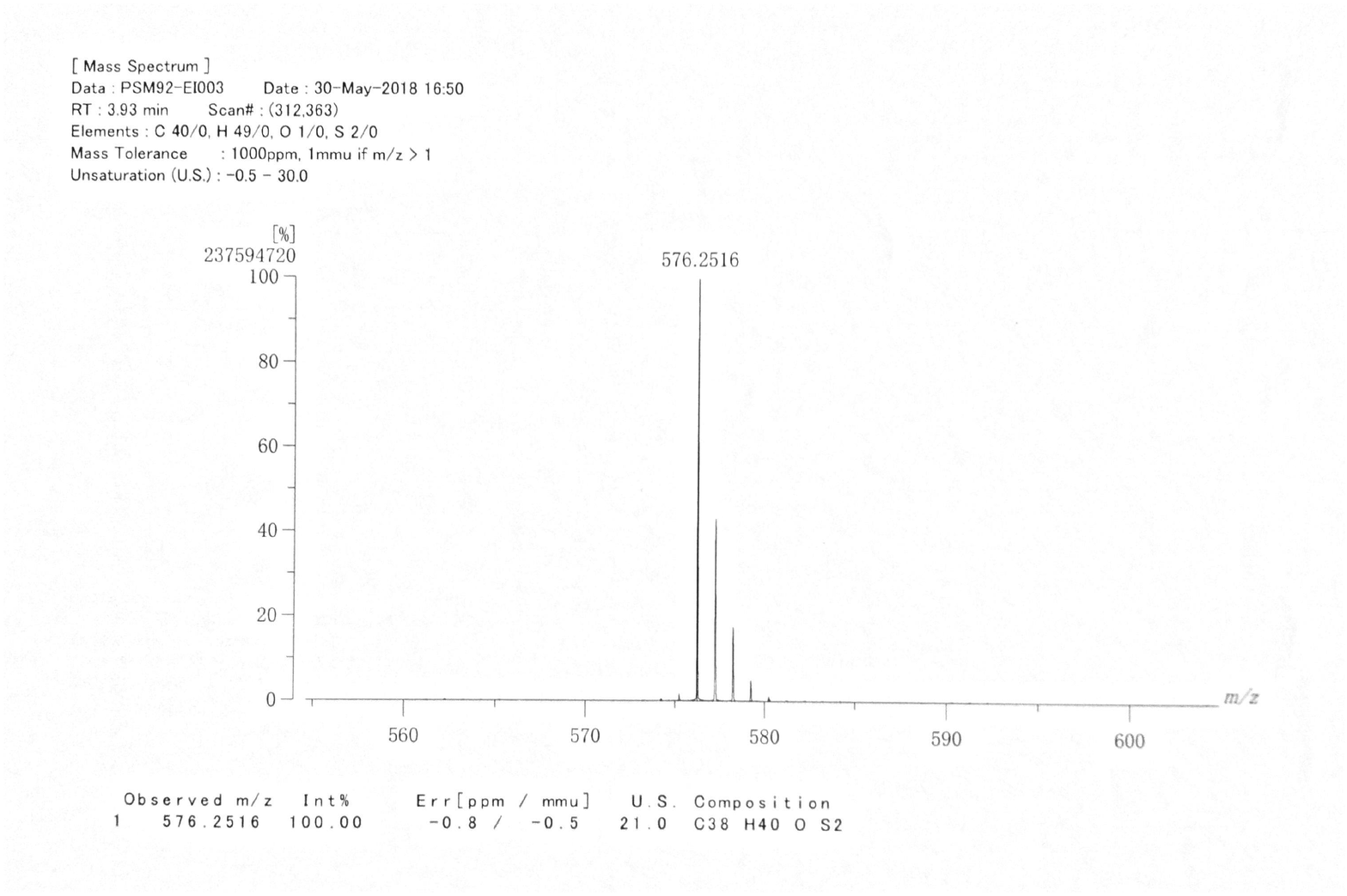}%
    \put(80,108){\Large(a)}%
    \put(80,45){\Large(b)}%
    \end{overpic}}
    \caption{(a) MS (EI$+$) and (b) HRMS (EI$+$) of \textbf{4}.}
    \label{SI_FigureS06}
\end{figure}

\begin{figure}[h]
    \centering
    \subfigure{\begin{overpic}%
    [trim = 0cm 7cm 0cm 0cm, clip, width=0.7\textwidth]%
    {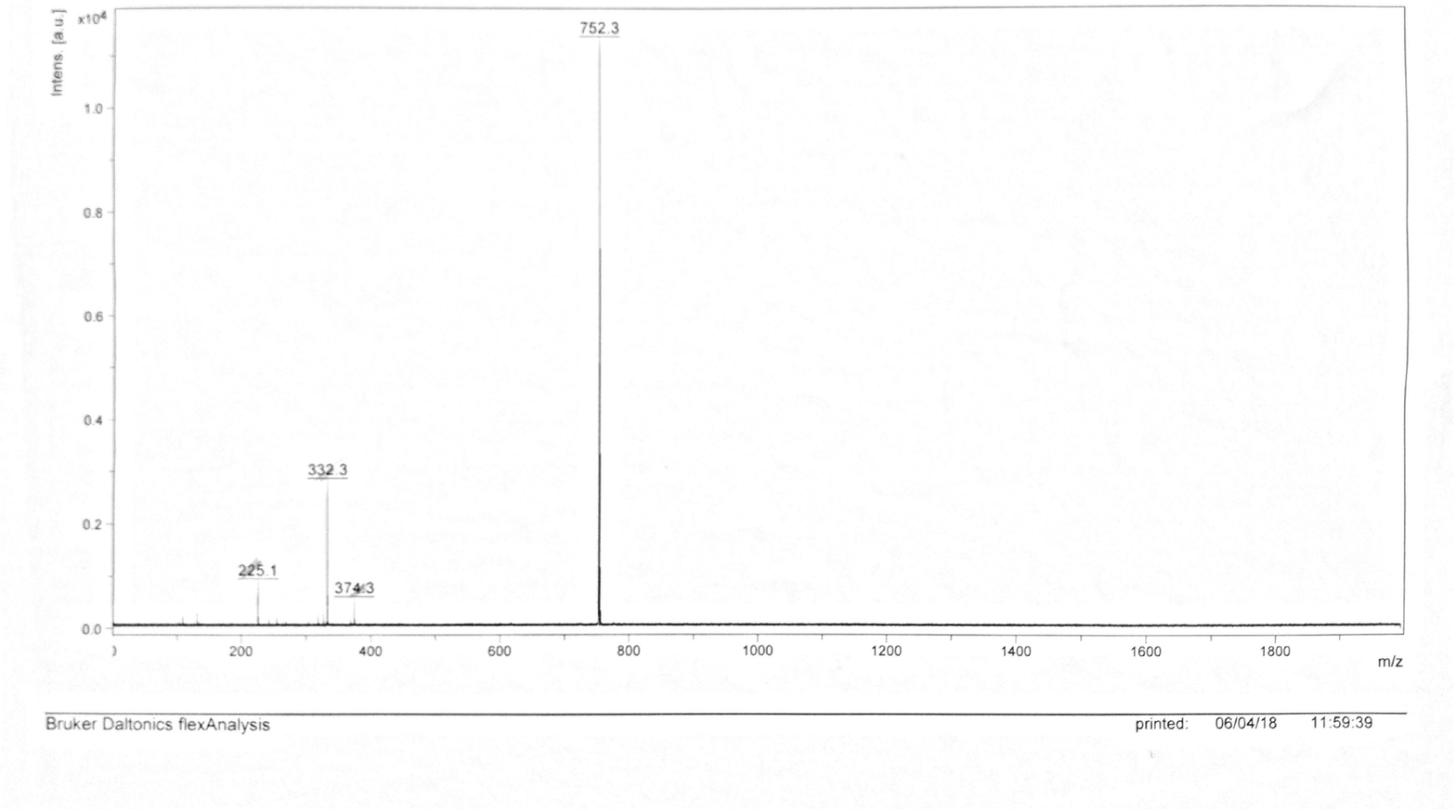}%
    \end{overpic}}
    \par\vfill
    \subfigure{\begin{overpic}%
    [trim = 1cm 2.5cm 0cm 1cm, clip, width=0.7\textwidth]%
    {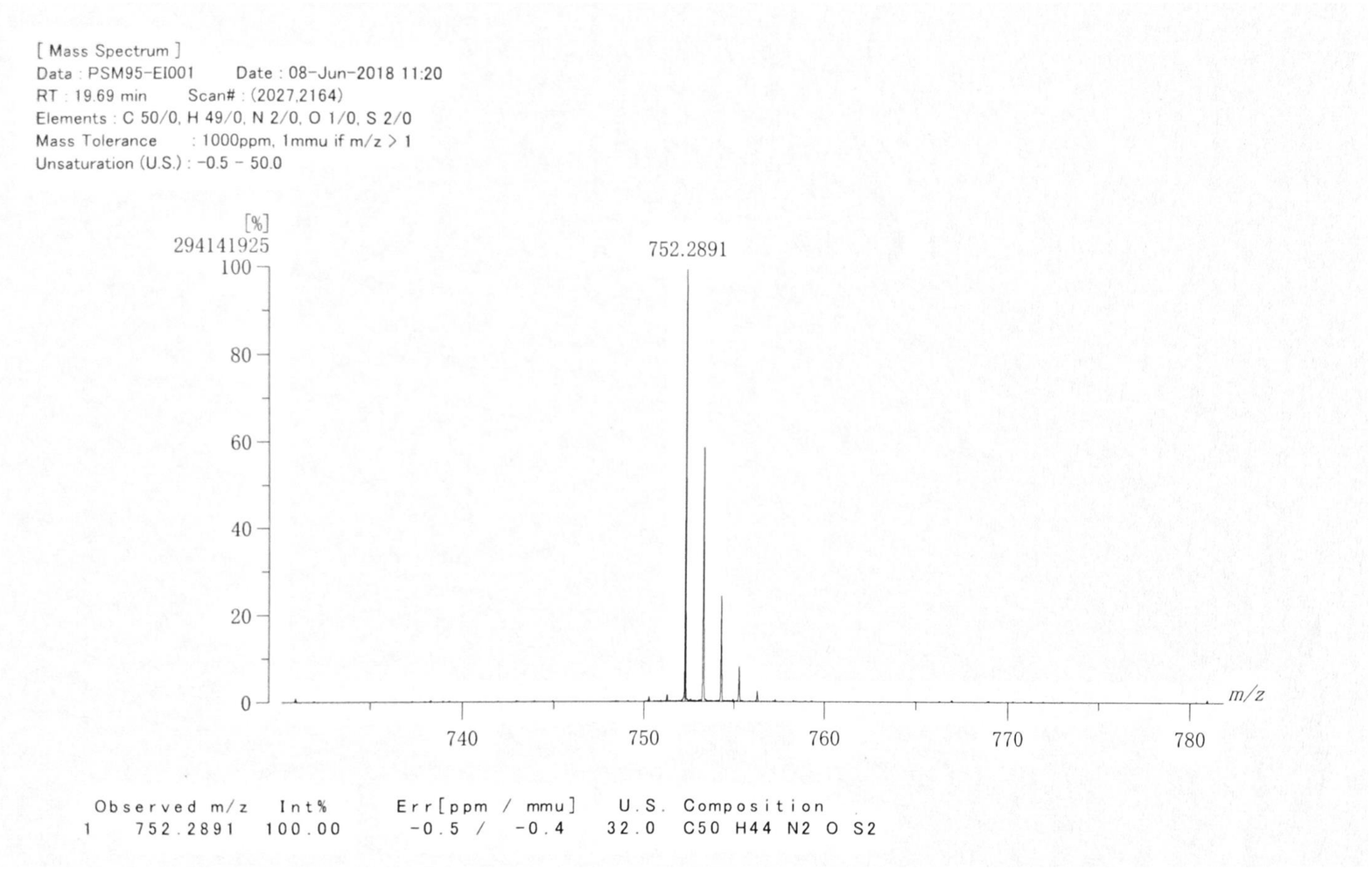}%
    \put(80,105){\Large(a)}%
    \put(80,42){\Large(b)}%
    \end{overpic}}
    \caption{(a) MALDI-TOF MS (dit$+$) and (b) HRMS (EI$+$) of \textbf{h-ITIC}.}
    \label{SI_FigureS07}
\end{figure}

\clearpage
\subsection{IR spectra}

\begin{figure}[h]
    \centering
    \subfigure{\begin{overpic}%
    [trim = 0cm 1cm 4cm 1cm, clip, width=0.7\textwidth]%
    {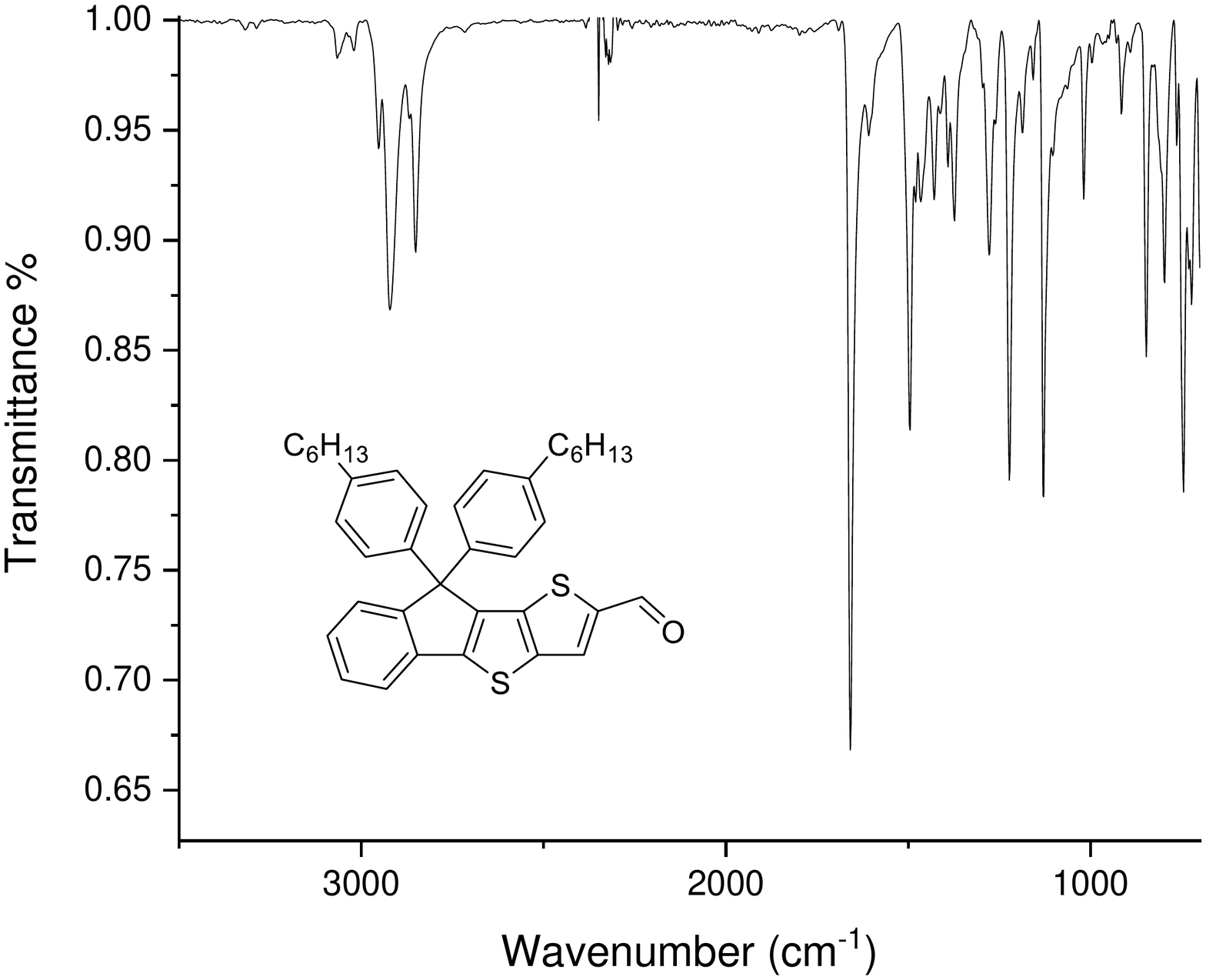}%
    \put(5,65){\Large(a)}%
    \end{overpic}}
    \par
    \subfigure{\begin{overpic}%
    [trim = 0cm 1cm 4cm 1cm, clip, width=0.7\textwidth]%
    {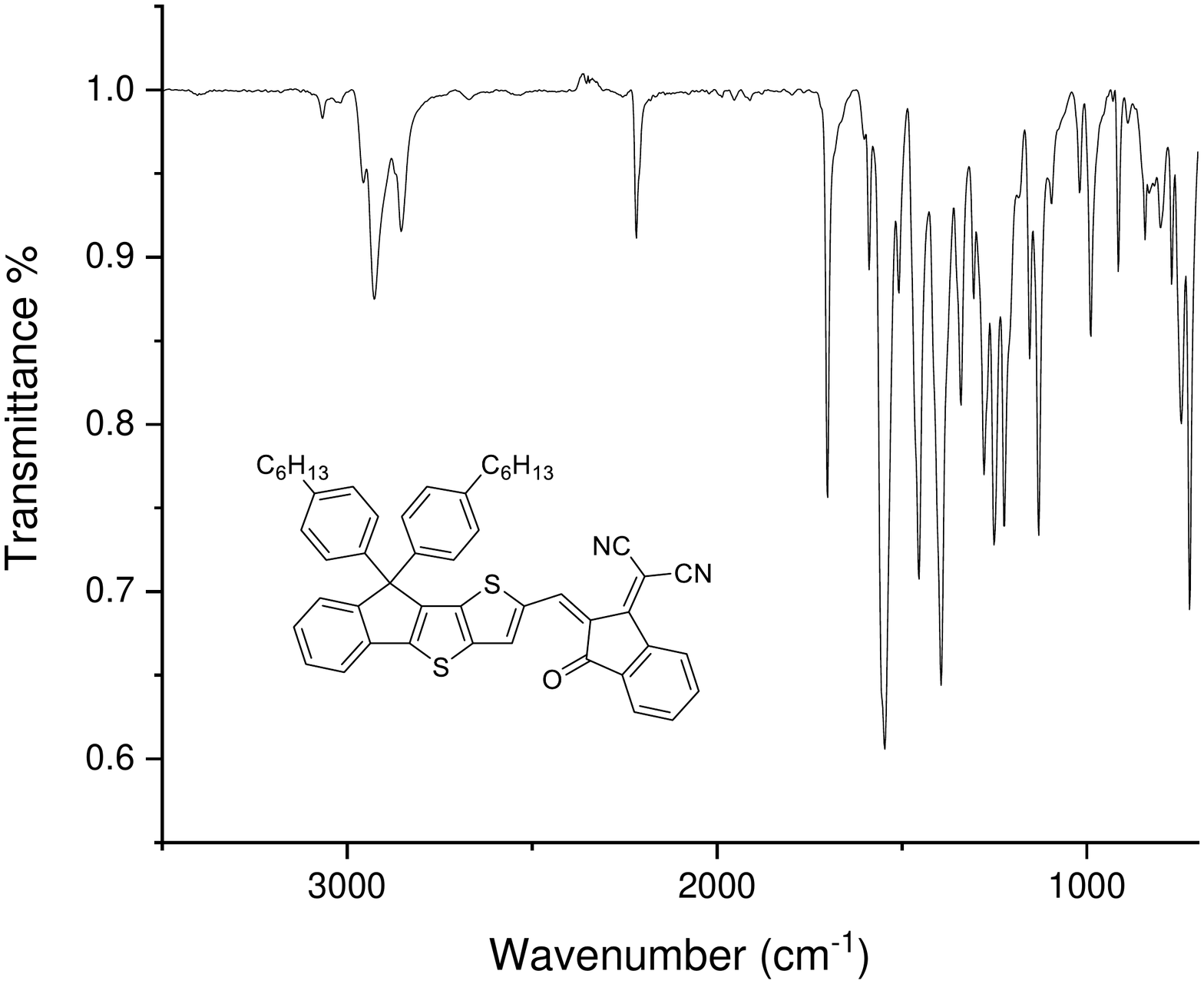}%
    \put(5,65){\Large(b)}%
    \end{overpic}}
    \caption{IR (neat product) spectra of (a) \textbf{4} and (b) \textbf{h-ITIC}.}
    \label{SI_FigureS08}
\end{figure}

\clearpage
\section{Crystallographic data}

\noindent X-ray single-crystal diffraction data were collected at 150~K on a Rigaku Oxford Diffraction SuperNova diffractometer equipped with an Atlas CCD detector and micro-focus Cu-K$\upalpha$ radiation ($\lambda = 1.54184$\,Å). The structure was solved by direct methods, expanded and refined on F$^2$ by full matrix least-squares techniques using SHELX programs (G. M. Sheldrick 2013-2016, SHELXS 2013/1 and SHELXL 2016/4). All non-H atoms were refined anisotropically and multiscan empirical absorption was corrected using CrysAlisPro program (CrysAlisPro, Agilent Technologies, V1.171.38.46, 2015). The H atoms were placed at calculated positions and refined using a riding model. The structure refinement showed disordered electron density which could not be reliably modeled and the program PLATON/SQUEEZE (A.L. Spek V290617 (1980-2019)) was used to remove the corresponding scattering contribution from the intensity data. This electron density can be attributed to missing atoms (on two terminal hexyl chains). The composition with missing atoms were used in the calculation of the empirical formula, formula weight, density, linear absorption coefficient, and F(000). CCDC-2020845 contains the supplementary crystallographic data for this paper.

\begin{figure}[!h]
    \centering
    \includegraphics[trim = 0cm 3cm 0cm 0cm, clip, width=0.8\textwidth]{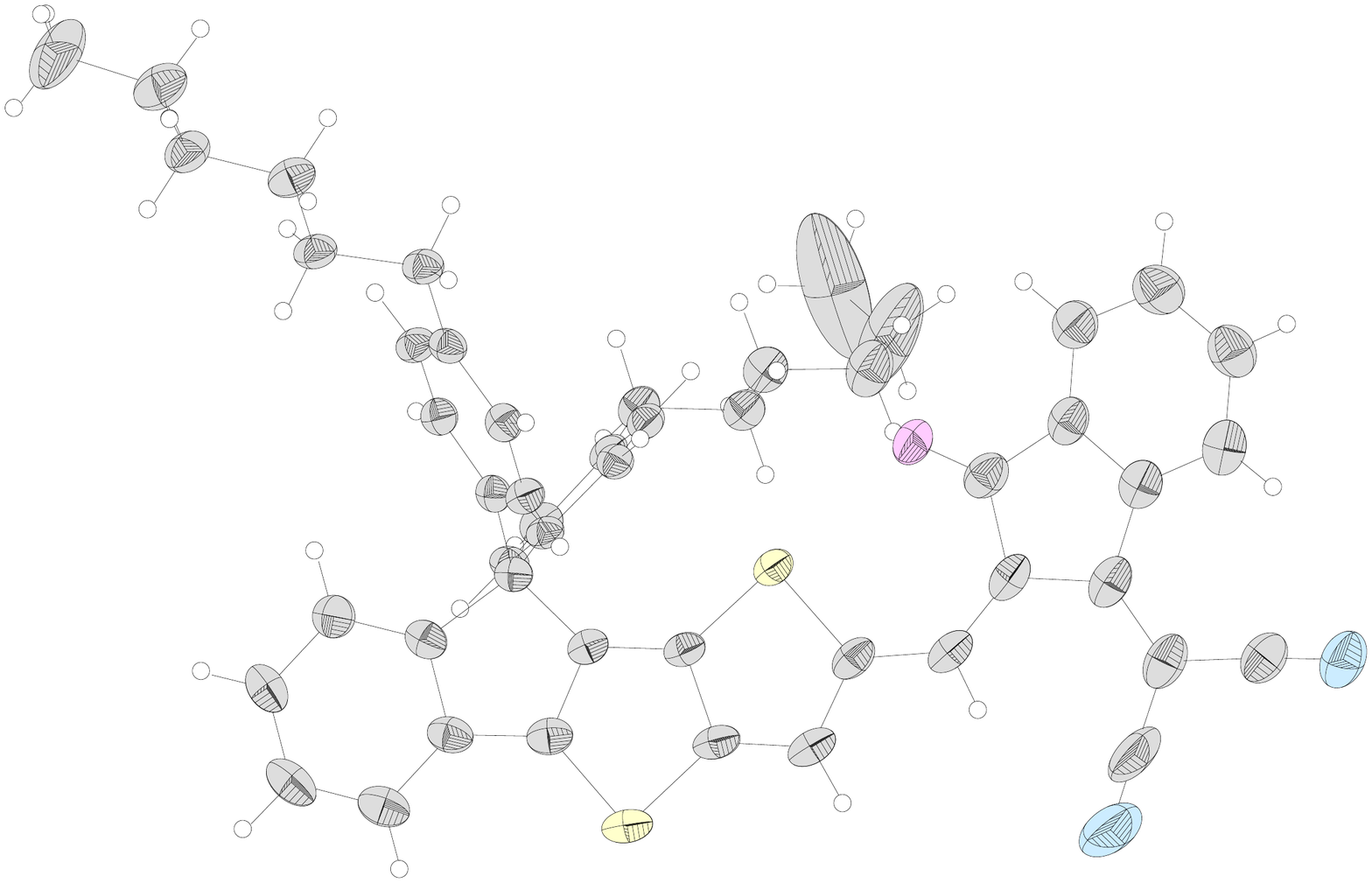}
    \caption{ORTEP-drawing of h-ITIC at 50\% probability level. Side chains are removed for simplicity.}
    \label{SI_FigureS09}
\end{figure}

\noindent \textbf{Crystallographic data for h-ITIC}: \ch{C200H176N8O4S8} or \ch{4 (C50H44N2OS2)}, M\,=\,3011.96, brown needle, $0.283 \times 0.073 \times 0.054$\,mm$^3$, triclinic, space group $P$-1, a = 14.6806(9)\,Å, b = 22.2918(9)\,Å, c = 24.804(1)\,Å, $\upalpha = 88.938(3)^\circ$, $\upbeta = 80.136(4)^\circ$, $\upgamma = 86.426(4)^\circ$, V = 7981.4(7)\,Å$^3$, Z = 2, $\rho_{\text{calc}} = 1.253$\,mg/cm$^3$, $\upmu = 1.515$\,mm$^{-1}$, F(000) = 3184, $\uptheta_{\text{min}} = 2.675^\circ$, $\uptheta_{\text{max}} = 72.977^\circ$, 40464 reflections collected, 40464 unique, parameters / restraints = 1908 / 18, R$_1 = 0.0622$ and wR$_2 = 0.1431$ using 24963 reflections with $\text{I}>2\upsigma(\text{I})$, R$_1 = 0.0931$ and wR$_2 = 0.1532$ using all data, GOF = 0.882, $-0.398 < \Delta\rho < 1.033$\,e$/$Å$^3$.

\clearpage
\section{Absorption and photoluminescence spectra}

\begin{figure}[h]
    \centering
    \subfigure{\begin{overpic}%
    [width=0.45\textwidth]%
    {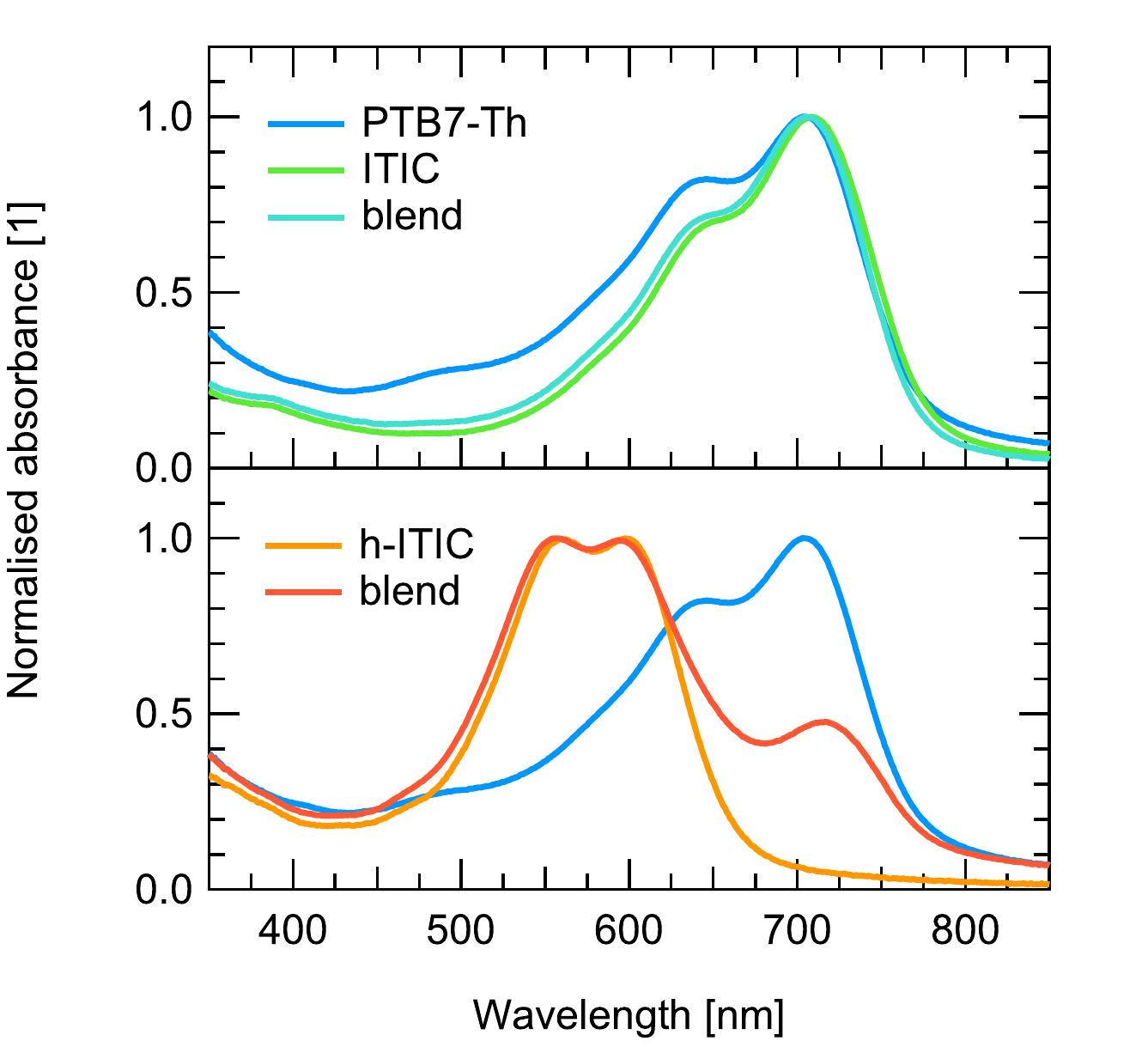}%
    \put(0,82){\Large(a)}%
    \put(109,82){\Large(b)}%
    \end{overpic}}
    \qquad
    \subfigure{\begin{overpic}%
    [width=0.46\textwidth]%
    {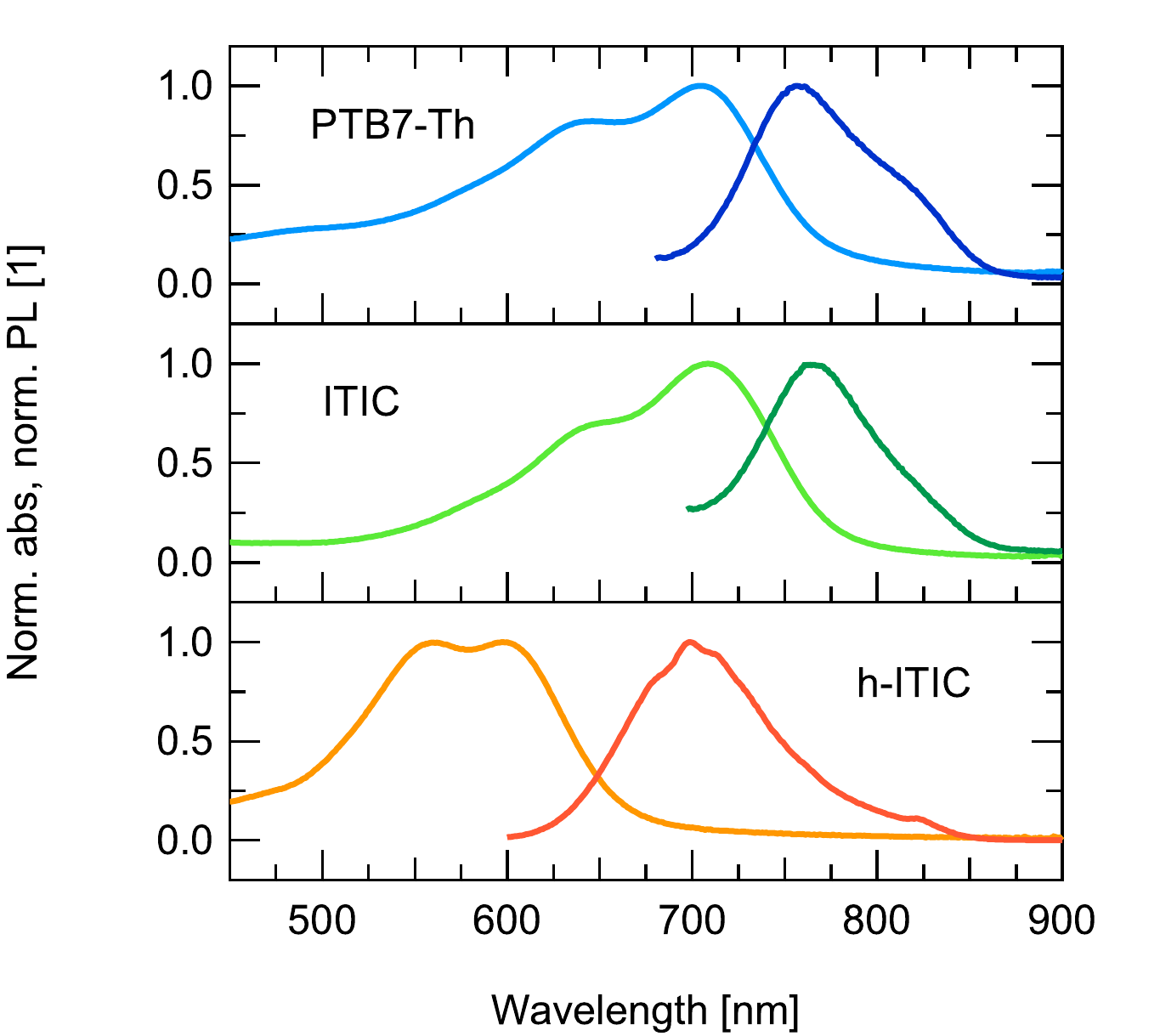}%
    \end{overpic}}
    \caption{(a) Normalised absorption spectra of PTB7-Th, ITIC, h-ITIC and blend films. (b) Normalised absorption and photoluminescence spectra of PTB7-Th, ITIC and h-ITIC films. The optical gap $E_\text{g}$ is determined as the crossing between the spectra. $E_\text{g}$ is 1.69\,eV for PTB7-Th, 1.68\,eV for ITIC and 1.91\,eV for h-ITIC.}
    \label{SI_FigureS10}
\end{figure}

\section{Singlet exciton losses}

\begin{figure}[h]
    \centering
    \subfigure{\begin{overpic}%
    [scale=.55]%
    {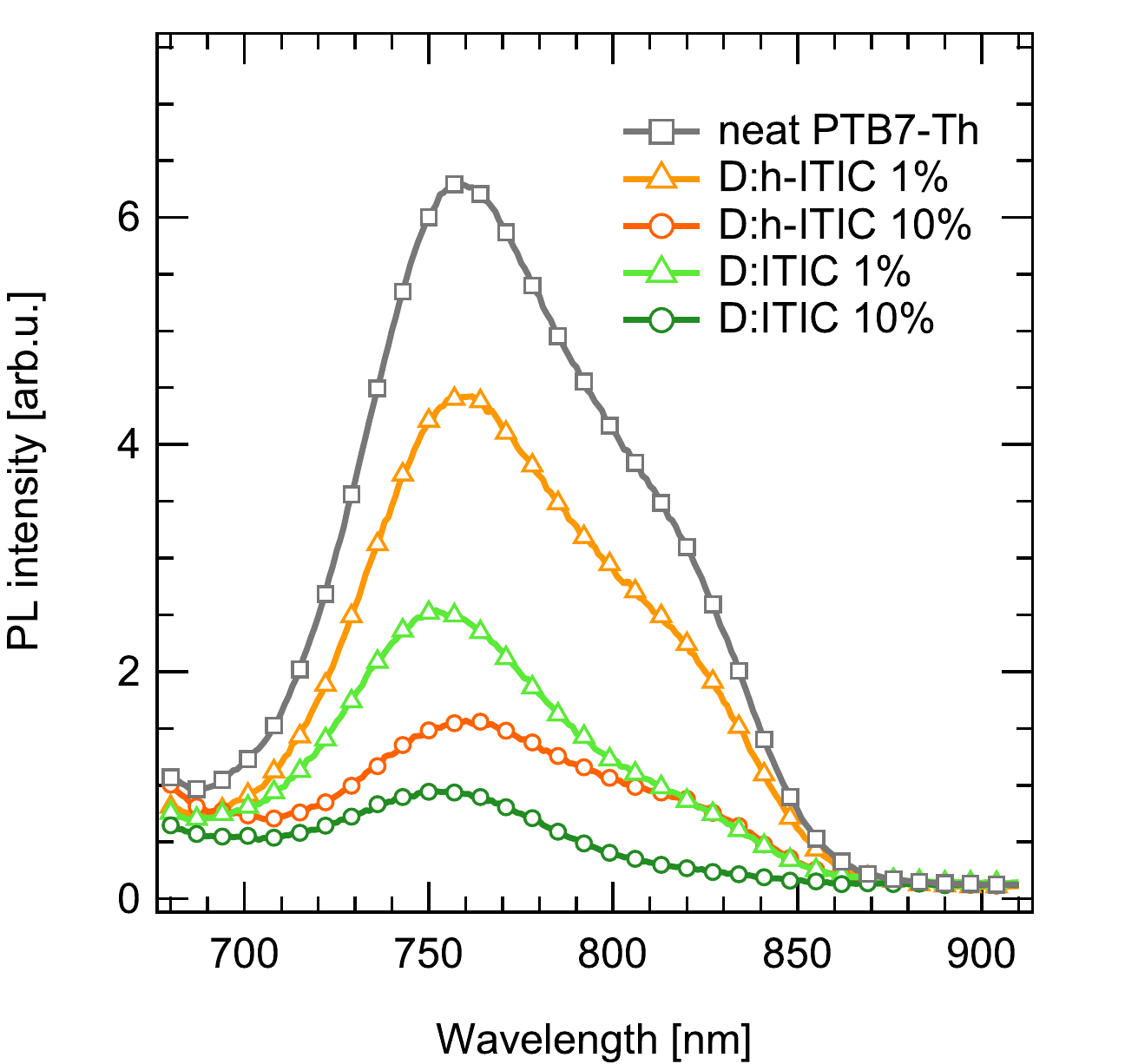}%
    \put(0,85){\Large(a)}%
    \put(109,85){\Large(b)}%
    \end{overpic}}
    \qquad
    \subfigure{\begin{overpic}%
    [scale=.55]%
    {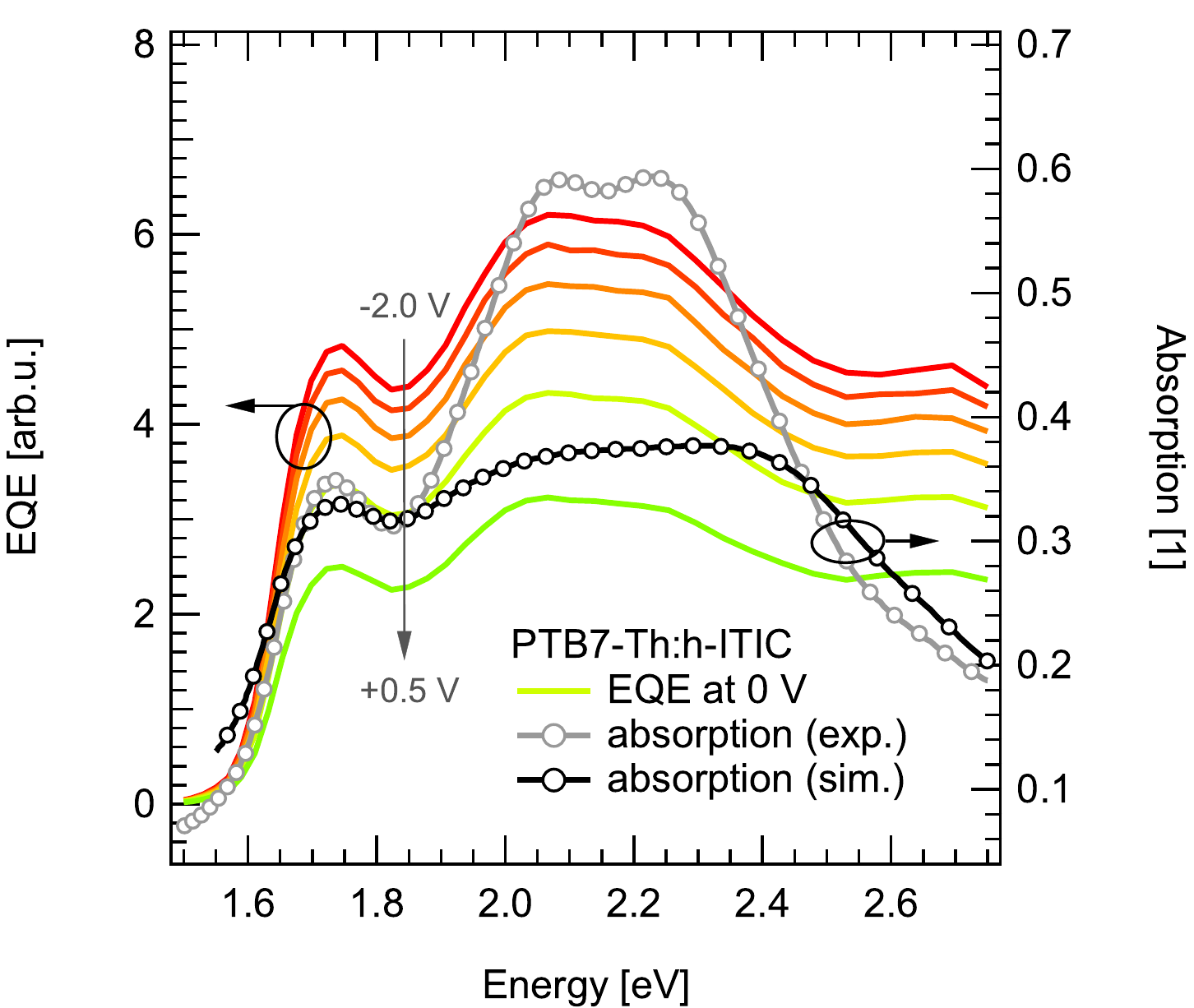}%
    \end{overpic}}
    \caption{(a) PL intensity of the neat PTB7-Th, PTB7-Th:ITIC and PTB7-Th:h-ITIC films with 1\% and 10\% quencher concentration at excitation wavelength of 540\,nm. Extrapolation of Stern-Volmer fits to the PL quantum yield gives an estimated quenching efficiency of ca.\ 98\% in PTB7-Th:ITIC and ca.\ 95\% in PTB7-Th:h-ITIC solar cells. (b) EQE of PTB7-Th:h-ITIC solar cells at various applied voltages and experimental and simulated absorption of PTB7-Th:h-ITIC film. EQE changes similarly with an applied external field at excitation energies relevant to donor and acceptor absorption, implying that the efficiencies of CT dissociation and charge transport are independent of excitation energy. EQE is not decreased in the absorption range of h-ITIC as compared to PTB7-Th, thus singlet exciton losses in the donor and acceptor phases are similar.}
    \label{SI_FigureS11}
\end{figure}

\clearpage
\section{Theoretical Calculations}
\subsection{Quantum chemical calculations and molecular configurations}
Using available crystal structures of h-ITIC and ITIC unit cells we have performed an analysis of all possible dimer configurations. In the case of both ITIC and h-ITIC there are two types of possible dimers. It turns out that only one dimer for each system is actually possible, for h-ITIC it is shown in blue (green dimer is not stable since it is leading to rising in the overall dipole moment). For ITIC two molecules forming a dimer with a closest intermolecular distance is shown in red. 

Dimer configurations extracted from the unit cell were used to calculate the quadrupole tensor of a single molecule ($Q_{20}^{mol}\approx51.3\,ea_0^2$ for ITIC and $Q_{20}^{mol}\approx9.7\,ea_0^2$ for h-ITIC) and a dimer ($Q_{20}^{tot}\approx94.2\,ea_0^2$ for ITIC and $Q_{20}^{tot}\approx38.8\,ea_0^2$ for h-ITIC). For both molecules we see more complex charge density distribution in the case of a unit cell, which leads to lower quadrupole moment per molecule as compared to the dimer configuration ($Q_{20}^{mol}\approx33.2\,ea_0^2$, $Q_{20}^{tot}\approx133.0\,ea_0^2$ for ITIC, $Q_{20}^{mol}\approx13.8\,ea_0^2$, $Q_{20}^{tot}\approx112.2\,ea_0^2$ for h-ITIC). 

\begin{figure}[h]
    \centering
    \includegraphics[width=0.95\textwidth]{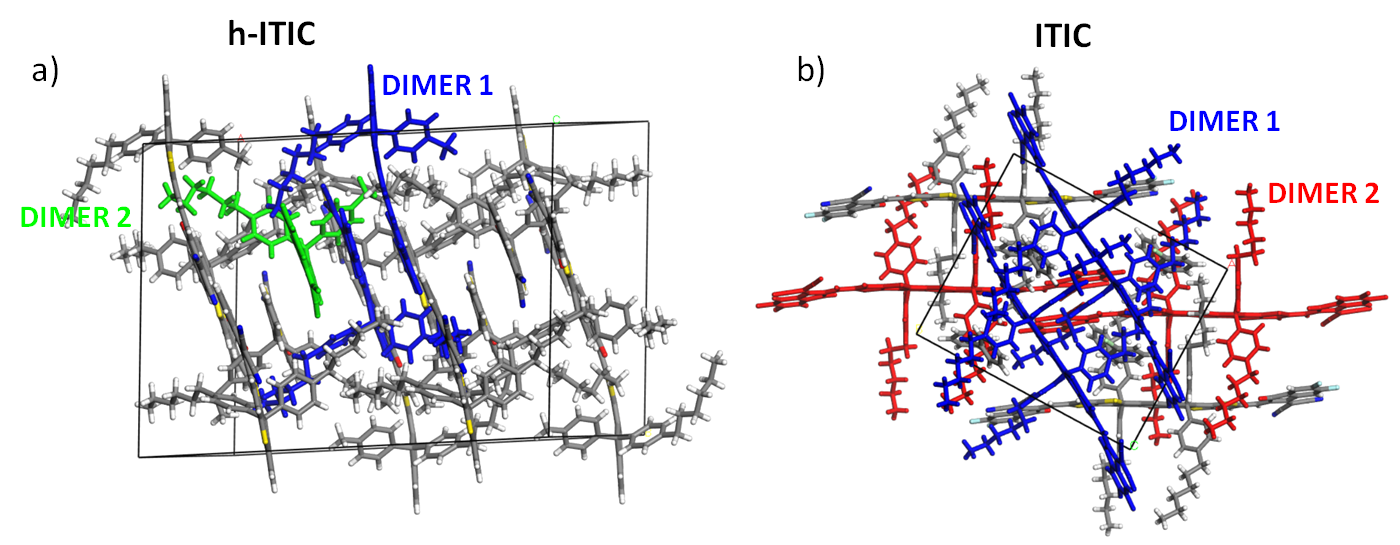}
    \caption{Crystal structures of a) h-ITIC (a possible dimer configuration is shown in blue, a green one - unfavourable) and b) ITIC (possible dimer configurations are shown in blue and green).}
    \label{SI_FigureS12}
\end{figure}

\clearpage
\subsection{Lattice model}
\begin{figure}[h]
    \centering
    \includegraphics[width=0.95\textwidth]{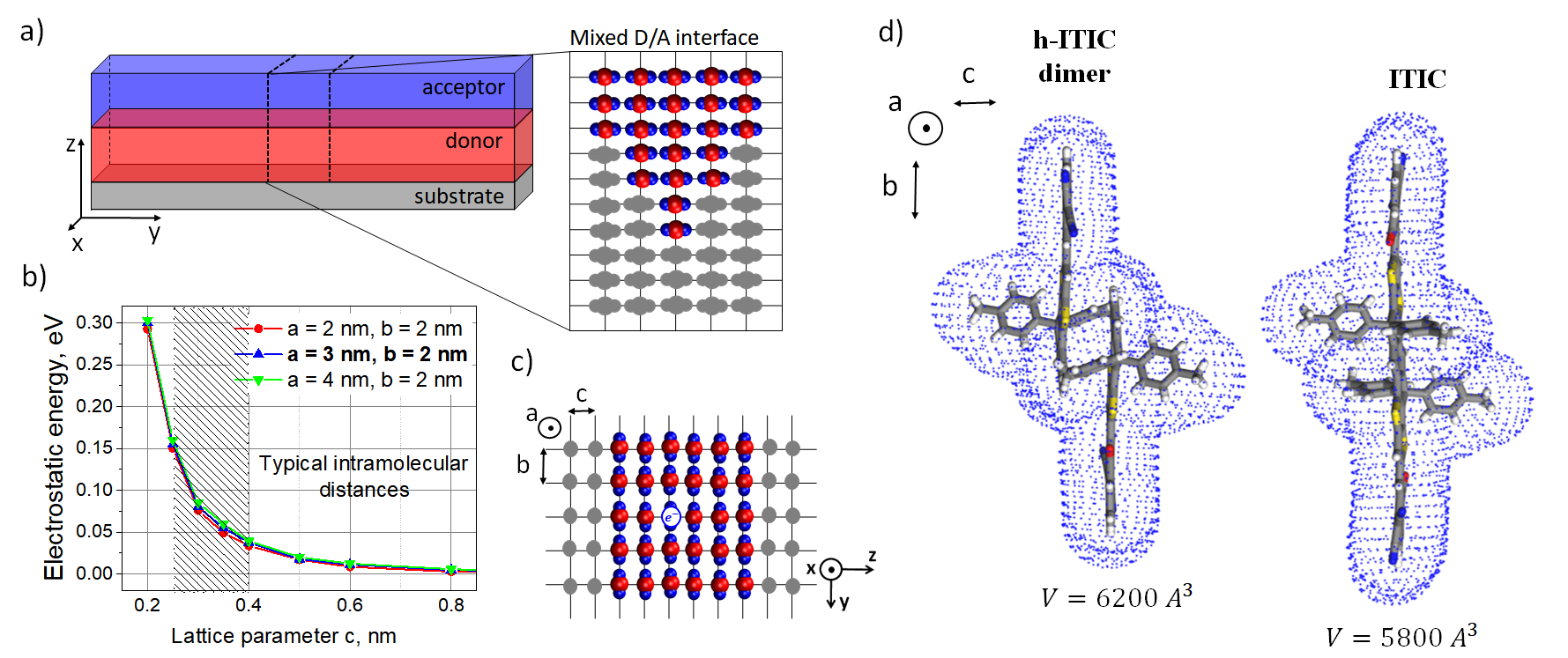}
    \caption{Crystal structures of a) Schematic representation of planar heterojunction thin film organic solar cell with rough mixed D/A interface, (zy-section is shown) a mixed thin film-quadrupoles interface. b) The total energy of charge-quadrupole interaction $E$ upon changing intramolecular distances $c$ along $z$ direction (which corresponds to the change of $\pi-\pi$ stacking distance). The different curves correspond to different lattice spacing, particularly blue lines correspond to the lattice spacing parameters $a = 3$\,nm, $b = 2$\,nm, $c = 0.4$\,nm (shown in bold in the legend), one of the typical NFAs packing parameters. c) A model lattice containing a charge (an electron) with lattice parameters $a$, $b$ and $c$. d) Van der Waals volumes of h-ITIC dimer ($V = 6200$\,\AA$^3$) and ITIC ($V = 5800$\,\AA$^3$).}
    \label{SI_FigureS13}
\end{figure}

We built a model system consisted of point quadrupoles (representing acceptor molecules) distributed on a lattice mimicking an interface with a vacuum. The vacuum was represented as a point without any charges or quadrupoles. Schematic representation of the model lattice is shown in Figure \ref{SI_FigureS13}(a). Stabilisation energy of a CT state is very sensitive to changing of the shortest intermolecular distance. In Figure~\ref{SI_FigureS13}(b) shadow region shows a range of possible $\pi-\pi$ staking (0.2-0.5\,nm), the energy drop within that region is 7.5 times. Parameters $a$ and $b$ are shown in Figure~\ref{SI_FigureS13}(c). 

Besides different quadrupole moments h-ITIC dimer and ITIC have various Van der Waals volumes, see Figure~\ref{SI_FigureS13}(d). This means that the lattice parameters while packing also are different ($\frac{c_d}{c_s}=1.2$, $\frac{b_d}{b_s}=0.93$). Therefore electrostatic driving force in the case of h-ITIC is lower than for ITIC.   

\clearpage
\section{Transient absorption}

\begin{figure}[h]
    \centering
    \subfigure{\begin{overpic}%
    [width=0.45\textwidth]%
    {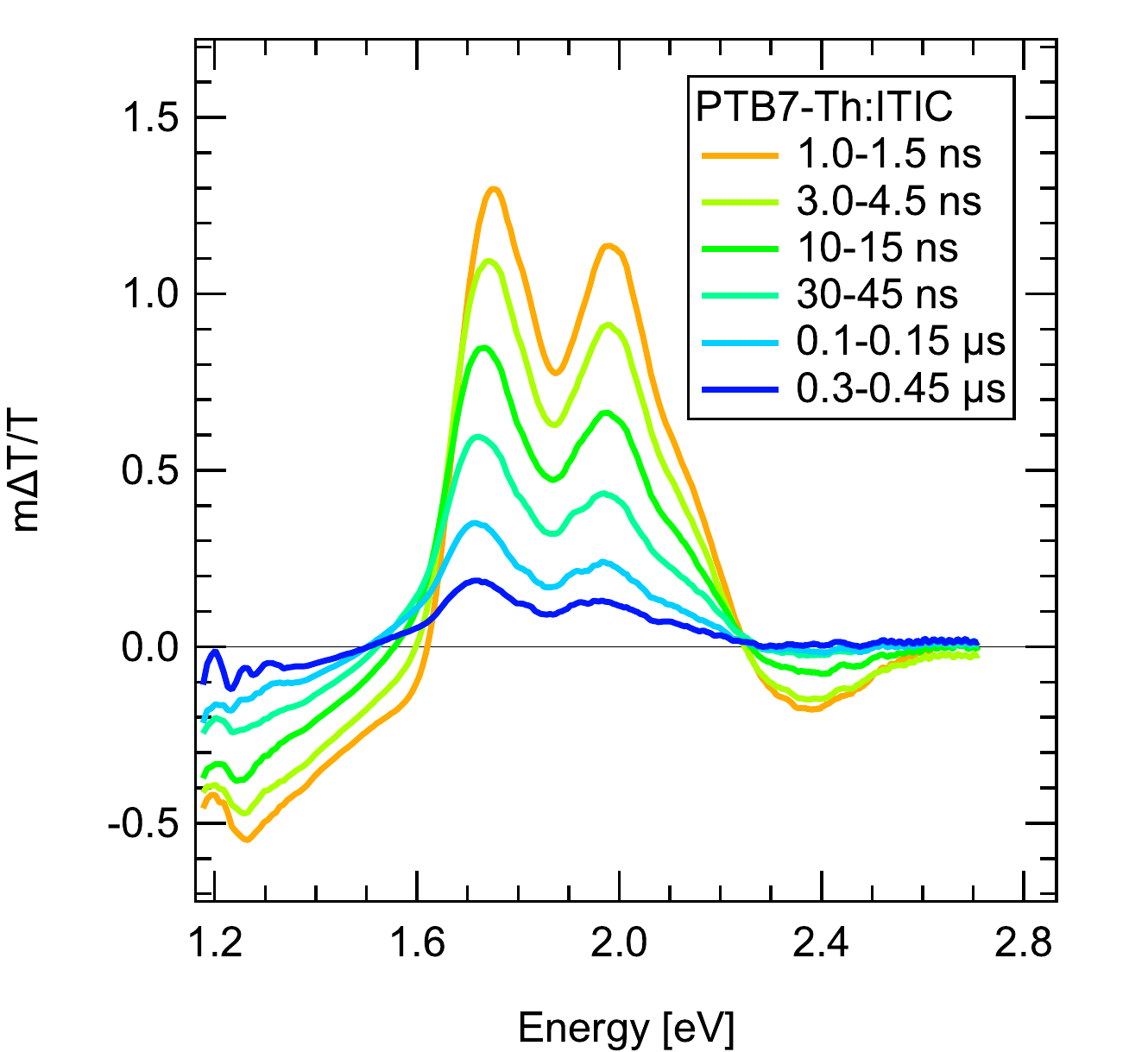}%
    \put(0,83){\Large(a)}%
    \put(109,83){\Large(b)}%
    \end{overpic}}
    \qquad
    \subfigure{\begin{overpic}%
    [width=0.45\textwidth]%
    {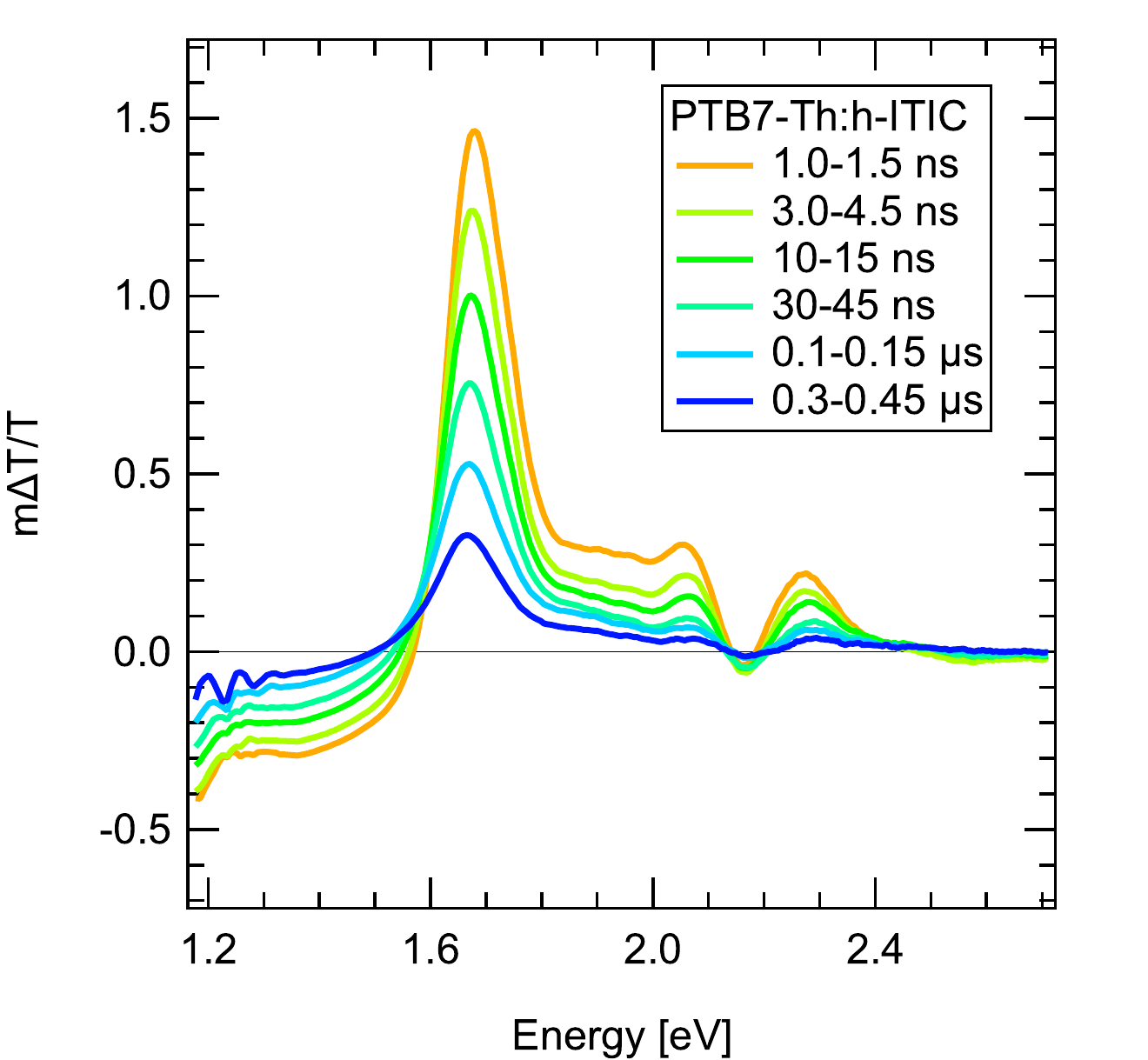}%
    \end{overpic}}
    \caption{ns-$\upmu$s transient absorption spectra of (a) PTB7-Th:ITIC film and (b) PTB7-Th:h-ITIC film upon excitation at 532\,nm with 2.1\,$\upmu$J/cm$^2$ and 1.6\,$\upmu$J/cm$^2$ fluence, respectively. The charge carrier density kinetics is monitored in the probe range of 1.6 - 2.2\,eV for the ITIC and 1.6 - 1.8\,eV for the h-ITIC blend film.}
    \label{SI_FigureS14}
\end{figure}

\begin{figure}[h]
    \centering
    \subfigure{\begin{overpic}%
    [width=0.47\textwidth]%
    {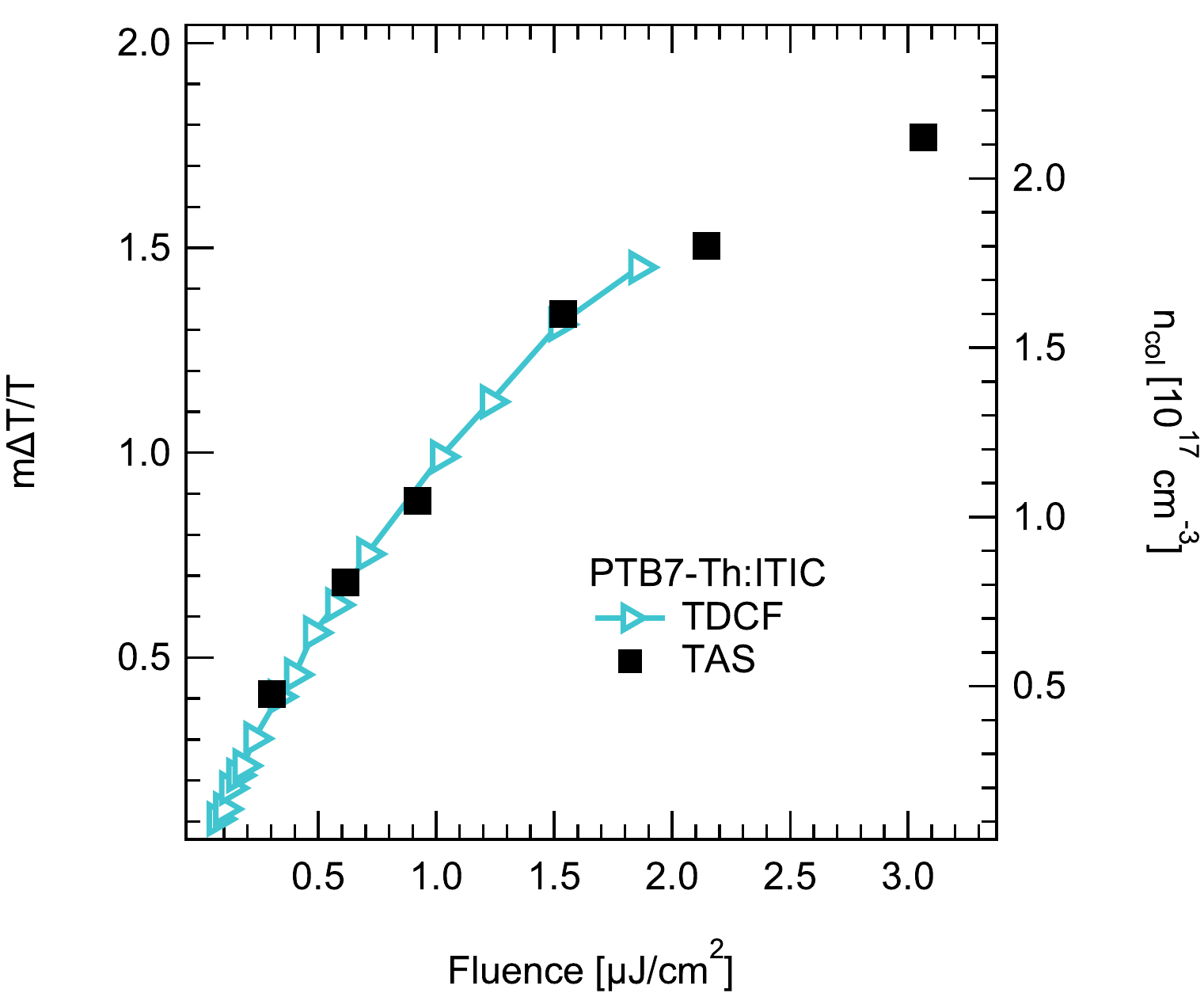}%
    \put(0,76){\Large(a)}%
    \put(109,76){\Large(b)}%
    \end{overpic}}
    \qquad
    \subfigure{\begin{overpic}%
    [width=0.47\textwidth]%
    {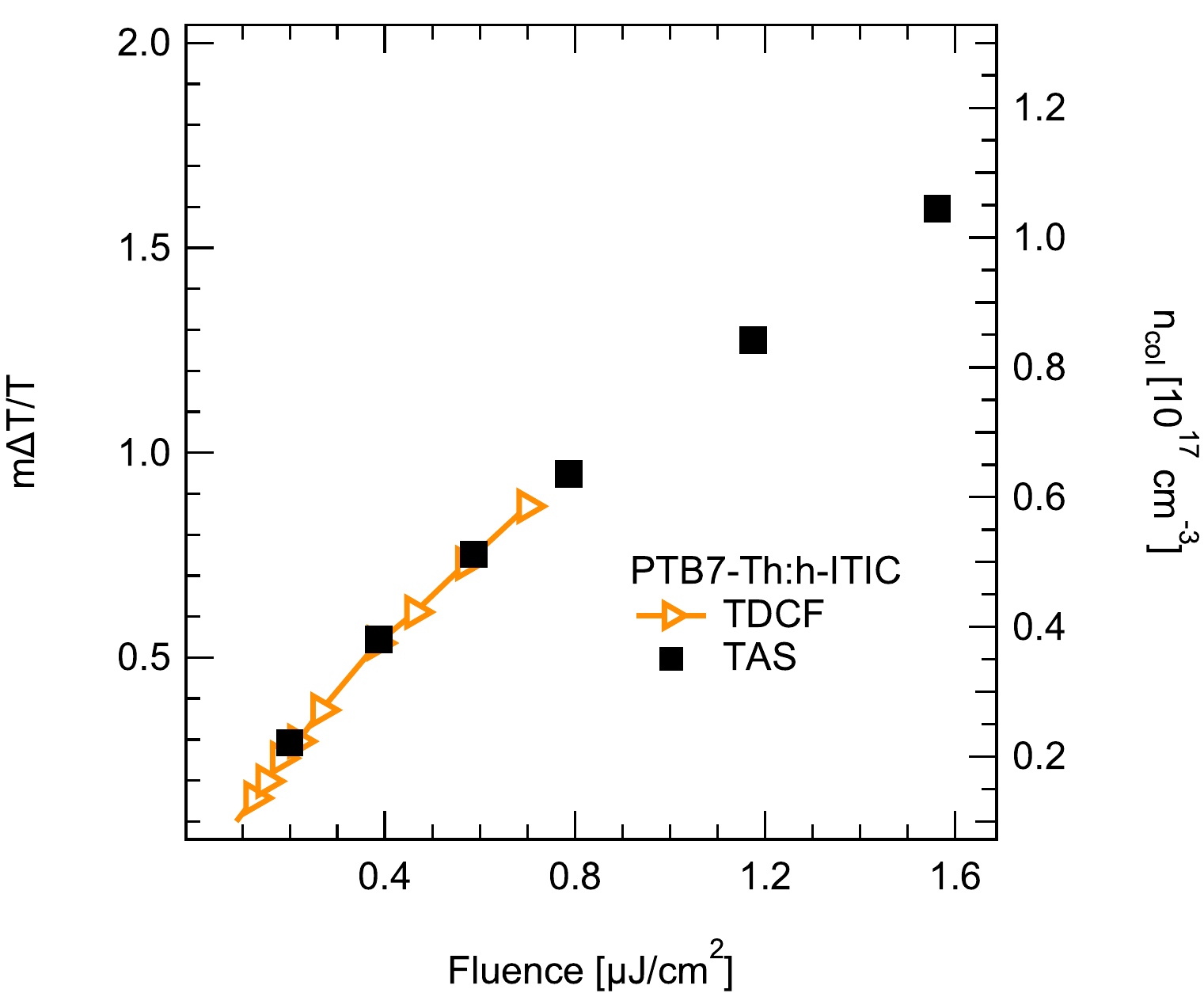}%
    \end{overpic}}
    \caption{Transient absorption signal $\Delta T/T$ at 5\,ns (full squares) and charge carrier density $n_\text{col}$ extracted from time delayed collection field at delay time of 5\,ns (open triangles) for (a) PTB7-Th:ITIC and (b) PTB7-Th:h-ITIC upon excitation at 532\,nm. $n_\text{col}$ from TDCF is used to estimate the initial charge carrier density $n_0$ in the global fit to the charge carrier density kinetics extracted from TAS.}
    \label{SI_FigureS15}
\end{figure}

\clearpage
\section{Charge carrier mobilities}

\begin{figure}[!h]
    \centering
    \includegraphics[width=0.95\textwidth]{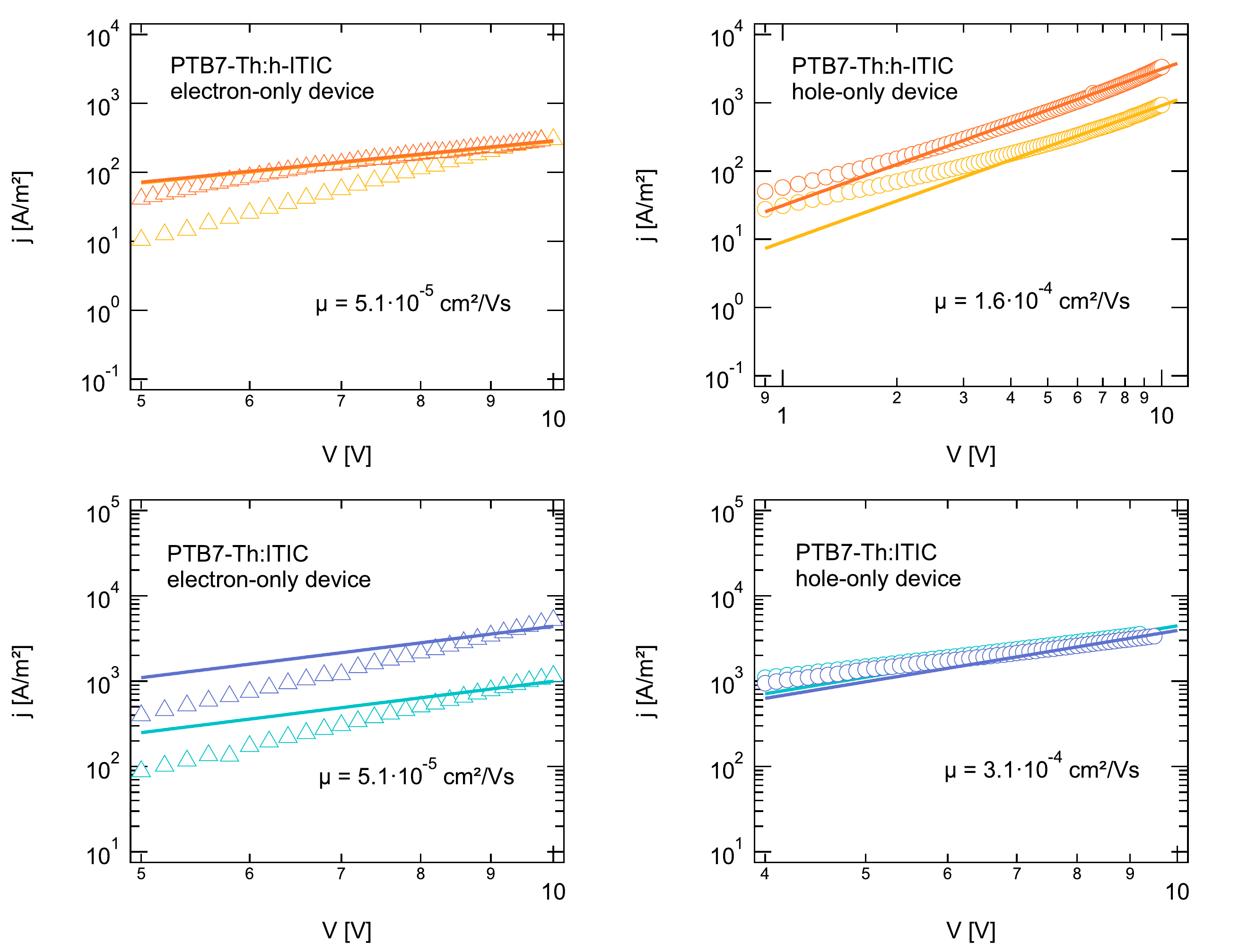}
    \caption{Charge carrier mobilities $\mu$ of PTB7-Th:ITIC and PTB7-Th:h-ITIC solar cells determined from the fits to SCLC data of the hole-only and electron-only devices according to $j_\text{sclc} = 9\epsilon\epsilon_0\mu V^2 / (8 L^3)$, where $\epsilon\epsilon_0$ is the permittivity of the organic semiconductor material, $V$ is the applied voltage and $L$ is the active layer thickness.}
    \label{SI_FigureS16}
\end{figure}

\clearpage
\section{Transport resistance}

\begin{figure}[!h]
    \centering
    \subfigure{\begin{overpic}%
    [width=0.45\textwidth]%
    {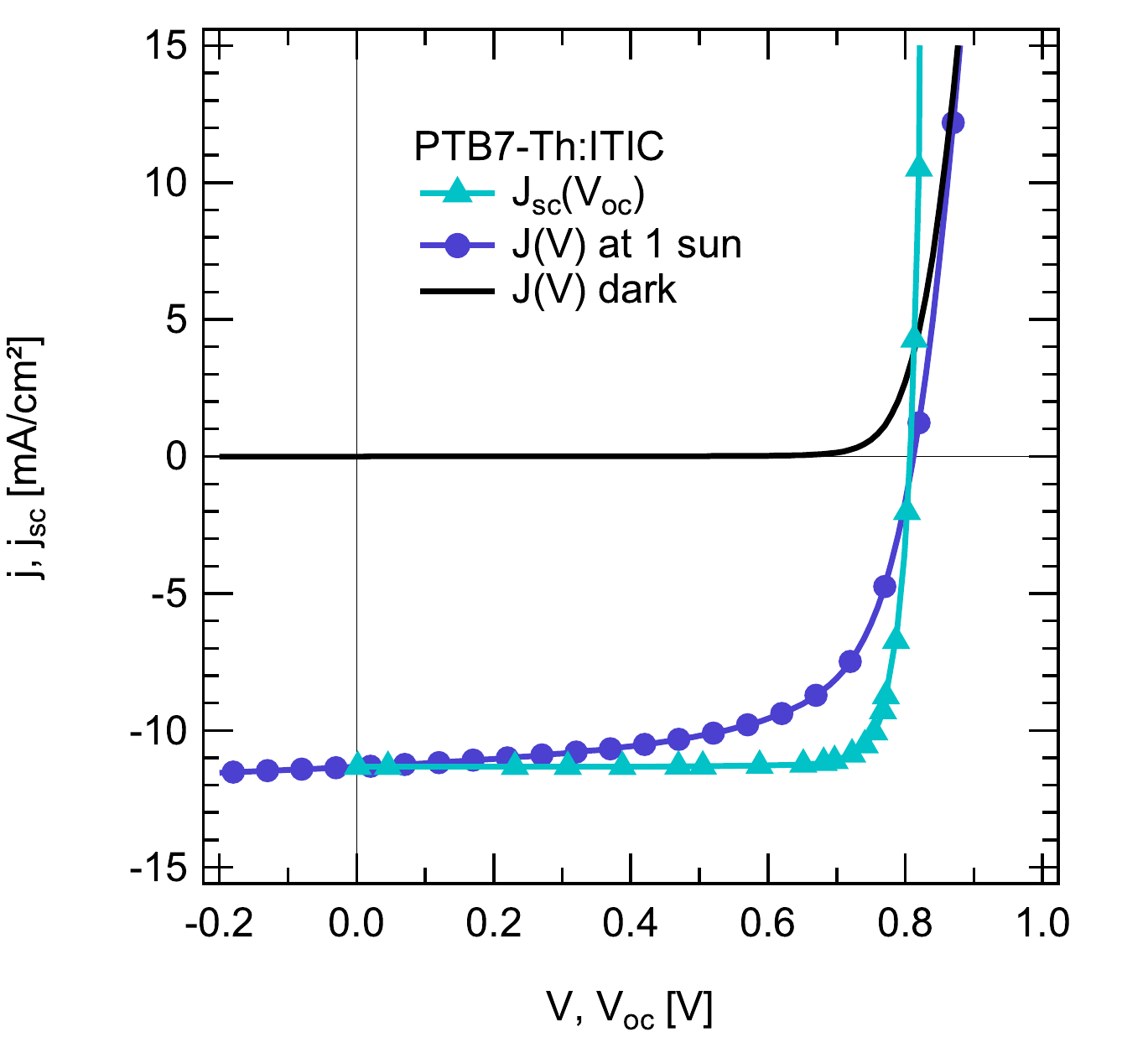}%
    \put(0,82){\Large(a)}%
    \end{overpic}}
    \subfigure{\begin{overpic}%
    [width=0.45\textwidth]%
    {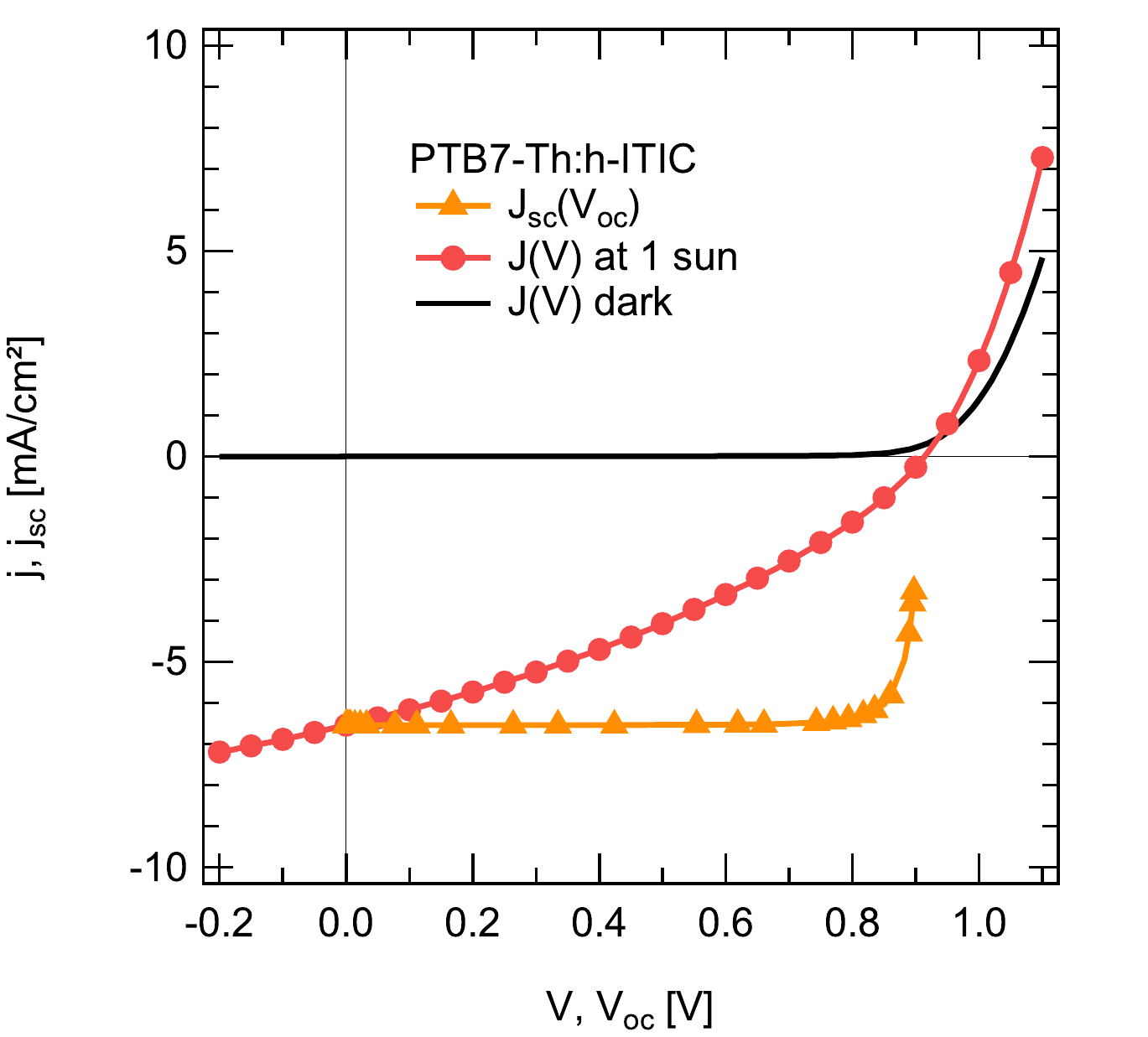}%
    \put(0,82){\Large(b)}%
    \end{overpic}}
    \caption{Pseudo-$J(V)$ curves of (a) PTB7-Th:ITIC and (b) PTB7-Th:h-ITIC constructed by shifting down the $J_\text{sc}(V_\text{oc})$, compared to $J(V)$ under 1 sun illumination.}
    \label{SI_FigureS17}
\end{figure}

\section{Charge Transfer State Energy}

\begin{figure}[!h]
    \centering
    \includegraphics[width=0.45\textwidth]{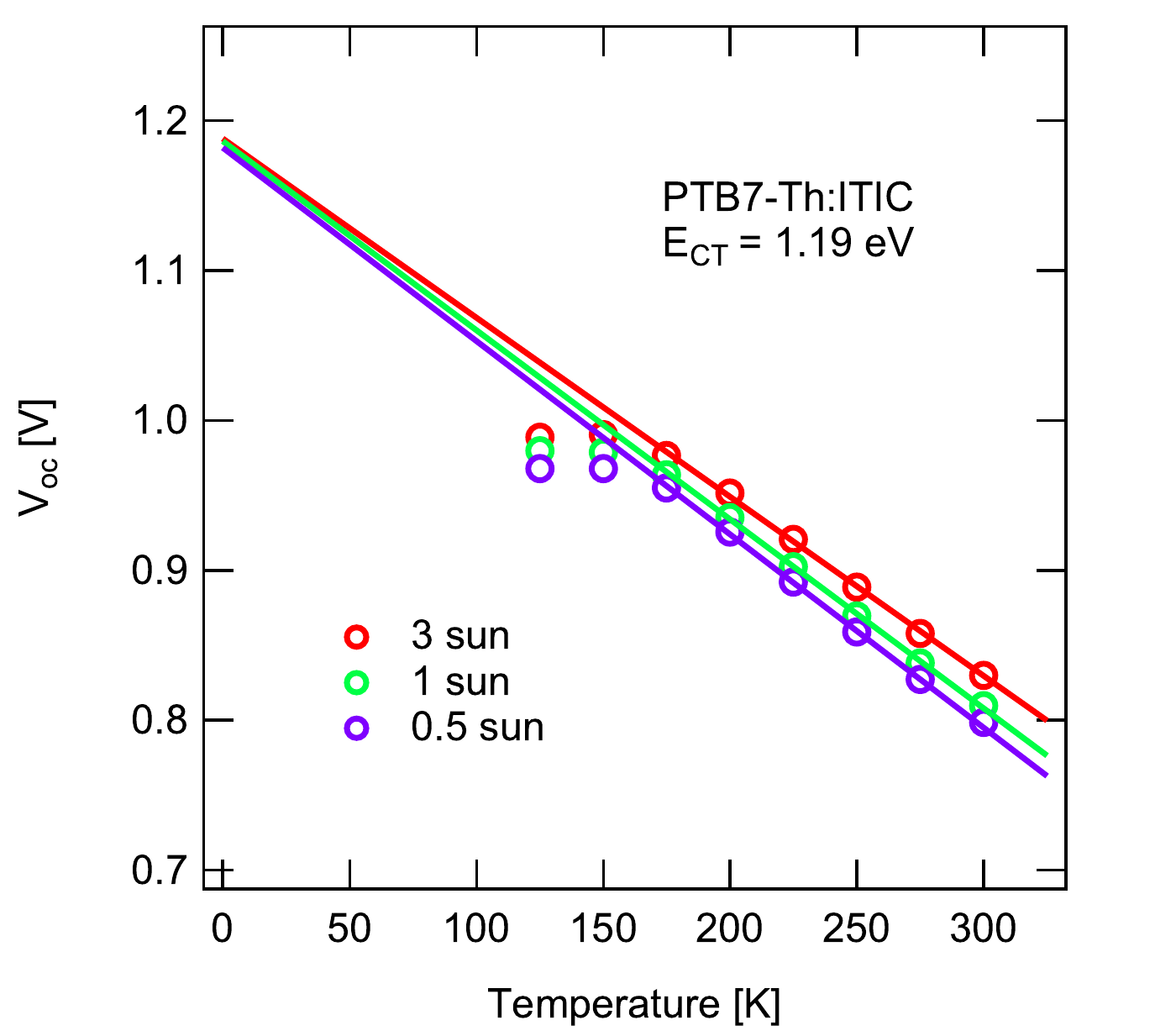}
    \caption{$V_\text{oc}$ of PTB7-Th:ITIC vs temperature. Extrapolation of linear fit to $0$\,K gives the lower limit of charge transfer state energy $E_\text{CT}$.}
    \label{SI_FigureS18}
\end{figure}

\clearpage
\section{Time Delayed Collection Field}

\subsection{Fluence dependence of the total extracted charge}

\begin{figure}[!h]
    \centering
    \includegraphics[width=0.45\textwidth]{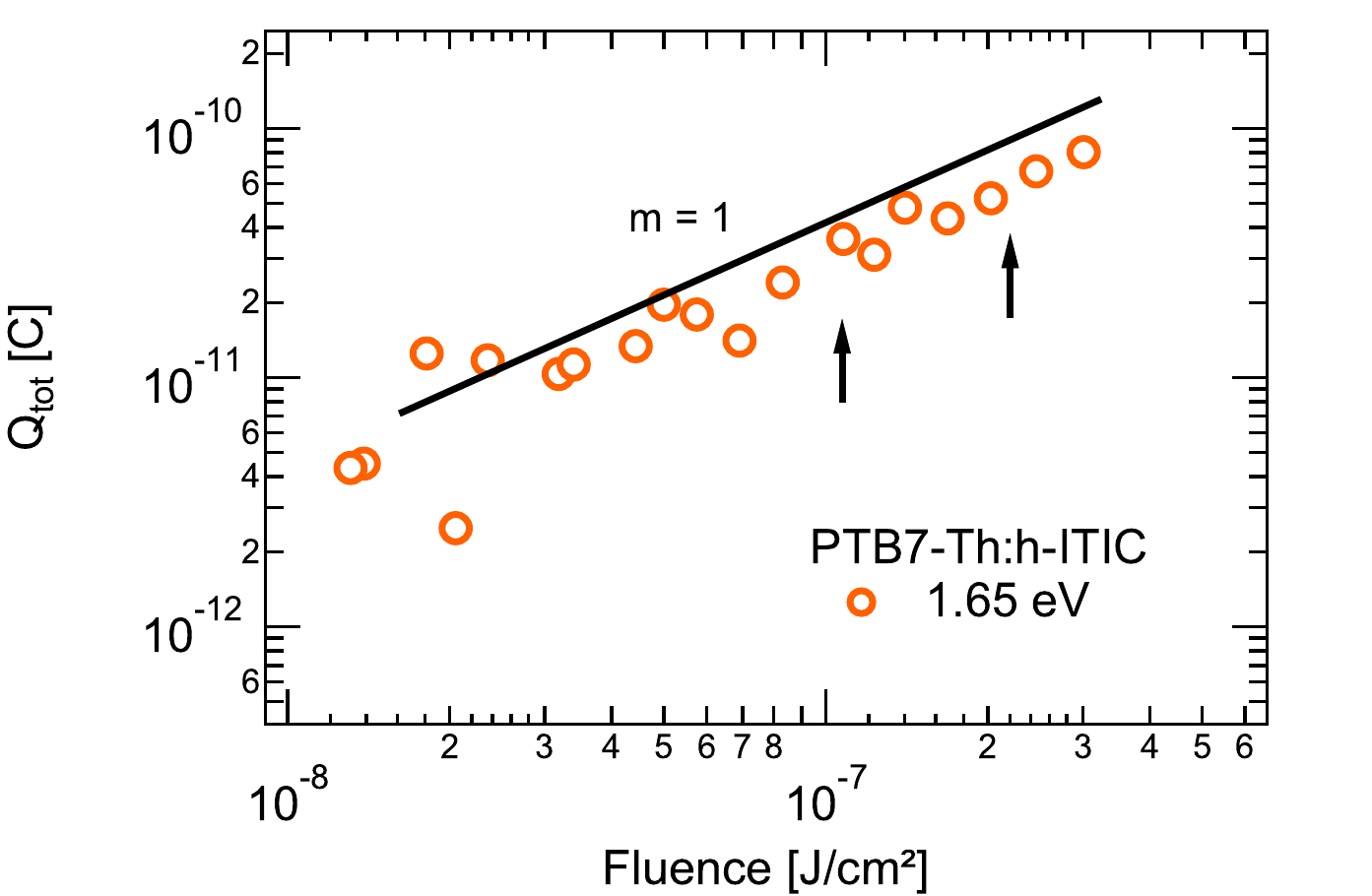}
    \includegraphics[width=0.45\textwidth]{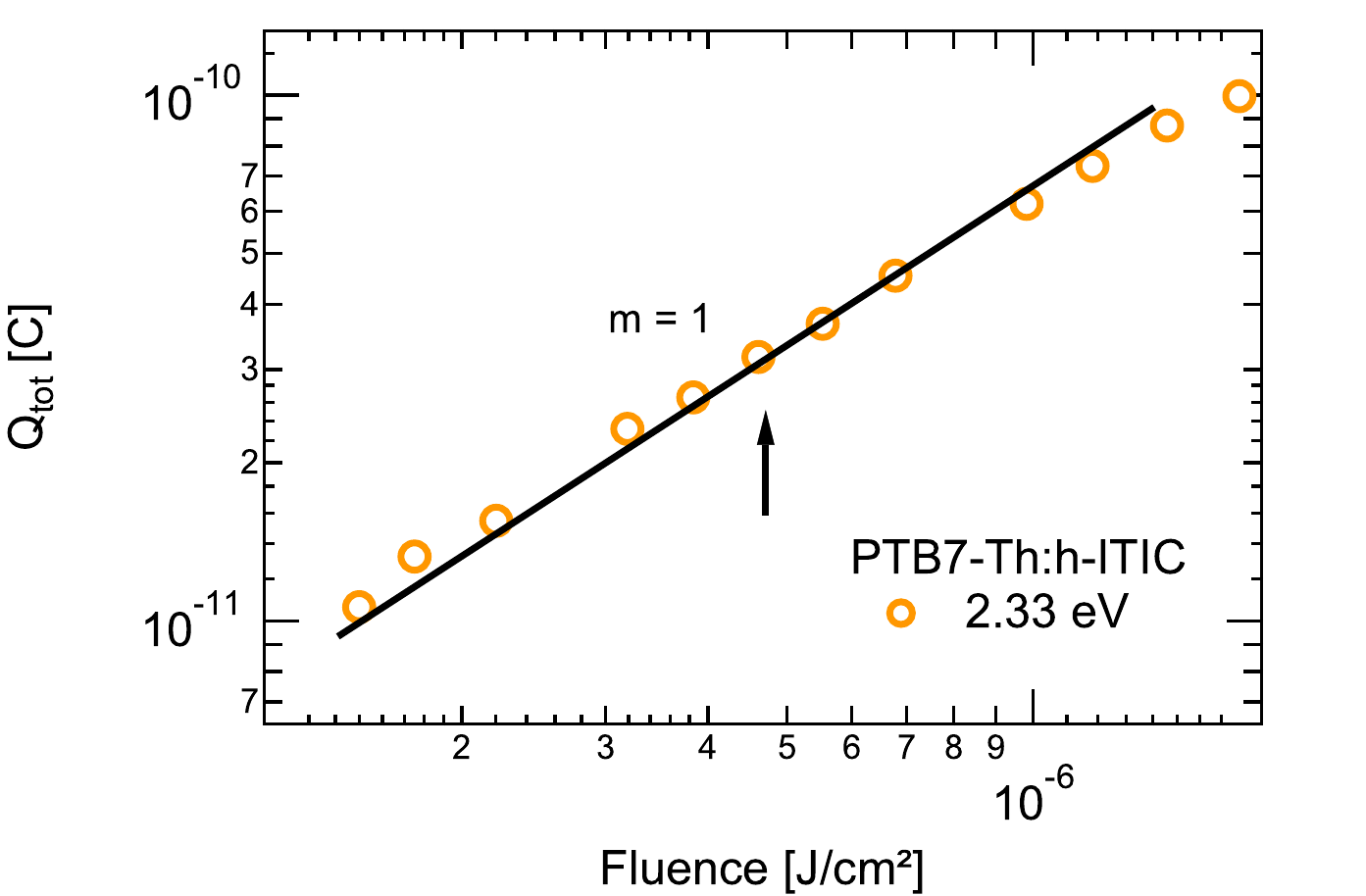}
    \includegraphics[width=0.45\textwidth]{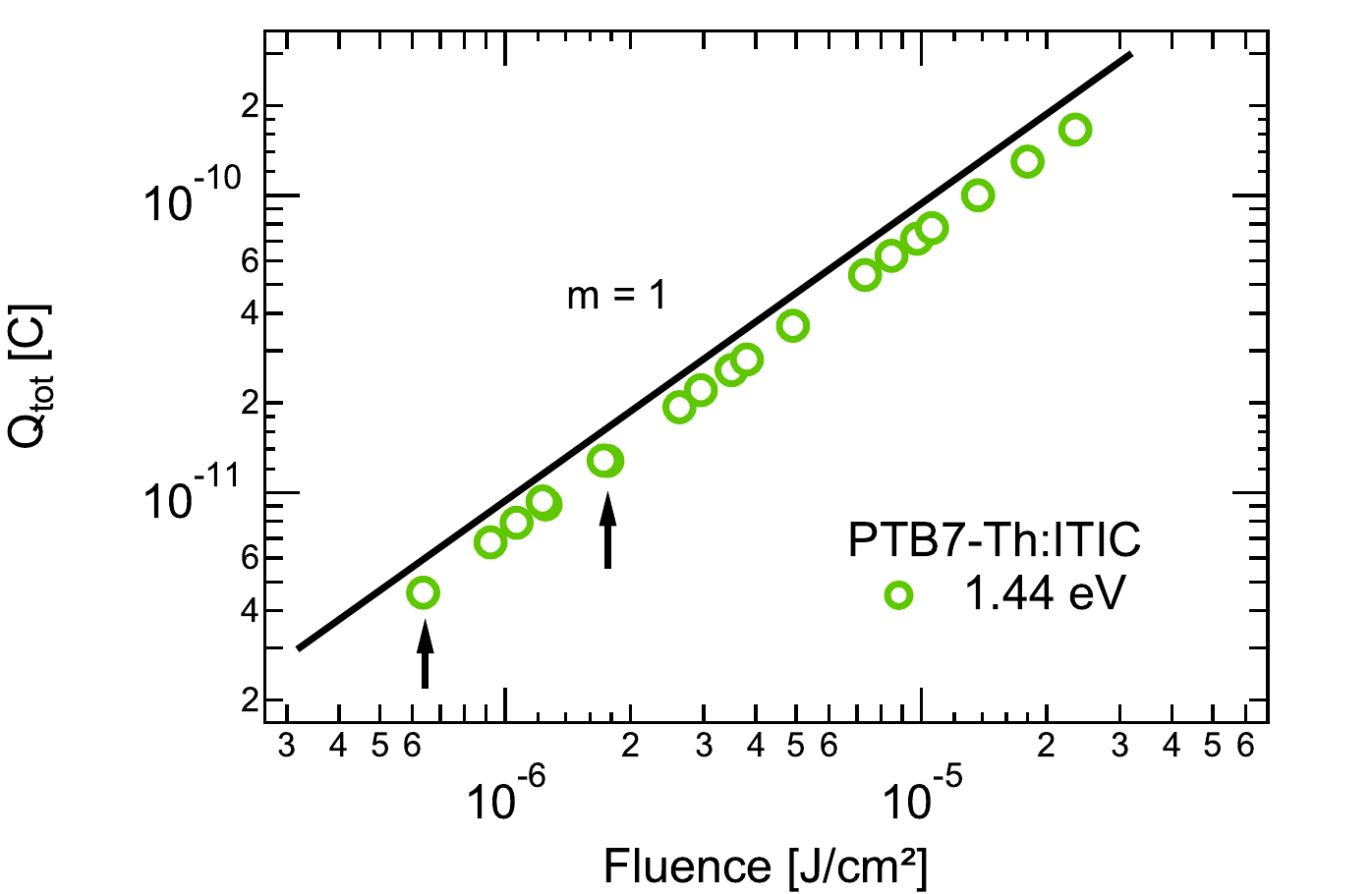}
    \includegraphics[width=0.45\textwidth]{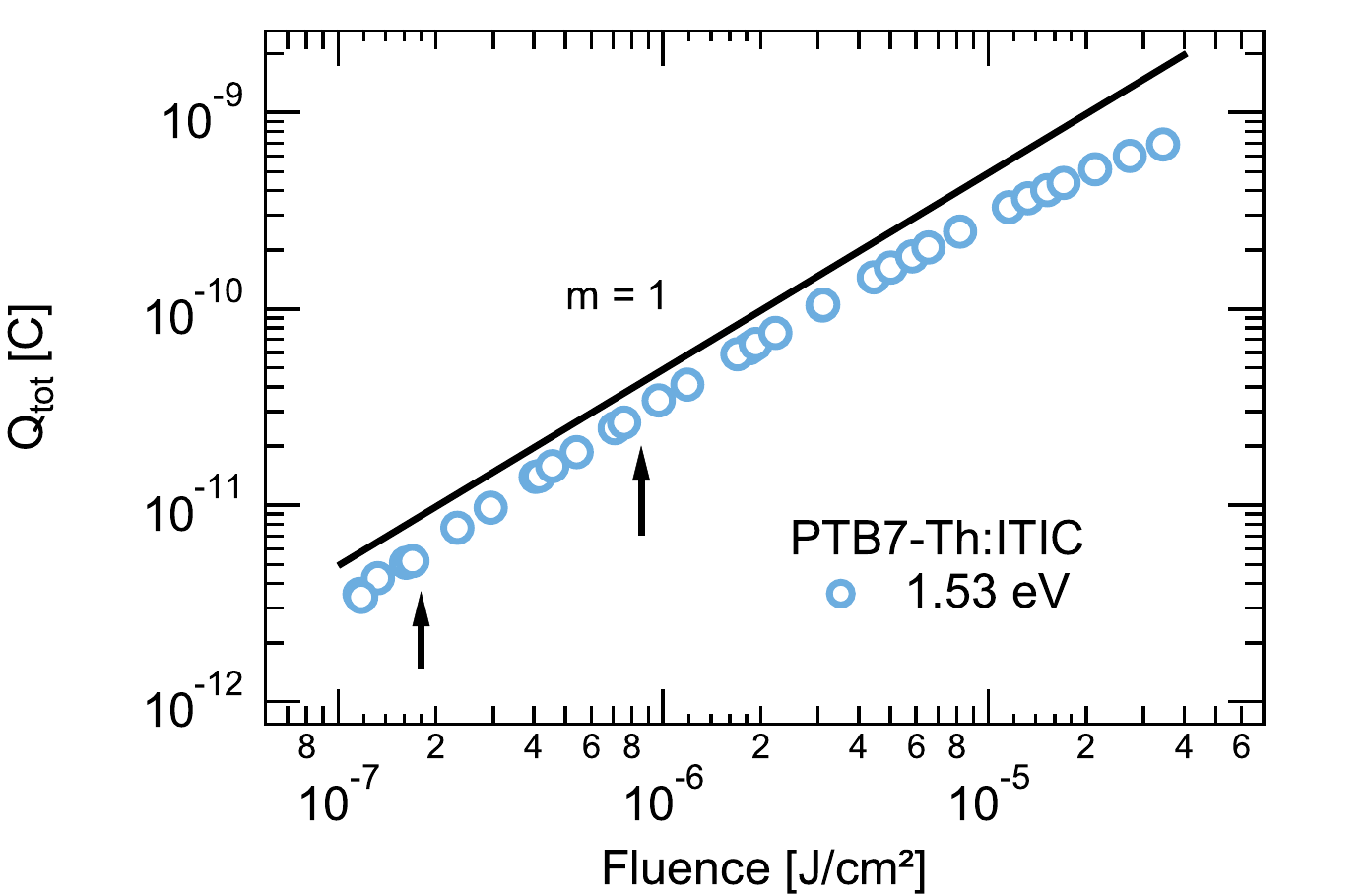}
    \includegraphics[width=0.45\textwidth]{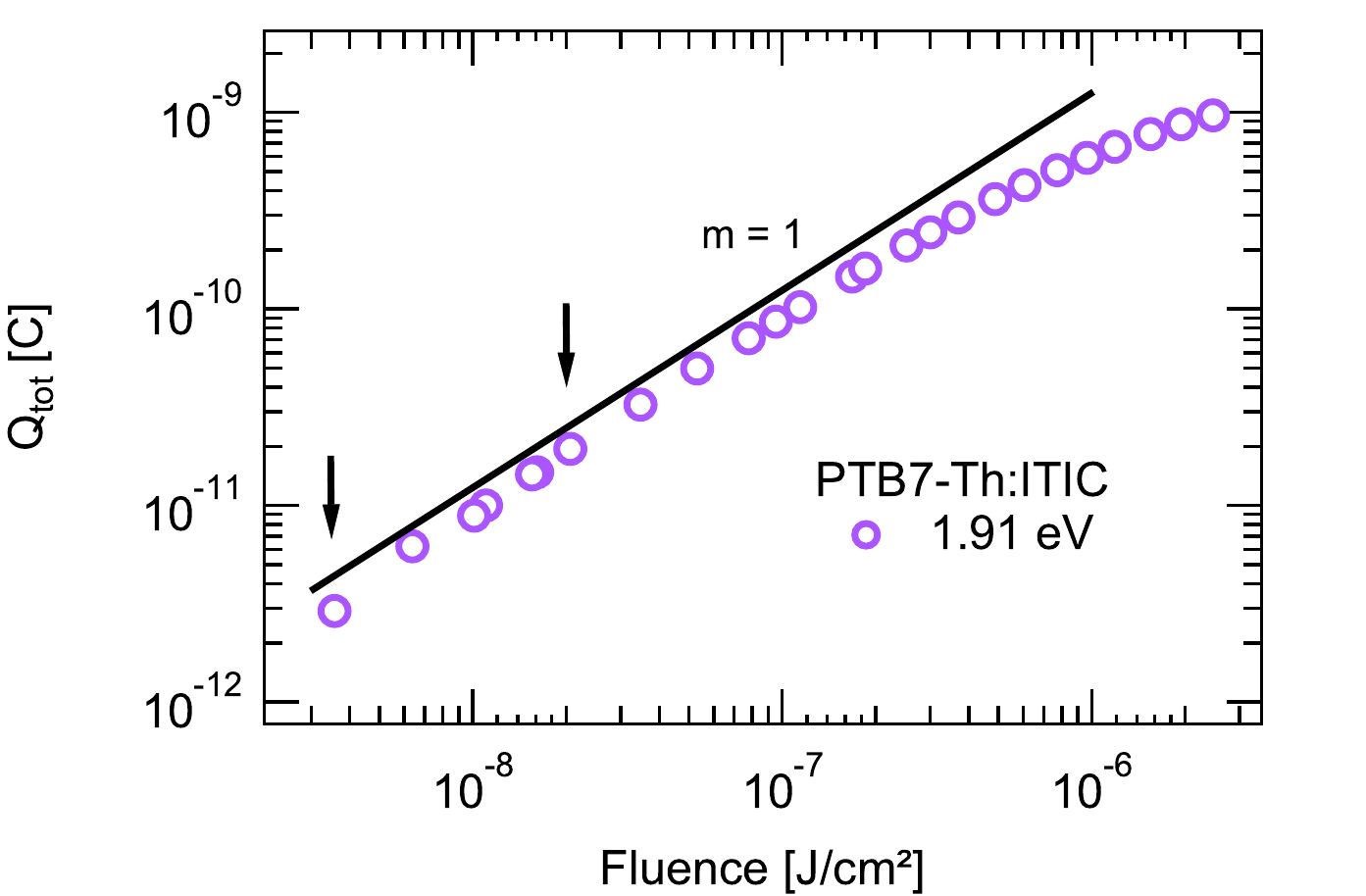}
    \caption{Total extracted charge $Q_\text{tot}$ from TDCF as a function of laser pulse fluence at excitation energies of 1.65\,eV and 2.33\,eV for the h-ITIC blend, and 1.44\,eV, 1.53\,eV and 1.91\,eV for the ITIC blend. Slope 1 is indicative of the absence of dominant 2\textsuperscript{nd} order recombination losses during extraction. Chosen fluences for photogeneration measurements are indicated by arrows.}
    \label{SI_FigureS19}
\end{figure}

\clearpage
\subsection{Delay time of electrical probe}

\noindent The delay time of electrical probe after the laser pulse in TDCF experiment is supposed to be longer than the effective CT state lifetime $\tau_\text{CT}$, yet short enough to minimise contribution from nongeminate recombination. $\tau_\text{CT} = 2.8$\,ns at zero applied field, as determined from TAS. Assuming exponential decay, ca.\ 17\% of the initial CT population survives at 5\,ns after the laser pulse at zero field. This value is expected to decrease in the presence of external field.
\begin{figure}[!h]
    \centering
    \includegraphics[width=0.45\textwidth]{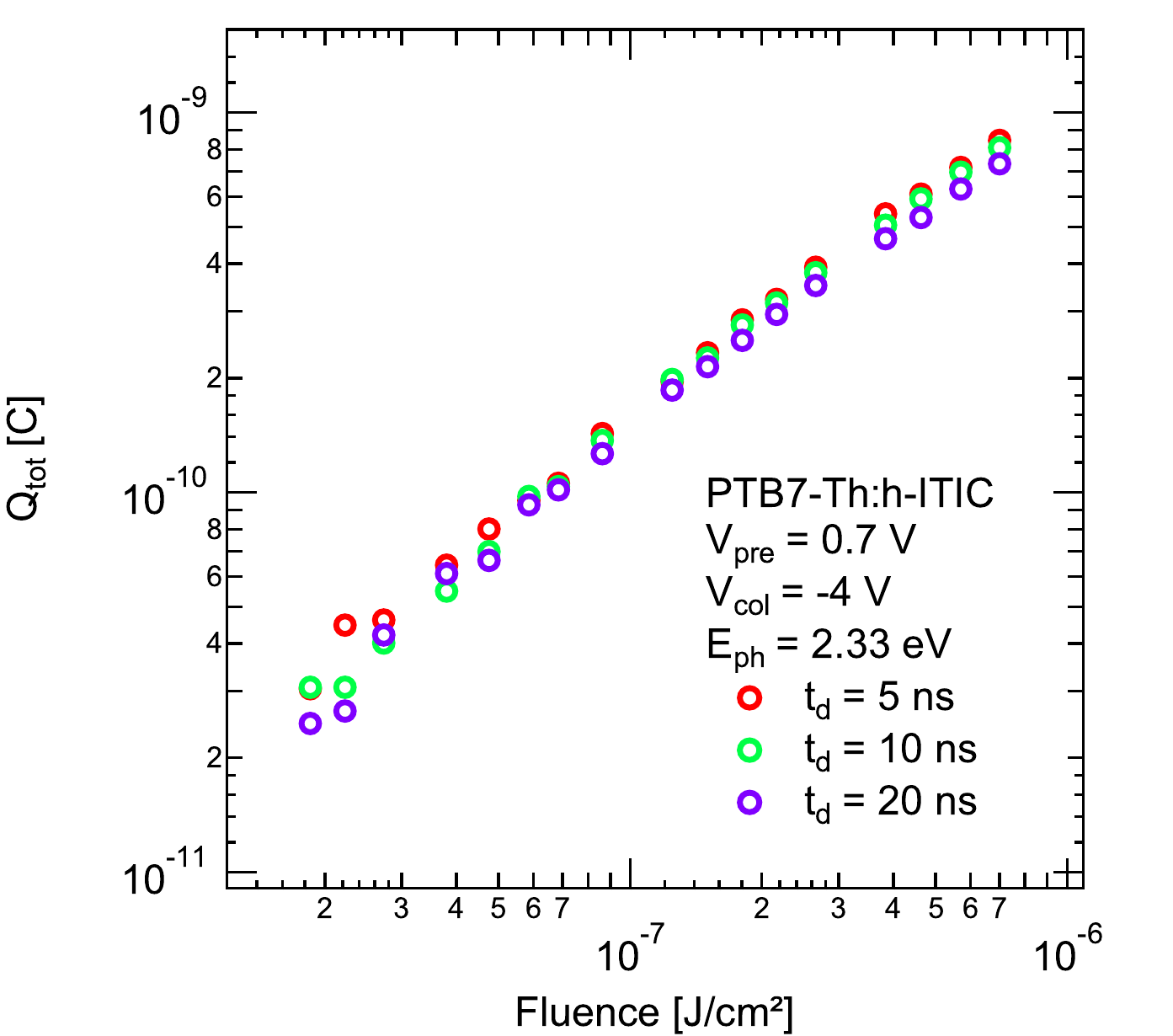}
    \caption{Total extracted charge $Q_\text{tot}$ from TDCF as a function of laser pulse fluence for PTB7-Th:h-ITIC at excitation energy of 2.33\,eV. $Q_\text{tot}$ and its fluence dependence changes little with $t_\text{d}$ between 5\,ns and 20\,ns, which confirms that the chosen delay time of 5\,ns for photogeneration measurements is adequate.}
    \label{SI_FigureS20}
\end{figure}
 
\subsection{Charge carrier densities}

\begin{figure}[h]
    \centering
    \begin{minipage}[c][7.5cm][t]{.5\textwidth}
    \vspace*{\fill}
    \includegraphics[trim = 7.5cm 0cm 0cm 0cm, clip,width=8cm]{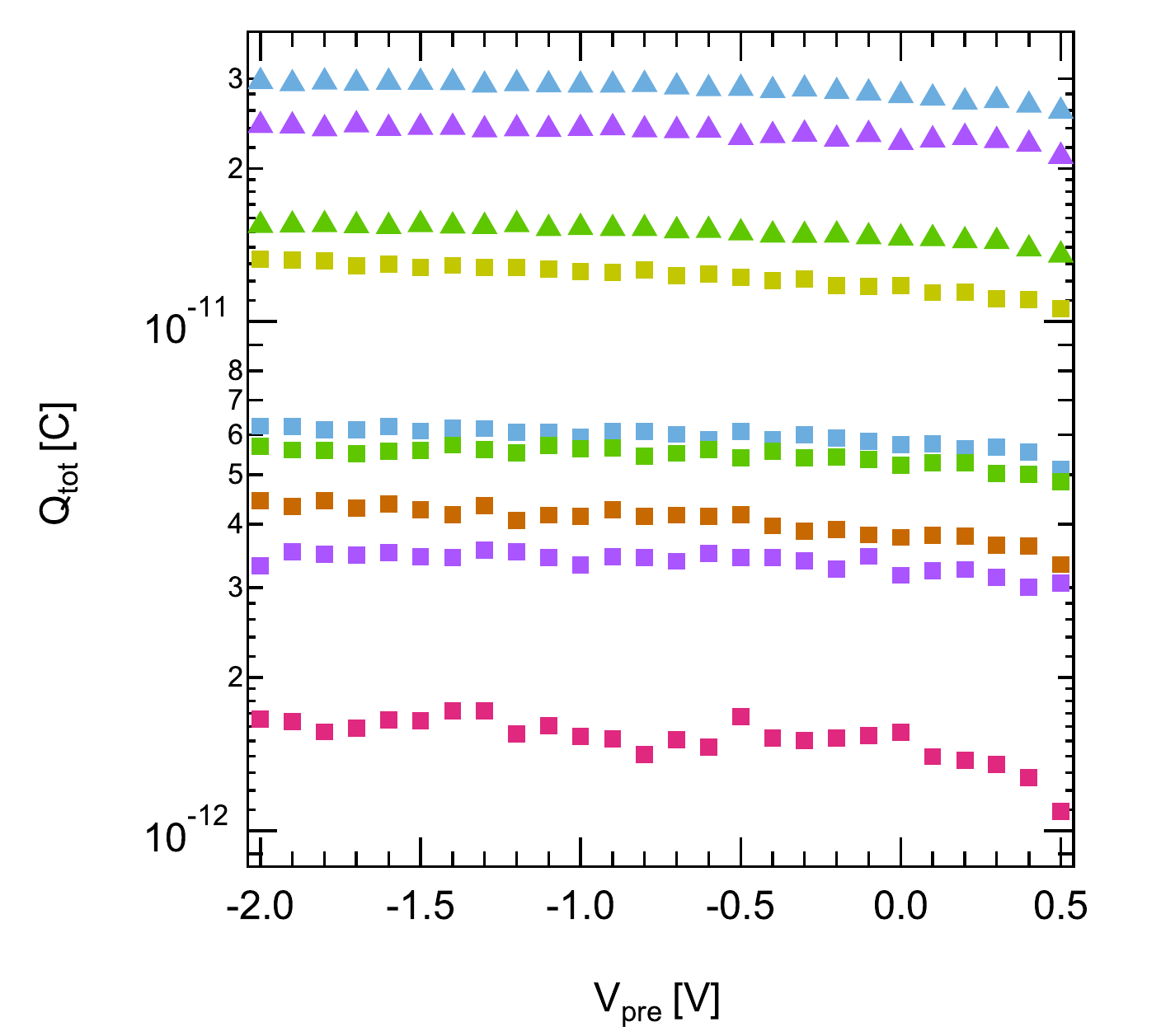}
    \end{minipage}%
    \begin{minipage}[c][7.5cm][t]{.25\textwidth}
    \vspace*{\fill}
    \includegraphics[trim = 0cm 0cm 14.5cm 0cm, clip, width=4cm]{figures/SI_FigureS21.pdf}
    \end{minipage}%
    \caption{Total extracted charge $Q_\text{tot}$ from TDCF as a function of prebias voltage $V_\text{pre}$ for PTB7-Th:ITIC. Similar charge carrier density ensures that the recombination mechanisms in the device are comparable at various excitation energies.}
    \label{SI_FigureS21}
\end{figure}

\clearpage
\section{Photogeneration yield from TDCF and EQE}

\noindent The photocurrent from biased EQE agrees well with EGE from TDCF at higher fields. The discrepancy at lower fields close to $V_\text{oc}$ most probably originates from increased nongeminate recombination during the EQE experiment. Thus, the data from biased EQE can be treated as the measure of CT dissociation yield at high enough applied field. The technique might be sensitive to monitor CT dissociation yield at lower photon energies below the optical gap.
\begin{figure}[ht]
    \centering
    \includegraphics[width=0.45\textwidth]{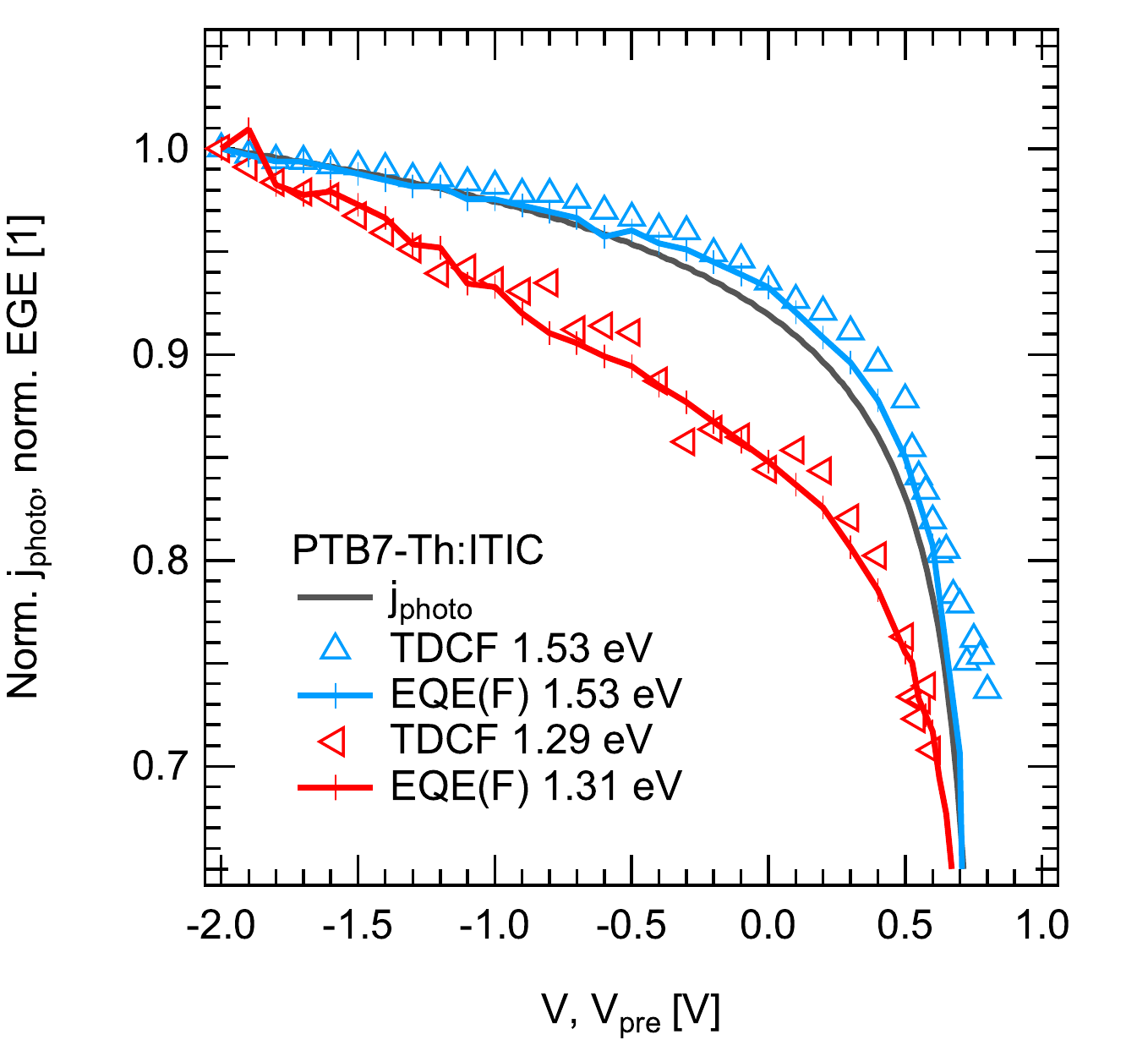}
    \caption{EGE from TDCF compared to the photocurrent from biased EQE.}
    \label{SI_FigureS22}
\end{figure}

\clearpage